\documentclass[]{aa} 

\usepackage{natbib}
\bibpunct{(}{)}{;}{a}{}{,} 
\usepackage{epsfig}

\newcommand{\beq}{\begin{equation}} 
\newcommand{\eeq}{\end{equation}} 
\newcommand{\beqa}{\begin{eqnarray}} 
\newcommand{\eeqa}{\end{eqnarray}} 
\newcommand{\bea}{\begin{array}} 
\newcommand{\ea}{\end{array}} 

\newcommand{\dd}{{\rm d}}

\newcommand{\lag}{\langle} 
\newcommand{\rag}{\rangle} 

\newcommand{\ii}{{\rm i}}

\newcommand{\rhob}{\overline{\rho}}

\newcommand{\vk}{{\bf k}}

\newcommand{\vq}{{\bf q}}

\newcommand{\vx}{{\bf x}}

\newcommand{\vPsi}{{\bf \Psi}}

\newcommand{\tdelta}{{\tilde{\delta}}}
\newcommand{\tF}{{\tilde{F}}}

\newcommand{\tu}{{\tilde{u}}}
\newcommand{\tW}{{\tilde{W}}}

\newcommand{\hn}{{\hat{n}}}

\newcommand{\cF}{{\cal{F}}}

\newcommand{\cO}{{\cal{O}}}

\newcommand{\simlt}{\hspace{0.3em}\raisebox{0.4ex}{$<$}\hspace{-0.75em}\raisebox{-.7ex}{$\sim$}\hspace{0.3em}}

\begin{document} 

\topmargin =0.3cm

\title{Combining perturbation theories with halo models for the matter bispectrum}    
\author{P. Valageas \and T. Nishimichi}   
\institute{Institut de Physique Th\'eorique, CEA Saclay, 91191 Gif-sur-Yvette, France \and Institute for the Physics and Mathematics of the Universe, University of Tokyo, Kashiwa, Chiba 277-8568, Japan}
\date{Received / Accepted } 
 
\abstract
{}
{We investigate how unified models should be built to be able to predict the
matter-density bispectrum (and power spectrum) from very large to small scales
and that are at the same time consistent with perturbation theory at low $k$
and with halo models at high $k$.}
{We use a Lagrangian framework to decompose the bispectrum into ``3-halo'',
``2-halo'', and ``1-halo'' contributions, related to ``perturbative'' and 
``non-perturbative'' terms. We describe a simple implementation of this approach
and present a detailed comparison with numerical simulations.}
{We show that the 1-halo and 2-halo contributions contain counterterms that
ensure their decay at low $k$, as required by physical constraints, and allow
a better match to simulations. Contrary to the
power spectrum, the standard 1-loop perturbation theory can be used for the
perturbative 3-halo contribution because it does not grow too fast at high $k$.
Moreover, it is much simpler and more accurate than two resummation schemes
investigated in this paper. We obtain a good agreement with numerical simulations
on both large and small scales, but the transition scales are poorly described
by the simplest implementation.
This cannot be amended by simple modifications to the halo parameters, but we show 
how it can be corrected for the power spectrum and the
bispectrum through a simple interpolation scheme that is restricted to this
intermediate regime. Then, we reach an accuracy on the order of $10\%$ on mildly
and highly nonlinear scales, while an accuracy on the order of $1\%$ is obtained
on larger weakly nonlinear scales. This also holds for the real-space two-point correlation
function.}
{}

\keywords{gravitation; cosmology: theory -- large-scale structure of Universe}

\maketitle

\section{Introduction} 
\label{Introduction}

According to standard cosmological scenarios, the large-scale structures of the
present Universe have formed through the amplification by gravitational
instability of small almost-Gaussian primordial fluctuations \citep{Peebles1980}.
Then, from observations of the recent Universe, through galaxy surveys
\citep{Tegmark2006,Cole2005}, weak-lensing studies 
\citep{Massey2007,Munshi2008}, measures of baryon acoustic oscillations
\citep{Eisenstein1998,Eisenstein2005}, one can derive constraints on the
cosmological parameters (such as the mean matter and dark energy contents)
and on the properties of the initial perturbations.
The main statistical quantity used in this context is the density two-point
correlation function, or power spectrum in Fourier space. This allows measuring the
scale dependence and the amplitude of the initial conditions, and for Gaussian
fields this fully determines all statistical properties of the matter distribution.
However, to constrain the possible non-Gaussianities of the
initial perturbations, to check the gravitational clustering scenario
(through the measure of the non-Gaussianities generated by the nonlinear
dynamics), and to
break parameter degeneracies, it is useful to study higher order statistics.
The three-point correlation function or bispectrum in Fourier space, which is
the lowest order statistics beyond the Gaussian, is the main quantity studied in this
context 
\citep{Peebles1980,Frieman1994,Frieman1999,Scoccimarro2001a,Bernardeau2002b,Verde2002,Takada2003,Kayo2004,Sefusatti2006,Sefusatti2007,Nishimichi2007,Sefusatti2010,Nishimichi2010}.

On large scales, or at early times, where the density fluctuations are small
within cold dark matter (CDM) scenarios \citep{Peebles1982}, one can
use perturbation theory \citep{Goroff1986,Scoccimarro1997,Scoccimarro1998b,Bernardeau2002}.
In order to extend the validity of perturbative predictions to somewhat smaller
scales and to increase their theoretical accuracy, many resummation schemes
have been proposed recently 
\citep{Crocce2006a,Crocce2006b,Valageas2007a,Matarrese2007,Pietroni2008,Taruya2008,Matsubara2008,Taruya2009}. 
Although these methods follow different routes and involve different expansion
schemes, they are consistent with each other and with the standard perturbation
theory up to the order of truncation, and only differ by higher order contributions.
Thus, they complete the standard perturbation theory contributions (which
consist of a finite number of Feynman diagrams) by (usually infinite) partial
resummations of higher order diagrams. Most of these studies have focused
on the matter power spectrum, except for \citet{Valageas2008}, who also
considers three-point (and higher) density correlations,
\citet{Bernardeau2008}, who study higher order response functions (propagators)
and the bispectrum,
and \citet{Pietroni2010}, who consider both the power spectrum and bispectrum
for Gaussian and non-Gaussian initial conditions.

On small scales, where the density fluctuations are large, one must use numerical
simulations or phenomenological models, such as the Lagrangian mapping
introduced in \citet{Hamilton1991} or the halo model
\citep{McClelland1977,Scherrer1991,Cooray2002}. Both approaches have already been
used for the bispectrum. However, while the first method has only had some
success on intermediate mildly nonlinear scales \citep{Pan2007}, the second
method is also able to describe the highly nonlinear scales \citep{Ma2000b,Scoccimarro2001,Wang2004,Fosalba2005}.

In order to compare theoretical predictions with observations, which go from
large linear scales to small highly nonlinear scales, it is useful to build
unified models, which combine for instance perturbation theories with halo
models. This combines the high accuracy on large scales ($\sim 1\%$) provided
by systematic perturbative expansions with the reasonably good accuracy
on small scales ($\sim 10\%$) provided by the halo model. 
This is particularly useful on the weakly nonlinear scales associated with the
baryon acoustic oscillations, where the matter power spectrum and bispectrum
show small ``wiggles'' that would be difficult to reproduce with a high accuracy
by a simple halo model or fits to simulations (unless one runs a simulation with the
required cosmological parameters).
As a first step in this direction, we have recently built such a unified model
in \citet{Valageas2010b}, focusing on the matter power spectrum.
In this paper, we extend this study to the matter bispectrum, using the same
halo model and perturbative schemes. As for the power spectrum, we explain
that the ``1-halo'' and ``2-halo'' contributions contain new counterterms that
ensure a physical behavior on large scales. Then, we show that we obtain a good
agreement with numerical simulations without introducing new parameters.
We also note that it is possible to improve the predictions for the power spectrum
on mildly nonlinear scales using the shape of the predicted bispectrum.

This paper is organized as follows. 
We first describe in Sect.~\ref{Decomposition} the computation of the
density bispectrum from a Lagrangian point of view, making the link with
halo models and perturbative expansions through the decomposition 
into 1-halo, 2-halo, and 3-halo contributions. 
Then, we present in Sect.~\ref{implementation} a simple implementation,
based on the halo model and the perturbative approaches used in
\citet{Valageas2010b} for the power spectrum. 
We compare our results for the bispectrum with numerical simulations
in Sect.~\ref{Comparison}, for equilateral and isosceles configurations, from
linear to highly nonlinear scales.
Next, we discuss the differences between our work and some previous studies in
Sect.~\ref{Comparison-wit-previous-models}, and
we investigate in Sect.~\ref{ingredients} the dependence of the predictions
on the choice of the perturbative scheme and on the halo parameters.
We explain in Sect.~\ref{Improving} how to use the shape of the predicted
reduced bispectrum to improve the predictions for the power spectrum and
the bispectrum, and we estimate the accuracy of our unified model in
Sect.~\ref{Typical-accuracy} before concluding in Sect.~\ref{Conclusion}.

\section{Decomposition of the density bispectrum from a
Lagrangian point of view}
\label{Decomposition}

We show in this section how the density bispectrum can be split into
3-halo, 2-halo, and 1-halo terms from a Lagrangian point of view, and we
make the connection with perturbation theory.

\subsection{Lagrangian-space formulation}
\label{Lagrangian-space}

In a Lagrangian framework, one considers the trajectories $\vx(\vq,t)$
of the particles, of initial Lagrangian coordinates $\vq$, and Eulerian coordinates
$\vx$ at time $t$. At any time $t$, this defines a mapping,
$\vq\mapsto\vx$, from Lagrangian to Eulerian space, which fully determines the
Eulerian density field $\rho(\vx)$ through the conservation of matter,
\beq
\rho(\vx) \, \dd \vx = \rhob \, \dd\vq ,
\label{continuity}
\eeq
where $\rhob$ is the mean comoving matter density of the Universe and we work
in comoving coordinates.
Then, defining the density contrast as
\beq
\delta(\vx,t) = \frac{\rho(\vx,t)-\rhob}{\rhob} ,
\label{deltadef}
\eeq
and its Fourier transform as 
\beq
\tdelta(\vk) = \int\frac{\dd\vx}{(2\pi)^3} \, e^{-\ii\vk\cdot\vx} \, \delta(\vx) ,
\label{tdelta}
\eeq
we obtain from Eq.(\ref{continuity})
\beq
\tdelta(\vk) = \int\frac{\dd\vq}{(2\pi)^3} \, \left( e^{-\ii\vk\cdot\vx(\vq)} 
- e^{-\ii\vk\cdot\vq} \right) .
\label{tdelta-q}
\eeq
We define the power spectrum, $P(k)$, and the bispectrum, $B(k_1,k_2,k_3)$,
which are the Fourier transforms of the two-point and three-point density
correlation functions, by
\beqa
\lag \tdelta(\vk_1) \tdelta(\vk_2) \rag & = & \delta_D(\vk_1+\vk_2) \, P(k_1) ,
\label{Pkdef} \\
\lag \tdelta(\vk_1) \tdelta(\vk_2) \tdelta(\vk_3) \rag & = & 
\delta_D(\vk_1+\vk_2+\vk_3) \, B(k_1,k_2,k_3) .
\label{Bkdef}
\eeqa
Here we used the fact that the system is statistically homogeneous,
which gives rise to the Dirac prefactors associated with statistical invariance
through translations, and statistically isotropic, which implies that
$P(k_1)$ and $B(k_1,k_2,k_3)$ only depend on the lengths $|\vk_1|$
and $\{|\vk_1|,|\vk_2|,|\vk_3|\}$.

We have already studied the power spectrum in a previous work
\citep{Valageas2010b}, so that we here focus on the bispectrum.
Our goal is to follow the same approach to combine the perturbation theory
and the halo model and to build a simple unified model.
Within this framework, we wish to decompose the bispectrum into ``perturbative''
and ``non-perturbative'' contributions. The former is associated with
configurations of particle triplets (or $p-$uplets for the $p-$point correlation
function) where the particles belong to different halos, whereas the latter is
associated with configurations where several particles belong to the same halo.
As in \citet{Valageas2010b}, our strategy is to evaluate the perturbative
contribution through perturbation theory (either with the standard expansion
or with resummation schemes) and the non-perturbative contribution through
a halo model. Then, the Lagrangian framework allows us to recover the
counterterms that arise in the halo-model contributions and that are necessary
to ensure a good behavior on large scales.
In addition, it is of interest to see how the standard expression for the bispectrum
obtained from an Eulerian point of view \citep{Cooray2002} can be recovered
(with the addition of these new counterterms) from a Lagrangian point of view.

In \citet{Valageas2010b}, where we studied the power spectrum, we could
factorize out the Dirac prefactor of Eq.(\ref{Pkdef}) and write for the power
spectrum the well-known expression \citep{Schneider1995,Taylor1996}
\beq
P(k) = \int\frac{\dd\vq}{(2\pi)^3} \, \lag e^{-\ii \vk \cdot \Delta\vx} - 
e^{\ii\vk\cdot\vq} \rag ,
\label{Pkxq}
\eeq
where we introduced the Eulerian-space separation $\Delta\vx = \vx(\vq) - \vx(0)$.
Thus, Eq.(\ref{Pkxq}) only depends on the relative displacements of the particles,
as it must because uniform translations do not affect the structural properties of the
density field.
Then, Eq.(\ref{Pkxq}) provides a convenient starting point because the Dirac
prefactor of Eq.(\ref{Pkdef}) has already been taken into account and it will not
be put in danger by approximations used in later steps.

For the bispectrum, it is still possible to factorize out the Dirac prefactor of
Eq.(\ref{Bkdef}) by introducing the relative displacements. This yields
\beqa
B(k_1,k_2,k_3) & = & \int\frac{\dd\vq_2\vq_3}{(2\pi)^6} \, 
\lag e^{-\ii \vk_2 \cdot \Delta\vx_2-\ii \vk_3 \cdot \Delta\vx_3} \rag 
\nonumber \\
&& \hspace{-1cm} - \delta_D(\vk_1) P(k_2) - \delta_D(\vk_2) P(k_3)
- \delta_D(\vk_3) P(k_1) \nonumber \\
&& \hspace{-1cm} - \delta_D(\vk_1) \delta_D(\vk_2) ,
\label{Bk-Dirac}
\eeqa
where we introduced $\Delta\vx_2 = \vx(\vq_2) - \vx(0)$ and
$\Delta\vx_3 = \vx(\vq_3) - \vx(0)$.
However, the symmetry over $\{k_1,k_2,k_3\}$ is no longer transparent in
Eq.(\ref{Bk-Dirac}) and it is not very convenient to count the triplets that are
within one, two, or three halos. Therefore, we prefer to work directly with the
third-order moment (\ref{Bkdef}), although this will require an additional
approximation to those needed to compute the power spectrum in
\citet{Valageas2010b}.
Thus, using Eq.(\ref{tdelta-q}) we write
\beqa
\lag \tdelta(\vk_1) \tdelta(\vk_2) \tdelta(\vk_3) \rag & = & \lag 
\int\frac{\dd\vq_1\dd\vq_2\vq_3}{(2\pi)^9} \, \left( e^{-\ii\vk_1\cdot\vx_1} 
- e^{-\ii\vk_1\cdot\vq_1} \right) \nonumber \\
&& \hspace{-1.9cm} \times \left( e^{-\ii\vk_2\cdot\vx_2} 
- e^{-\ii\vk_2\cdot\vq_2} \right) \left( e^{-\ii\vk_3\cdot\vx_3} 
- e^{-\ii\vk_3\cdot\vq_3} \right) \rag ,
\label{Bk-x3q3}
\eeqa
where $\vx_j=\vx(\vq_j)$.
Next, we split the average (\ref{Bk-x3q3}) into three contributions, associated
with the cases where the triplet $\{\vq_1,\vq_2,\vq_3\}$ belongs to either
one, two, or three halos:
\beqa
\lag \tdelta(\vk_1) \tdelta(\vk_2) \tdelta(\vk_3) \rag & = & 
\lag \tdelta(\vk_1) \tdelta(\vk_2) \tdelta(\vk_3) \rag_{1\rm H} \nonumber \\
&& \hspace{-2.3cm} + \lag \tdelta(\vk_1) \tdelta(\vk_2) \tdelta(\vk_3)
\rag_{2\rm H} + \lag \tdelta(\vk_1) \tdelta(\vk_2) \tdelta(\vk_3) \rag_{3\rm H} .
\label{B-123H}
\eeqa
This assumes that all the matter is contained within halos, so that these three
contributions sum up to Eq.(\ref{Bk-x3q3}), counting all particles.

\subsection{``1-halo'' contribution}
\label{1-halo}

Let us first consider the 1-halo contribution. It can be written as
\beqa
\lag \tdelta(\vk_1) \tdelta(\vk_2) \tdelta(\vk_3) \rag_{1\rm H} & \! = \! & \lag \int
\prod_{j=1}^3 \frac{\dd\vq_j}{(2\pi)^3} \, \left( e^{-\ii\vk_j\cdot\vx_j} 
- e^{-\ii\vk_j\cdot\vq_j} \right) \nonumber \\
&& \times \sum_{\alpha} \prod_{j=1}^3 \theta( \vq_{j} \in M_{\alpha}) \; \rag ,
\label{B-1H-1}
\eeqa
where the sum runs over all halos, labeled by the index $\alpha$, and the
top-hat factors $\theta( \vq_{j} \in M_{\alpha})$ are unity if the particle $\vq_j$
belongs to the halo $\alpha$ and zero otherwise. This clearly gives the contribution
to Eq.(\ref{Bk-x3q3}) of the configurations where the triplet $\{\vq_1,\vq_2,\vq_3\}$
belongs to a single halo.
Equation (\ref{B-1H-1}) also reads as
\beqa
\lag \tdelta(\vk_1) \tdelta(\vk_2) \tdelta(\vk_3) \rag_{1\rm H} & \! = \! & \lag \int
\prod_{j=1}^3 \frac{\dd\vq_j}{(2\pi)^3} \, \left( e^{-\ii\vk_j\cdot\vx_j} 
- e^{-\ii\vk_j\cdot\vq_j} \right) \nonumber \\
&& \hspace{-2.5cm} \times \int \dd\vq^{\rm c} \dd M \,
\hn(\vq^{\rm c},M) \, \prod_{j=1}^3
\theta( \vq_{j} \in M) \; \rag ,
\label{B-1H-2}
\eeqa
where $\hn(\vq^{\rm c},M)$ is the halo number density in a given realization
of the initial conditions (as denoted by the hat, in contrast with its mean),
\beq
\hn(\vq^{\rm c},M) = \sum_{\alpha} 
\delta_D(\vq^{\rm c}-\vq^{\rm c}_{\alpha}) \, \delta_D(M-M_{\alpha}) .
\label{hn-def}
\eeq
Here $\vq^{\rm c}_{\alpha}$ is the ``center'' of the halo $\alpha$, which may be
defined as the halo center of mass, in Lagrangian space (i.e., in the initial or linear
density field). Note that in Eq.(\ref{B-1H-2}) we have chosen
to count halos in Lagrangian space, by their Lagrangian center
$\vq^{\rm c}_{\alpha}$, but we could as well count them in Eulerian space
by using the Eulerian-space center of mass $\vx^{\rm c}_{\alpha}$.
The mean of the halo number density does not depend on location, thanks to
statistical homogeneity, and it is given by the halo mass function,
\beq
\lag  \hn(\vq^{\rm c},M) \rag = n(M) ,
\label{nM-def}
\eeq
which we write as
\beq
n(M) \dd M = \frac{\rhob}{M} f(\nu) \frac{\dd\nu}{\nu} , \;\; \mbox{with} \;\;
\nu = \frac{\delta_L}{\sigma(M)} .
\label{nM-fnu}
\eeq
As usual, we have introduced in Eq.(\ref{nM-fnu}) the scaling function $f(\nu)$ and
the reduced variable $\nu$, where $\sigma(M)$ is the rms linear density contrast
at scale $M$, or Lagrangian radius $q_M$, with
\beq
\sigma(M) = \sigma(q_M)  \;\; \mbox{with} \;\; M= \rhob \frac{4\pi}{3} q_M^3 ,
\label{Mq}
\eeq
and
\beq
\sigma^2(q_M) = 4\pi \int_0^{\infty} \dd k \, k^2 P_L(k) \tW(kq_M)^2 ,
\label{sigma2-def}
\eeq
where $\tW(kq_M)$ is the Fourier transform of the top-hat of radius $q_M$,
defined as
\beq
\tW(kq_M) \! = \! \int_{V_M}\!\frac{\dd \vq}{V_M} \, e^{\ii \vk\cdot\vq}
= 3 \, \frac{\sin(kq_M) \! - \! kq_M\cos(kq_M)}{(kq_M)^3} .
\label{tWdef}
\eeq
In the second Eq.(\ref{nM-fnu}), the linear density contrast $\delta_L$ is related
to the nonlinear density threshold $\delta$ that defines the halos through the
spherical collapse dynamics, as $\delta = \cF(\delta_L)$, see \citet{Valageas2009d}.
Thus, Eq.(\ref{B-1H-2}) also writes as
\beqa
\lag \tdelta(\vk_1) \tdelta(\vk_2) \tdelta(\vk_3) \rag_{1\rm H} & = &
\int \dd\vq^{\rm c} \frac{\dd\nu}{\nu} \, \frac{\rhob}{M} f(\nu) \nonumber \\
&& \hspace{-2.5cm} \times \lag  \int_{V_M} \prod_{j=1}^3
\frac{\dd\vq_j}{(2\pi)^3} \, \left( e^{-\ii\vk_j\cdot\vx_j} -
e^{-\ii\vk_j\cdot\vq_j} \right) \rag_{\vq^{\rm c},M} ,
\label{B-1H-3}
\eeqa
where the average $\lag .. \rag_{\vq^{\rm c},M}$ is the conditional average
under the constraint that the three particles $\vq_j$ belong to the same
halo of center $\vq^{\rm c}$, mass $M$, and Lagrangian volume $V_M$.
As in \citet{Valageas2010b}, we make the approximation of fully virialized halos,
that is, the particle $\vq$ has lost all memory of its initial location and velocity
and it is located at random within the halo. This means that in Eulerian space
the average is defined by the density profile of the halo,
$\rho_{\vx^{\rm c},M}(\vx)$, as
\beq
\lag e^{-\ii\vk\cdot\vx_j} \rag_{\vq^{\rm c},M} = 
\frac{1}{M} \int \dd \vx \, e^{-\ii\vk\cdot\vx} \, \rho_M(|\vx-\vx^{\rm c}|) ,
\label{mean-x-def}
\eeq
because $\vq_j$ can be identified with a uniform probability to all the particles
that make up the halo (approximation of complete relaxation).
Here we also made the approximation of spherically symmetric halos, and
introducing the normalized Fourier transform of the halo radial profile,
\beq
\tu_M(k) = \frac{\int\dd\vx \, e^{-\ii\vk\cdot\vx} \rho_M(x)}{\int\dd\vx \,
\rho_M(x)} = \frac{1}{M} \int\dd\vx \, e^{-\ii\vk\cdot\vx} \rho_M(x) ,
\label{uM-k-def}
\eeq
we obtain
\beq
\lag e^{-\ii\vk\cdot\vx_j} \rag_{\vq^{\rm c},M} = e^{-\ii\vk\cdot\vx^{\rm c}}
\, \tu_M(k) .
\label{mean-x-1}
\eeq
Making the approximation that these halos are also spherically symmetric
in Lagrangian space, we can use Eq.(\ref{tWdef}) for the factors
$e^{-\ii\vk_j\cdot\vq_j}$ and Eq.(\ref{B-1H-3}) reads as
\beqa
\lag \tdelta(\vk_1) \tdelta(\vk_2) \tdelta(\vk_3) \rag_{1\rm H} & \! = \! &
\int \dd\vq^{\rm c} \frac{\dd\nu}{\nu} \, \frac{\rhob}{M} f(\nu) \,
e^{-\ii(\vk_1+\vk_2+\vk_3)\cdot\vq^{\rm c}} \nonumber \\
&& \hspace{-2.9cm} \times \left( \frac{M}{\rhob(2\pi)^3} \right)^{\!3}
\prod_{j=1}^3 \left( e^{-\ii\vk_j\cdot\vPsi^{\rm c}} \tu_M(k_j)  -
\tW(k_jq_M) \right) ,
\label{B-1H-4}
\eeqa
where we introduced the displacement, $\vPsi^{\rm c}= \vx^{\rm c} - \vq^{\rm c}$,
of the center of mass of the halo.
We have written Eqs.(\ref{mean-x-def}) and (\ref{mean-x-1})-(\ref{B-1H-4}) as if
the relation $\vx^{\rm c}(\vq^{\rm c})$ were deterministic, but a priori
we should take the average over the displacement $\vPsi^{\rm c}$ in
Eq.(\ref{B-1H-4}). First, we note that $\lag e^{-\ii\vk_j\cdot\vPsi^{\rm c}} \rag$
does not depend on $\vq^{\rm c}$, thanks to statistical homogeneity, so that
the integral over $\vq^{\rm c}$ in Eq.(\ref{B-1H-4}) gives the expected
Dirac factor $\delta_D(\vk_1+\vk_2+\vk_3)$, in agreement with Eq.(\ref{Bkdef}).
Then, the displacement $\vPsi^{\rm c}$ only appears in mixed terms, such
as $e^{-\ii(\vk_1+\vk_2)\cdot\vPsi^{\rm c}}\tu_M(k_1)\tu_M(k_2)\tW(k_3q_M)$.
Going back to Eq.(\ref{tdelta-q}), we can see that the term $\tW(k_3q_M)$
arises from a factor $\delta_D(\vk_3)$, so that in the regime where the
1-halo contribution is dominant, that is $\lag \tdelta(\vk_1) \tdelta(\vk_2)
\tdelta(\vk_3) \rag \simeq \lag \tdelta(\vk_1) \tdelta(\vk_2) \tdelta(\vk_3)
\rag_{1\rm H}$, it should reduce to $\delta_D(\vk_3)$. In Eq.(\ref{B-1H-4})
this corresponds to the fact that $|\tW(k_3q_M)| \ll 1$ for $k_3 \gg 1/q_M$.
Then, thanks to the prefactor $\delta_D(\vk_1+\vk_2+\vk_3)$ we can
see that in this regime $|\vk_1+\vk_2|\rightarrow 0$ and
$e^{-\ii(\vk_1+\vk_2)\cdot\vPsi^{\rm c}} \rightarrow 1$.
In order to satisfy these properties, we simply make the approximation
$\vPsi^{\rm c}=0$, that is, we neglect the displacements of halos.
Then, performing the integral over $\vq^{\rm c}$, which allows the factorization of
the Dirac factor $\delta_D(\vk_1+\vk_2+\vk_3)$, we obtain the 1-halo
contribution to the bispectrum as
\beqa
B_{1\rm H}(k_1,k_2,k_3) & = & \int \frac{\dd\nu}{\nu} \, f(\nu) \,
\left( \frac{M}{\rhob(2\pi)^3} \right)^{\!2} \nonumber \\
&& \times \prod_{j=1}^3 \left( \tu_M(k_j)  - \tW(k_jq_M) \right) .
\label{B-1H-5}
\eeqa

Thus, as for the power spectrum studied in \citet{Valageas2010b}, we recover
the standard expression of the 1-halo contribution to the bispectrum
derived within an Eulerian framework, see \citet{Cooray2002,Scoccimarro2001},
with the addition of the new counterterms associated with the factors
$\tW$. Note that for the power spectrum we obtained in \citet{Valageas2010b}
a factor $(\tu_M(k)^2-\tW(kq_M)^2)$, instead of the factor
$[\tu_M(k)-\tW(kq_M)]^2$ that we would obtain using the approach described
above. The difference arises because for the power spectrum we
used Eq.(\ref{Pkxq}), where the Dirac prefactor $\delta_D(\vk_1+\vk_2)$ has
already been factorized, whereas for the bispectrum we used Eq.(\ref{Bk-x3q3})
instead of Eq.(\ref{Bk-Dirac}). 
The advantage of Eqs.(\ref{Pkxq}) and (\ref{Bk-Dirac}) is that they only involve
relative positions, $\Delta\vx$, so that Eulerian position $\vx^{\rm c}$ of the
halo never appears (for particles located in the same halo) and we do not need
to introduce any approximation for the halo displacement $\vPsi^{\rm c}$.

However, for higher-order functions, such as the bispectrum or the trispectrum,
the symmetric equation (\ref{Bk-x3q3}) (and its $N$-point generalization)
provides a simpler starting point. In particular, the counting of the volumes
over $\{\vq_1,..,\vq_N\}$ associated with a single halo simply factorizes as
$V_M^N$, whereas introducing relative positions makes the geometrical
countings slightly more intricate.

The factors $[\tu_M(k_j)  - \tW(k_jq_M)]$ cannot be interpreted as
modifications of the halo profiles $\rho_M(x)$ under the form of compensated
profiles (the result $\rho_M(x)-\rhob$ would actually be negative at large $x$).
Indeed, both quantities do not have the same radius in real space, and
actually apply to two different spaces, namely the Eulerian and Lagrangian spaces.

A second important feature is that the expression (\ref{B-1H-1})
automatically ensures that the 1-halo contribution decays at least as
$B_{1\rm H}(k_1,k_2,k_3) \propto k_j^2$ for any $k_j\rightarrow 0$,
since no linear dependence on $k_j$ can remain because of statistical
homogeneity and isotropy.
This can be checked on the final expression (\ref{B-1H-5}), since
$\tu_M(0)=\tW(0)=1$ and the functions $\tu_M(k)$ and $\tW(kq_M)$
are regular functions of $k^2$ at the origin,
\beq
k_j\rightarrow 0 : \;\;\; B_{1\rm H}(k_1,k_2,k_3) \propto k_j^2 .
\label{kj-0}
\eeq
This agrees with the requirements associated with a small-scale redistribution
of matter. Here, there is a simplification as compared with the matter
power spectrum $P(k)$. Indeed, as discussed in \citet{Peebles1974},
small-scale redistributions of matter generically give a low-$k$ tail
$\tdelta(\vk) \propto k$, whence  $P(k) \propto k^2$, while taking into
account momentum conservation implies the steeper decay
$\tdelta(\vk) \propto k^2$, whence  $P(k) \propto k^4$.
Thus, in \citet{Valageas2010b} we recovered the expected $k^2$ tail, through the
factor $(\tu_M(k)^2-\tW(kq_M)^2)$, since our analysis did not enforce momentum
conservation. Using the approach described above, we would actually obtain
a $k^4$ tail, through the factor $[\tu_M(k)-\tW(kq_M)]^2$, once we make the
approximation $\vPsi^{\rm c}=0$. This clearly satisfies momentum
conservation (the peculiar momentum of each halo is zero) but comes at
the expense of an additional approximation and one should not give too much
weight to this property.
By contrast, for the bispectrum (and higher-order correlations) small-scale
redistributions of matter lead to the same $k_j^2$ tail, whether we take into
account momentum conservation or not, because the linear term $k_j$ vanishes
by symmetry.
Note that when all wavenumbers go to zero at the same rate, we have
from Eq.(\ref{B-1H-5}) the decrease
\beq
\lambda\rightarrow 0 : \;\;\; B_{1\rm H}(\lambda k_1,\lambda k_2,\lambda k_3)
\propto \lambda^6 .
\label{lambda-0}
\eeq
As noticed in previous works \citep{Cooray2002}, in the usual version of the
halo model the 1-halo contribution to the bispectrum does not contain the
counterterms $\tW(k_jq_M)$, so that it goes to a strictly positive constant
at large scales and dominates over the lowest-order prediction of perturbation
theory. This led to rather inaccurate predictions on large scales
(an overestimate of $20\%$ at $k\sim 0.01h$Mpc$^{-1}$), because for CDM
cosmologies the perturbative prediction scales as $B_{\rm pert} \sim 
P_L(k)^2 \propto k^{2n_s}$ with $n_s\simeq 1$ (for $k_j\sim k$ for all $j$).
As seen from Eqs.(\ref{kj-0})-(\ref{lambda-0}), our approach solves this problem
(we will return to this point in Sect.~\ref{Comparison-wit-previous-models}).

\subsection{``2-halo'' contribution}
\label{2-halo}

We now turn to the 2-halo contribution to the bispectrum, following the
procedure described in the previous section for the 1-halo term.
Thus, as in Eq.(\ref{B-1H-2}) we can write the 2-halo contribution as
\beqa
\lag \tdelta(\vk_1) \tdelta(\vk_2) \tdelta(\vk_3) \rag_{2\rm H} & \! = \! & \lag \int
\prod_{j=1}^3 \frac{\dd\vq_j}{(2\pi)^3} \, \left( e^{-\ii\vk_j\cdot\vx_j} 
- e^{-\ii\vk_j\cdot\vq_j} \right) \nonumber \\
&& \hspace{-2.5cm} \times \int_{\vq^{\rm c}_1\neq\vq^{\rm c}_2}
\dd\vq^{\rm c}_1 \dd M_1 \, \dd\vq^{\rm c}_2 \dd M_2 \, 
\hn(\vq^{\rm c}_1,M_1) \, \hn(\vq^{\rm c}_2,M_2) \nonumber \\
&& \hspace{-2.5cm} \times \theta( \vq_1 \in M_1) \, \theta( \vq_2 \in M_2)
\, \theta( \vq_3 \in M_2)  \; \rag + 2 \, {\rm cyc.} ,
\label{B-2H-1}
\eeqa
where we sum over all pairs of distinct halos, of mass $M_1$ and $M_2$,
which contain one and two of the particles $\vq_j$.
Here we have written the contribution where the halo $M_1$
contains the particle $\vq_1$, and the two additional contributions, noted
``2 cyc.'', correspond to the cases where $M_1$ contains the particle $\vq_2$
or $\vq_3$. Using again the approximation (\ref{mean-x-1}) of fully relaxed
halos, the average of Eq.(\ref{B-2H-1}) writes as
\beqa
\lag \tdelta(\vk_1) \tdelta(\vk_2) \tdelta(\vk_3) \rag_{2\rm H} \! & \! = \! & \!
\int\! \dd\vq^{\rm c}_1 \frac{\dd\nu_1}{\nu_1} \, \dd\vq^{\rm c}_2
\frac{\dd\nu_2}{\nu_2} \, \frac{\rhob^2}{M_1M_2} f(\nu_1) f(\nu_2)
\nonumber \\
&& \hspace{-2.5cm} \times \, \left( 1 + \xi_{12} \right)
\frac{M_1 M_2^2}{\rhob^3(2\pi)^9} \, e^{-\ii\vk_1\cdot\vq^{\rm c}_1} \,
e^{-\ii(\vk_2+\vk_3)\cdot\vq^{\rm c}_2} \nonumber \\
&& \hspace{-2.5cm} \times
\left( e^{-\ii\vk_1\cdot\vPsi^{\rm c}_1} \, \tu_{M_1}(k_1)  - \tW(k_1q_{M1}) \right)
\nonumber \\
&& \hspace{-2.5cm} \times \prod_{j=2}^3 
\left( e^{-\ii\vk_j\cdot\vPsi^{\rm c}_2} \, \tu_{M_2}(k_j)  - \tW(k_jq_{M2}) \right)
 + 2 \, {\rm cyc.} ,
\label{B-2H-2}
\eeqa
where $\xi_{12}(|\vq^{\rm c}_2-\vq^{\rm c}_1|)$ is the two-point correlation
function of halos of mass $M_1$ and $M_2$, in Lagrangian space.
As a first step, let us neglect halo motions, $\vPsi^{\rm c}=0$. Then,
as expected, the contribution associated with the factor $1$ in $(1+\xi_{12})$
vanishes (because the integral over $\vq^{\rm c}_1$ yields $\delta_D(\vk_1)$
and $[\tu_{M_1}(k_1)  - \tW(k_1q_{M1})]=0$ for $k_1=0$), so that only the
term associated with the large-scale correlation of halos remains.
Writing this two-point correlation in terms of the halo power spectrum,
\beq
\xi_{12}(|\vq^{\rm c}_2-\vq^{\rm c}_1|) = \int\dd\vk \,
e^{\ii\vk\cdot(\vq^{\rm c}_2-\vq^{\rm c}_1)} \, P_{12}(k) ,
\label{P12-def}
\eeq
the integrals over $\vq^{\rm c}_1$ and $\vq^{\rm c}_2$ give the Dirac factors
$\delta_D(\vk+\vk_1) \delta_D(\vk-\vk_2-\vk_3)$. Therefore, we recover the
Dirac prefactor $\delta_D(\vk_1+\vk_2+\vk_3)$, which we can factorize out
to write the 2-halo contribution to the bispectrum as
\beqa
B_{2\rm H}(k_1,k_2,k_3) & = & \int \frac{\dd\nu_1}{\nu_1} \frac{\dd\nu_2}{\nu_2}
\frac{M_2}{\rhob(2\pi)^3} \, f(\nu_1) f(\nu_2) \, P_{12}(k_1)  \nonumber \\
&& \hspace{-2.5cm} \times \left( \tu_{M_1}(k_1)  - \tW(k_1q_{M1}) \right)
\prod_{j=2}^3 \left( \tu_{M_2}(k_j)  - \tW(k_jq_{M2}) \right) \nonumber \\
&& \hspace{-2.5cm} + 2 \, {\rm cyc.} 
\label{B-2H-3}
\eeqa
Then, as is customary, we could write the halo power spectrum as
$P_{12}(k) = b(M_1) b(M_2) P(k)$, where $b(M)$ is a mass-dependent bias
factor (of order unity for typical halos) that we approximate as scale-independent
(thereby neglecting exclusion constraints between halos).

However, the expression (\ref{B-2H-3}) is not really satisfactory.
Indeed, for $k_1\rightarrow 0$ it decreases as $k_1^2P(k_1)$ because of the
prefactor $[\tu_{M_1}(k_1)  - \tW(k_1q_{M1})]$, while we would expect to recover
a behavior proportional to $P(k_1)$. Indeed, in this regime this 2-halo contribution
should describe the power that arises from the large-scale correlation between
density fluctuations at a position $\vq_1$ and at a position $\vq_2$, with
$|\vq_2-\vq_1| \sim 2\pi/k_1$, the fluctuations at $\vq_2$ being furthermore
decomposed over smaller-scale fluctuations within single halos of mass $M_2$.
In other words, the large-scale power $P(k_1)$ should not be damped by the
prefactor $[\tu_{M_1}(k_1)  - \tW(k_1q_{M1})]$.
Going back to Eq.(\ref{B-2H-2}), we can see that the product
$\tu_{M_1}(k_1)\tu_{M_2}(k_2)\tu_{M_2}(k_3)$ involves a prefactor
$e^{\ii\vk_1\cdot(\vPsi^{\rm c}_2-\vPsi^{\rm c}_1)}$, which only depends on
relative displacements and as such is physically meaningful and should be
taken into account.
This should be contrasted with the behavior obtained for the 1-halo contribution
(\ref{B-1H-4}), where the product of three terms $\tu_M$ gave the prefactor
$e^{\ii(\vk_1+\vk_2+\vk_3)\cdot\vPsi^{\rm c}}$, which had to disappear
because it does not depend on relative displacements, and was indeed
irrelevant thanks to the Dirac prefactor $\delta_D(\vk_1+\vk_2+\vk_3)$.
This means that the approximation $\vPsi^{\rm c}=0$ is not as good for the
2-halo contribution as for the 1-halo contribution, because it neglects
relative halo displacements, which did not appear in the latter case
(i.e. its consequences are stronger in the former case).

In order to have a 2-halo contribution that behaves in a reasonable fashion,
we choose to use instead of Eq.(\ref{B-2H-3}) the simple expression
\beqa
B_{2\rm H}(k_1,k_2,k_3) & = &  P_L(k_1) \int \frac{\dd\nu}{\nu}
\frac{M}{\rhob(2\pi)^3} \, f(\nu)  \nonumber \\
&& \hspace{-1.cm} \times 
\prod_{j=2}^3 \left( \tu_M(k_j)  - \tW(k_jq_M) \right) + 2 \, {\rm cyc.} 
\label{B-2H-4}
\eeqa
Thus, we have made the approximation $P_{12}(k)\simeq P_L(k)$, where
$P_L(k)$ is the linear matter density power spectrum, and we have set
the spurious prefactor $[\tu_{M_1}(k_1)  - \tW(k_1q_{M1})]$ to unity.
In principle, instead of the linear power $P_L(k)$ we could include higher orders
of perturbation theory. However, in view of the approximate nature of
Eq.(\ref{B-2H-4}) this is not really necessary, especially since the 2-halo contribution
is subdominant on most scales of interest, as shown by the numerical
results in Sect.~\ref{Comparison}.

In the large-scale limit we obtain from Eq.(\ref{B-2H-4}) the asymptotic behaviors
\beq
k_j\rightarrow 0 : \;\;\; B_{2\rm H}(k_1,k_2,k_3) \propto P_L(k_j) ,
\label{B-2H-kj}
\eeq
when only one wavenumber goes to zero, and
\beq
\lambda\rightarrow 0 : \;\;\; B_{2\rm H}(\lambda k_1,\lambda k_2,\lambda k_3)
\propto \lambda^4 P_L(\lambda) ,
\label{B-2H-lambda}
\eeq
when all wavenumbers decrease at the same rate. In Eq.(\ref{B-2H-kj}) we used the
fact that for CDM cosmologies we have $P_L(k) \sim k^{n_s}$ with $n_s\simeq 1$
at low $k$.
It is interesting to note that the behavior (\ref{B-2H-kj}) agrees with perturbation
theory, as shown in App.~\ref{bispectrum-for-one-low-wavenumber}
and Eq.(\ref{B3-pert-F-2}).
There, we note that this scaling, which applies when only one of the wavenumbers
$k_j$ goes to zero, for instance $k_1$, is valid at all orders of perturbation theory.
In this regime, the 2-halo contribution, which is non-perturbative (as discussed in
Sect.~\ref{3-halo} below), is therefore on the same order (i.e., $\sim P_L(k_1)$) as
the perturbative contribution. This can be understood from the arguments
developed above
to obtain the expression (\ref{B-2H-4}), as the non-perturbative effects associated
with the formation of small-scale nonlinear structures around the nearby points  
$\vx_2$ and $\vx_3$ (or $\vq_2$ and $\vq_3$ in Lagrangian space) do not modify
the larger-scale correlation with the very distant point $\vx_1$.
That is, the non-perturbative effects associated with the formation of a halo
correspond to a local redistribution of matter.
Then, at leading order, non-perturbative terms simply lead to a ``renormalization''
of the coefficient that multiplies the factor $P_L(k_1)$, by an amount that depends
on how far in the nonlinear regime the wavenumbers $k_2$ and $k_3$ are.

When all wavenumbers go to zero, the property (\ref{B-2H-lambda}) again
ensures that the 2-halo contribution becomes negligible as compared with
the lowest order term of perturbation theory, which scales as $P_L(\lambda)^2$.
Thus, as for the 1-halo contribution, the counterterms $\tW$ of Eq.(\ref{B-2H-4})
solve the problem encountered with the usual implementation of the halo
model, which does not recover perturbation theory on very large scales.

\subsection{``3-halo'' contribution as a perturbative contribution}
\label{3-halo}

\begin{figure}
\begin{center}
\epsfxsize=7.8 cm \epsfysize=5.6 cm {\epsfbox{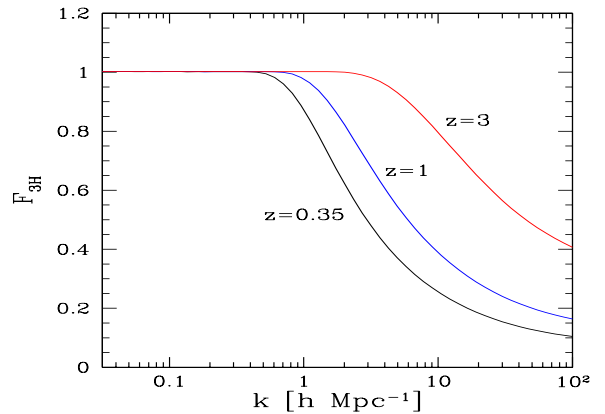}}
\end{center}
\caption{Probability $F_{3\rm H}$ that the triplets associated with the
equilateral bispectrum $B_{\rm eq}(k)$ belong to three different halos, from
Eq.(\ref{F3H-def}), at redshifts $z=0.35, 1,$ and $3$.}
\label{fig_F3H}
\end{figure}

For the 3-halo contribution we follow the spirit of the approach described
in \citet{Valageas2010b}, as we bypass the halo model to make contact with
standard perturbation theory. Indeed, as we have seen above for the 2-halo
term, it is not straightforward to take into account large-scale correlations
within this Lagrangian implementation of the halo model, starting with
the expression (\ref{Bk-x3q3}). This would require some modeling of relative
displacements, which may not be worth the effort (if one wishes to improve
in a significant manner over the approximation (\ref{B-3H-1}) used below).
Thus, we take a much more direct and simpler route, noting that the
1-halo and 2-halo contributions vanish at all orders of perturbation theory.
Here, we mean by perturbation theory any expansion over powers of the
linear power spectrum $P_L$ (for the Gaussian initial conditions that we
consider in this article), such as the standard perturbation theory derived
within the fluid approximation to the equations of motion
\citep{Goroff1986,Bernardeau2002}.
Indeed, the halo mass function shows a large-mass tail of the form
$e^{-\nu^2/2} = e^{-\delta_L^2/(2\sigma^2(M))}$, which vanishes
exponentially for $P_L\rightarrow 0$ and has a Taylor expansion over powers
of $P_L$ that is identically zero.
Therefore, the remaining 3-halo contribution to Eq.(\ref{B-123H}) must be consistent
with perturbation theory up to all orders, and we simply use the approximation
\beq
B_{3\rm H}(k_1,k_2,k_3) = B_{\rm pert}(k_1,k_2,k_3) ,
\label{B-3H-1}
\eeq
where $B_{\rm pert}$ is the bispectrum obtained from perturbation theory.
To follow more closely the approach used in \citet{Valageas2010b} for the
power spectrum, we should multiply $B_{\rm pert}$ in Eq.(\ref{B-3H-1}) by
a factor $F_{3\rm H}$, with $0<F_{3\rm H}<1$, which counts the fraction
of triplets that are located within three distinct halos. From the same arguments
we have $F_{3\rm H}=1$ at all orders of perturbation theory, and it only differs
from unity by non-perturbative terms such as $e^{-\delta_L^2/(2\sigma^2(M))}$,
where $M$ is typically the mass scale associated with the maximum of
$\{2\pi/k_j\}$.
For illustration purposes, we display in Fig.~\ref{fig_F3H} the following estimate of
$F_{3\rm H}$,
\beq
F_{3\rm H}(k) = \left( \int_0^{\nu_k} \frac{\dd\nu}{\nu} \, f(\nu) \right)^3 ,
\label{F3H-def}
\eeq
where $\nu_k=\delta_L/\sigma(q_k)$ with $q_k=2\pi/k$. From Eq.(\ref{nM-fnu}),
this is the probability that three particles belong to halos of size smaller than
$q_k$, neglecting halo correlations, and it should give an estimate of the
range where $F_{3\rm H}\simeq 1$ (for the equilateral bispectrum, which
involves a single wavenumber $k$).
As expected, the comparison with Fig.~\ref{fig_Bk_eq} below shows that the
departure from unity of $F_{3\rm H}$ roughly corresponds to the scale where
the 1-halo and 2-halo terms become of the same order as the 3-halo term.
In view of the approximations involved in these 1-halo and 2-halo contributions,
we do not try to include the deviations from unity of $F_{3\rm H}$, which only
play a role in the transition regime.
Therefore, we make the approximation $F_{3\rm H} \simeq 1$ and use the
simple prescription (\ref{B-3H-1}).

In practice, the perturbative term $B_{\rm pert}$ is not exactly known 
(assuming the perturbative series is convergent), and one must truncate the
perturbative expansion. There are many ways to do so, because a priori one can
use the standard perturbation theory or any alternative expansion scheme,
which corresponds to partial resummations of the standard diagrams. 
As shown in \citet{Valageas2010b}, for the power spectrum it happens that
standard perturbation theory is not a good choice, because higher order terms
grow increasingly fast on small scales and prevent a good description of the
highly nonlinear regime. Then, one must use resummation schemes that agree
with standard perturbation theory up to the required order while remaining
well-behaved on small scales.
As we will check in Sect.~\ref{Comparison}, for the bispectrum it turns out
that at one-loop
order the standard perturbation theory remains small at high $k$ compared with
the 1-halo and 2-halo contributions, which makes this an efficient choice; but we
will also study the direct steepest-descent resummation scheme described in
detail in \citet{Valageas2010b} for the computation of the power spectrum.

In most studies that are based on the halo model, the 3-halo term is rather
written as \citep{Ma2000b,Scoccimarro2001,Cooray2002}
\beq
B^{\rm h.m.}_{3\rm H}(k_1,k_2,k_3) \! = \left[ \prod_{i=1}^3\lag b(M) \tu_M(k_i)\rag
\! \right] B_{\rm tree}(k_1,k_2,k_3) ,
\label{B3H-hm}
\eeq
where $\lag b(M) \tu_M\rag$ is the average over mass, weighted by the halo mass
function, of the bias parameter $b$ and halo profile $\tu_M$; $B_{\rm tree}$ is
the matter bispectrum obtained at lowest order through perturbation theory,
given by Eq.(\ref{Ba-def}) below.
In our approach, we do not introduce this halo bias parameter because we
simply write the 3-halo term as the perturbative matter bispectrum (\ref{B-3H-1}).
This implies that we focus on the matter bispectrum because our approach does
not contain the halo bispectrum
$B^{\rm h.m.}_{3\rm H}(k_1,k_2,k_3;M_1,M_2,M_3)$ as an intermediate tool.
Therefore, to compute the bispectrum of halos of given masses we would
need to add another prescription. On the other hand, this makes our model
for the matter bispectrum simpler and more robust because it does not rely
on bias parameters that are fairly difficult to compute in a systematic and
well-controlled manner (because halos themselves do not appear in a natural
fashion in the equations of motion). Moreover, Eq.(\ref{B-3H-1}) ensures that
our results agree with perturbation theory up to the order of truncation.
In this paper, we will go up to 1-loop order, while most previous studies
that involved the halo model only used the tree-level contribution, as
in Eq.(\ref{B3H-hm}).

\subsection{Comparison with the Eulerian framework}
\label{Eulerian}

As seen in the previous sections, the derivation of the bispectrum from
a Lagrangian implementation of the halo model is not as straightforward
nor as ``clean'' as for the power spectrum \citep{Valageas2010b}.
Indeed, we had to introduce some additional approximation regarding the halo
displacements $\vPsi^{\rm c}$, and for the 2-halo term that mixes larger-scale
and intra-halo wavenumbers we had to partially bypass the naive prediction
of this halo model to recover the large-scale behavior (\ref{B-2H-kj}).

More generally, it is not surprising that the Lagrangian implementation of the halo
model requires more steps than the usual Eulerian version.
Indeed, within an Eulerian framework we only need to describe the density
field at the time of interest $t$. Within a Lagrangian framework, we need to
add some information on the building of this density field, that is, the mapping
between the initial coordinates $\vq$ and the current positions $\vx$ of the
particles. This necessarily implies that a practical Lagrangian implementation
requires more hypothesis or approximations (because there are more
intermediate quantities, even though in principle the results would be the same
if we made no approximation at all).
However, we think it remains of interest to describe how $N$-point correlation
functions or poly-spectra can be constructed within such a Lagrangian
framework, based on the halo model, as we have done above.
First, it is useful to see how identical or similar approximate models can be
derived from different points of view, since this gives more weight to the
results. Second, it provides additional insight on the underlying approximations
and it could offer an alternative route to more precise modeling.
Third, it provides a natural derivation of the counterterms $\tW$ in the 1-halo
contribution (\ref{B-1H-5}), which ensure the large-scale behaviors
(\ref{kj-0})-(\ref{lambda-0}) that were missed in previous Eulerian implementations.

As explained above, for the 2-halo contribution we had to take some liberty
with a naive implementation of the halo model to recover the
asymptotic scaling (\ref{B-2H-kj}). However, in view of the approximate nature
of these Lagrangian halo models, we think it is best to be guided by physical
arguments and to make sure physical constraints are satisfied, instead of
strictly adhering to approximate models that can show a varying degree of
accuracy, depending on the quantities under scrutiny.

However, as in \citet{Valageas2010b}, it is interesting to note that the counterterms
of Eqs.(\ref{B-1H-5}) and (\ref{B-2H-4}) might also be guessed within an Eulerian
framework, from the requirement that the bispectrum
should vanish for a perfectly homogeneous universe. Indeed, with the usual
halo-model approximations, we can still split a constant-density medium over the
same set of ``halos'', or cells, of radius $q_M$ and center $\vq^{\rm c}$
(neglecting geometrical constraints associated with the impossibility of covering
a 3D space with spherical cells). The only difference is that because these
``halos'' have not changed and have remained at the constant density
$\rhob$, their Eulerian and Lagrangian properties are the same.
In particular, the Eulerian radius $x_M$ is equal to $q_M$, and the halo profiles
are simply given by $\tu_M(k)=\tW(kq_M)$ for this uniform system.
Then, the counterterms in Eqs.(\ref {B-1H-5}) and (\ref{B-2H-4}) clearly satisfy the
constraint $B(k_1,k_2,k_3)=0$ for this homogeneous universe. 
However, by itself this argument is not sufficient to imply the precise form
of Eq.(\ref {B-1H-5}) (nor of Eq.(\ref{B-2H-4})), since the choice
$\prod_j\tu_M(k_j) - \prod_j\tW(k_jq_M)$ would also satisfy this constraint
(but not the properties (\ref{kj-0})).

Although the main reason for taking into account the counterterms $\tW$
in the 1-halo and 2-halo contributions is to satisfy physical requirements,
this is also needed to reach high accuracy on weakly nonlinear scales.
Thus, \citet{Guo2009} find that on scales on the order of $0.1 h$Mpc$^{-1}$
their 1-halo and 2-halo terms (which do not include these counterterms) are
still too large and spoil the agreement of (tree-order) perturbation theory with
numerical simulations.
Giving a faster decay on large scales of these 1-halo and 2-halo contributions,
the counterterms $\tW$ improve the agreement of the analytical
predictions with the simulations, as we will check in
Sect.~\ref{Comparison-wit-previous-models}.

\section{A simple implementation}
\label{implementation}

We briefly describe in this section the numerical implementation of our
model for the matter density bispectrum.
Further details can be found in \citet{Valageas2010b}.

\subsection{Halo properties}
\label{Halo-properties}

For numerical computations we use the same halo model as the one
used in \citet{Valageas2010b} for the power spectrum. Thus, the halo mass
function is given by \citep{Valageas2009d}
\beq
f(\nu) = 0.502 \left[ (0.6\,\nu)^{2.5}+(0.62\,\nu)^{0.5} \right] \, e^{-\nu^2/2} ,
\label{f-fit}
\eeq
which is normalized to unity and provides a good match to numerical simulations.
We choose the usual NFW profile for the halo density profile $\rho_M(x)$
\citep{NFW1997}, which also has an explicit form for its Fourier transform
$\tu_M(k)$ \citep{Scoccimarro2001}.
In particular, we truncate the halo profiles at the density contrast
$\delta(<r_{200})=200$, which defines their radius $r_{200}$.
For the concentration parameter we take
\beq
c(M_{200}) = 10.04 \, \left( \frac{M_{200}}
{2\times 10^{12} h^{-1} M_{\odot}}\right)^{-0.1} \, (1+z)^{-0.8} ,
\label{cM-1}
\eeq
which is similar to the behaviors measured in numerical simulations
\citep{Dolag2004,Duffy2008}, and was found in \citet{Valageas2010b}
to provide a good tool for the density power spectrum (in this sense, it
would also describe to some degree the effect of halo substructures).

\subsection{Perturbative contribution}
\label{Perturbative-contribution}

\begin{figure}[htb]
\begin{center}
\epsfxsize=7 cm \epsfysize=8 cm {\epsfbox{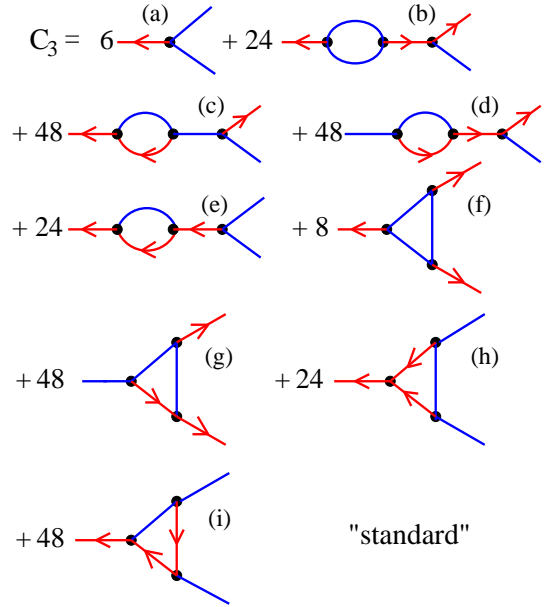}}
\end{center}
\caption{Diagrams associated with the standard perturbation theory
as obtained from a path-integral formalism up to 1-loop order for the
three-point correlation $C_3$. Although they are different from
those obtained by the standard approach, their sum at each order is identical
to the sum of the standard diagrams of the same order over $P_L(k)$.
The lines are the linear two-point correlation $C_L$ (blue solid line) and the
linear response $R_L$ (red solid line with an arrow).
The black dots are the three-leg vertex $K_s$ that enters the quadratic term
of the equation of motion.
The numbers are the multiplicity factors of each diagram.
The tree-order diagram (a) gives Eq.(\ref{Ba-def}), for the density bispectrum,
while the 1-loop diagrams (b),..,(i), give the contribution of order $P_L^3$.}
\label{fig_C3-1loop}
\end{figure}

For the perturbative contribution (\ref{B-3H-1}) to the bispectrum, we investigate
both the standard perturbation theory and the ``direct steepest-descent''
resummation scheme introduced (among other perturbative expansions)
in \citet{Valageas2007a}.
We only go up to 1-loop order for both methods, so that the first approach
only contains all 1-loop diagrams, while the second approach also includes
partial resummations of diagrams at all orders.

A detailed description of these methods for two-point and three-point
functions is given in \citet{Valageas2008}. To facilitate the theoretical
comparison between both approaches it is convenient to write them
with the same diagrammatic language. Then, the diagrams associated with
standard perturbation theory are shown in Fig.~\ref{fig_C3-1loop}, where the
solid lines are the linear two-point functions, either the linear correlation
(lines without arrows) or the linear response (lines with an arrow that shows
the flow of time). The three-leg vertex is the kernel $K_s$ that appears in the
equation of motion, which can be written as $\cO\cdot\psi=K_s\cdot\psi\psi$,
where $\cO$ is a linear operator and $\psi$ is a two-component vector that
describes both the density and velocity fields.
The diagrams shown in Fig.~\ref{fig_C3-1loop} are not those
associated with the usual description of standard perturbation theory,
which is described in App.~\ref{bispectrum-for-one-low-wavenumber},
but of course they are equivalent.
The expansion of Fig.~\ref{fig_C3-1loop} applies to the three-point
correlation $C_3=\lag\psi\psi\psi\rag$, which contains the density
and velocity three-point functions, as well as their cross correlations,
but in this article we only consider the density three-point correlation,
that is, the matter density bispectrum in Fourier space.

The usual description of standard perturbation theory arises
from the expansion of the density and velocity fields over powers of
the linear field $\delta_L$, as in Eqs.(\ref{delta_n})-(\ref{Fn-def}). Then,
the $N$-point correlations are obtained by taking the Gaussian average of
the product of these expansions, as in Eq.(\ref{B-def-pert}), using Wick's theorem
as in Eq.(\ref{B3-pert-F-1}). This yields diagrams with vertices $F_n$ that
have an increasing number $n$ of legs as one goes to higher orders, see
also Fig.~\ref{fig_B_pert}.
The diagrams of Fig.~\ref{fig_C3-1loop} are obtained from a path-integral
formalism, where the averages over the initial conditions are already taken
and one directly works with correlation and response functions. Then, expanding
over powers of the nonlinear interaction vertex $K_s$ gives the expansion
of Fig.~\ref{fig_C3-1loop}.
It looks quite different from the standard diagrams, since only one vertex
appears, i.e., the three-leg kernel $K_s$, whatever the order of the expansion.
On the other hand, in addition to the linear power spectrum these diagrams
also involve the linear response function.
Then, the order of the expansion corresponds to the number of loops of the
diagrams.

Since both expansions can also be written as expansions over powers of the
linear power spectrum $P_L(k)$, they are actually equivalent, at each order.
However, at a given order, there can be a different number of diagrams
between both expansions, and it is only the two sums over all diagrams
of that order over $P_L(k)$ that coincide.
In particular, it can be seen that diagrams (h) and (i) in Fig.~\ref{fig_C3-1loop}
show an ultraviolet divergence for linear power spectra with a large-$k$ tail
$P_L(k) \propto k^n$ with $n\geq -3$. However, these two divergences
cancel out and the sum is finite for $n<-1$, as in the case of the standard
diagrams (see \citet{Valageas2008} for details).

The lowest-order contribution, or ``tree-order'' term, is given by diagram (a),
which yields the well-known result
\beqa
B^{(a)}(k_1,k_2,k_3) & \! = \! & P_L(k_2) P_L(k_3)
\left[\frac{10}{7}+\left(\frac{k_2}{k_3}\!+\!\frac{k_3}{k_2}\right) 
\frac{\vk_2\cdot\vk_3}{k_2 k_3} \right.  \nonumber \\
&& \left.  + \frac{4(\vk_2\cdot\vk_3)^2}{7 k_2^2 k_3^2} \right] + 2 \, {\rm cyc.}
\label{Ba-def}
\eeqa
The explicit expressions of the 1-loop order diagrams (b),..,(i), which are on the
order of $P_L^3$ (as shown by the three blue solid lines which they contain),
are given in App.A of \citet{Valageas2008}.
For numerical computations it is convenient to perform analytically integrations
over angles instead of directly implementing these expressions into
the codes.
Although this requires some care (since angular integrations may yield
trigonometric or hyperbolic functions, depending on the amplitude of the
wavenumbers), there is no fundamental difficulty.

\begin{figure}[htb]
\begin{center}
\epsfxsize=7 cm \epsfysize=6 cm {\epsfbox{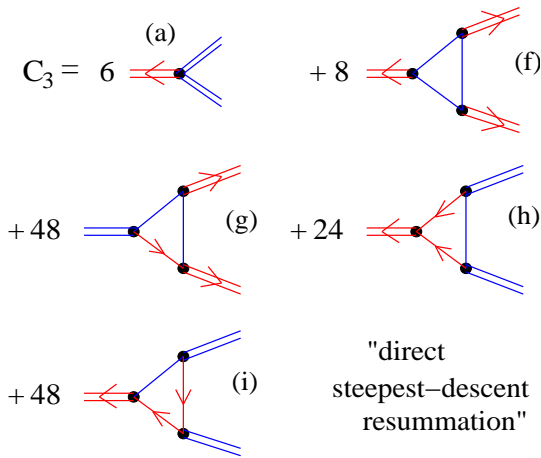}}
\end{center}
\caption{Diagrams obtained at 1-loop order for the three-point correlation
$C_3$ by the ``direct steepest-descent'' resummation scheme.
The double lines are the nonlinear two-point functions $C$ and $R$,
while the single lines are the linear two-point functions $C_L$ and $R_L$
as in Fig.~\ref{fig_C3-1loop}. The black dots are again the three-leg vertex
$K_s$.}
\label{fig_C3-sd}
\end{figure}

\begin{figure}[htb]
\begin{center}
\epsfxsize=7 cm \epsfysize=6 cm {\epsfbox{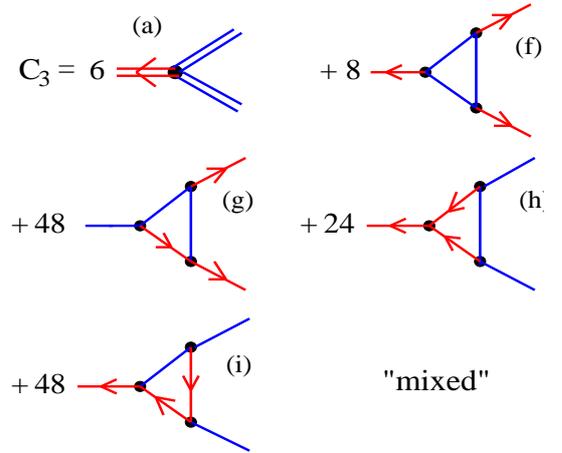}}
\end{center}
\caption{Diagrams obtained at 1-loop order for the three-point correlation
$C_3$ within a ``mixed'' case. We keep the resummed diagram (a) of
Fig.~\ref{fig_C3-sd} but we replace diagrams (f), (g), (h), and (i) of
Fig.~\ref{fig_C3-sd} by their leading contributions, given by diagrams (f), (g), (h),
and (i) of Fig.~\ref{fig_C3-1loop}.}
\label{fig_C3-mixed}
\end{figure}

Of course, it is also possible (and more common) to use the standard diagrams
for the computation of the bispectrum within standard perturbation theory
\citep{Scoccimarro1997,Scoccimarro1998b}.
As noticed above, for our purposes the interest of the description of
Fig.~\ref{fig_C3-1loop} is that it clarifies the link with the
``direct steepest-descent'' resummation scheme that we also investigate
in this article. We focus on this specific resummation scheme to be consistent
with our previous study for the power spectrum in \citet{Valageas2010b}.
In addition, as discussed in that article (in particular in its Appendix A), this
method allows a fast numerical implementation.
Moreover, its predictions for the bispectrum (and higher order correlations),
and more precisely the structure of its diagrammatic expansion,
have already been studied in \citet{Valageas2008}, so that no further theoretical
work is required.
Then, within this ``direct steepest-descent'' resummation scheme, the
diagrams for the bispectrum up to 1-loop order are those shown in
Fig.~\ref{fig_C3-sd}. The difference with Fig.~\ref{fig_C3-1loop} is that
we now have double-line two-point functions. They correspond to the nonlinear
two-point correlation and response functions, as given by the same method
up to the same 1-loop order. The single lines are the linear two-point functions,
as in Fig.~\ref{fig_C3-1loop}.

Within this resummation scheme, nonlinear two-point functions at ``1-loop''
order actually contain an infinite number of ``bubble'' diagrams, as shown
in Figs.~8 and 9 of \citet{Valageas2008}. There, the label ``1-loop'' does not
mean that we only include 1-loop diagrams, as in the standard perturbation
theory of Fig.~\ref{fig_C3-1loop}, but that the diagrammatic expansion
is only complete up to 1-loop (i.e., although we include some diagrams at all orders,
we miss some contributions at 2-loop and higher orders).
Then, substituting the expression of the nonlinear two-point functions
in terms of the linear functions, we can check that the diagram (a)
of Fig.~\ref{fig_C3-sd} actually contains the five diagrams (a),..,(e), of 
Fig.~\ref{fig_C3-1loop}, as well as an infinite number of higher order diagrams.
On the other hand, the diagrams (f),..,(i), of Fig.~\ref{fig_C3-sd} contain
their counterparts of Fig.~\ref{fig_C3-1loop}: each diagram among
(f),..,(i), of Fig.~\ref{fig_C3-1loop} is the lowest order contribution to the
corresponding diagram in Fig.~\ref{fig_C3-sd}, where nonlinear two-point
functions are replaced by their linear counterparts.

At the order $P_L^3$, which corresponds to the 1-loop diagrams of
Fig.~\ref{fig_C3-1loop}, we can also consider the ``mixed'' case shown
in Fig.~\ref{fig_C3-mixed}, where we keep the resummed diagram (a) of
Fig.~\ref{fig_C3-sd}, but we use for diagrams (f),..,(i), their lowest order terms
shown in Fig.~\ref{fig_C3-1loop}.
Of course, these three choices, drawn in Figs.~\ref{fig_C3-1loop}, \ref{fig_C3-sd},
and \ref{fig_C3-mixed}, agree up to order $P_L^3$, and only differ by the
number of higher order terms that are included.

Thus, for the perturbative (3-halo) contribution (\ref{B-3H-1}) we will
consider the three alternatives shown in Figs.~\ref{fig_C3-1loop}, \ref{fig_C3-sd},
and \ref{fig_C3-mixed}. Contrary to the numerical computations used
in \citet{Valageas2008}, we do not introduce any approximation for the
computation of these diagrams. Therefore, the perturbative term
$B_{\rm pert}(k_1,k_2,k_3)$ contains no free parameter and is exact,
up to 1-loop order (or to the truncation order of the perturbative scheme).
This is an improvement over most previous studies involving the halo model
\citep{Ma2000b,Fosalba2005},
which only used the tree-order bispectrum (\ref{Ba-def}) for the 3-halo term,
with the addition of bias factors (which we do not introduce in our approach).

\section{Comparison with numerical simulations}
\label{Comparison}

\subsection{Numerical procedure}
\label{simulations}

We use a set of large $N$-body simulations in a $\Lambda$CDM universe,
consistent with the five-year observation of the WMAP satellite \citep{Komatsu2009}, 
which are described in \citet{Valageas2010b}.
We analyze the four main simulations ($N=2048^3$), together with supplementary
smaller simulations ($N=1024^3, 512^3$ and $256^3$) for convergence tests.

We measure the bispectrum from a fast Fourier transformation of the density field
obtained by the cloud-in-cells interpolation of the $N$-body particles, 
at $z=0.35$, $1$ and $3$. 
In doing so, we use the folding scheme to speed up the measurements at large
wavenumber bins without systematic error from the interpolation 
(see \citet{Valageas2010b} and also \citet{Colombi2009}). 
We set the wavenumber bins for $k_1$, $k_2$, and $k_3$, logarithmically
with $2$ bins per factor $2$.

The statistical uncertainties (at $1$-$\sigma$ level) of the measurements are
estimated assuming that they mainly come from their Gaussian part
\citep{Scoccimarro1998b},
\beqa
\left[\Delta B(k_1,k_2,k_3)\right]^2 =  \frac{s_{123} V}{(2\pi)^3 2 N_{\rm tri}}
P(k_1)P(k_2)P(k_3) ,
\label{errorbar}
\eeqa
where the factor $s_{123}$ is $6$ for equilateral triangles, $2$ for isosceles
triangles and $1$ otherwise.  
In the above, $V$ denotes the volume of the simulations and $N_{\rm tri}$ is the
number of Fourier space triangles that fall into the bin of $(k_1,k_2,k_3)$.

We do not correct for the shot-noise contributions to the power spectrum and the
bispectrum because the naive expectation for the shot noise obtained by assuming
a Poisson process does not seem very accurate, especially on large scales.
Because we start the simulation from a regular lattice, the discreteness noise is
boxed in on scales smaller than the inter-particle distance at the beginning.
After gravitational evolution, the density fluctuations neatly follow the linear growth
rate at $k\simlt 0.1h$Mpc$^{-1}$, hence we conclude that we do not have any sign
of shot noise on those large scales.
On small scales, however, we can see that our measurements from simulations
approach the Poisson noise \citep[e.g.,][]{Verde2002},
\beqa
P_{\rm shot}(k) &=& \frac{1}{(2\pi)^3} \frac{L_{\rm box}^3}{N} ,
\label{shotnoise_P}\\
B_{\rm shot}(k_1,k_2,k_3) &=& \frac{1}{(2\pi)^6}
\left[ \frac{L_{\rm box}^3}{N} \left(P(k_1)+P(k_2)+P(k_3)\right)\right.\nonumber\\
&&\left.-2\left(\frac{L_{\rm box}^3}{N}\right)^2\right],
\label{shotnoise_B}
\eeqa
for the power spectrum and the bispectrum, respectively.
Note that the shot noise is included in $P(k_1)$, $P(k_2)$, and $P(k_3)$ of
Eq.~(\ref{shotnoise_B}).
Instead of subtracting these contributions from the measured spectra, we assess the
shot-noise level from Eqs.~(\ref{shotnoise_P}) and (\ref{shotnoise_B}), and then plot
their relative importance in Figs.~\ref{fig_dP} and \ref{fig_dBk_eq} below.
For the reduced bispectrum, we estimate the shot noise level from
\beqa
Q_{\rm eq,\,shot}(k) &=& \left|\frac{B_{\rm eq}(k)-B_{\rm eq,\,shot}(k)}
{3\left(P(k)-P_{\rm shot}\right)^2} - \frac{B_{\rm eq}(k)}{3P(k)^2}\right|,
\label{shotnoise_Q}
\eeqa
which is plotted in Fig.~\ref{fig_dQk_eq} below, where again $P(k)$ and
$B_{\rm eq}(k)$ are the measured power spectrum and bispectrum (and thus
include the shot-noise contribution and other numerical errors, as compared with
the theoretical values).

We now compare our model, described in Sects.~\ref{Decomposition}
and \ref{implementation}, with these numerical simulations.

\subsection{Bispectrum $B(k_1,k_2,k_3)$}
\label{Comp-Bispec}

\subsubsection{Equilateral triangles}
\label{Equilateral}

\begin{figure*}
\begin{center}
\epsfxsize=6.1 cm \epsfysize=5.4 cm {\epsfbox{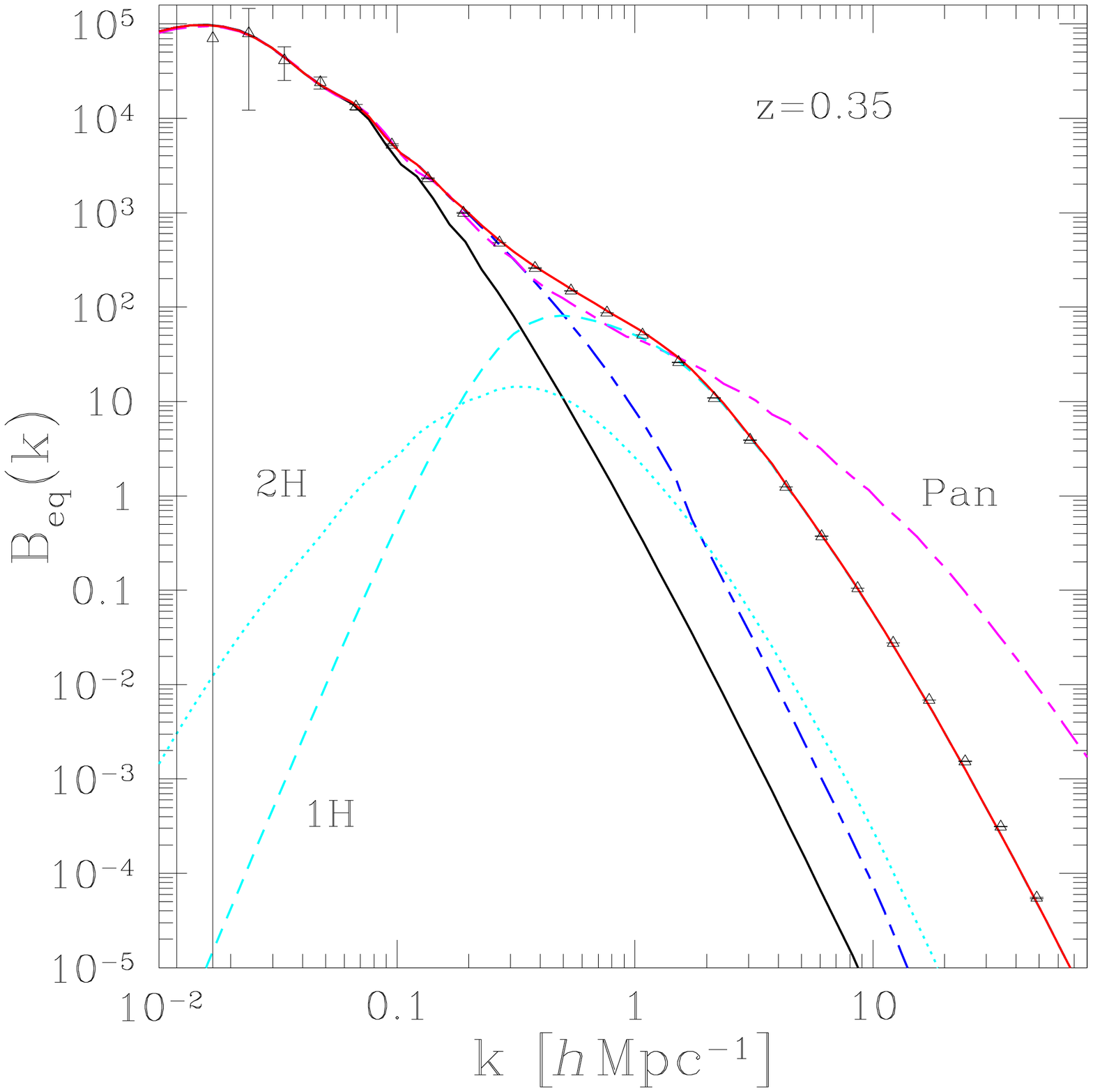}}
\epsfxsize=6.05 cm \epsfysize=5.4 cm {\epsfbox{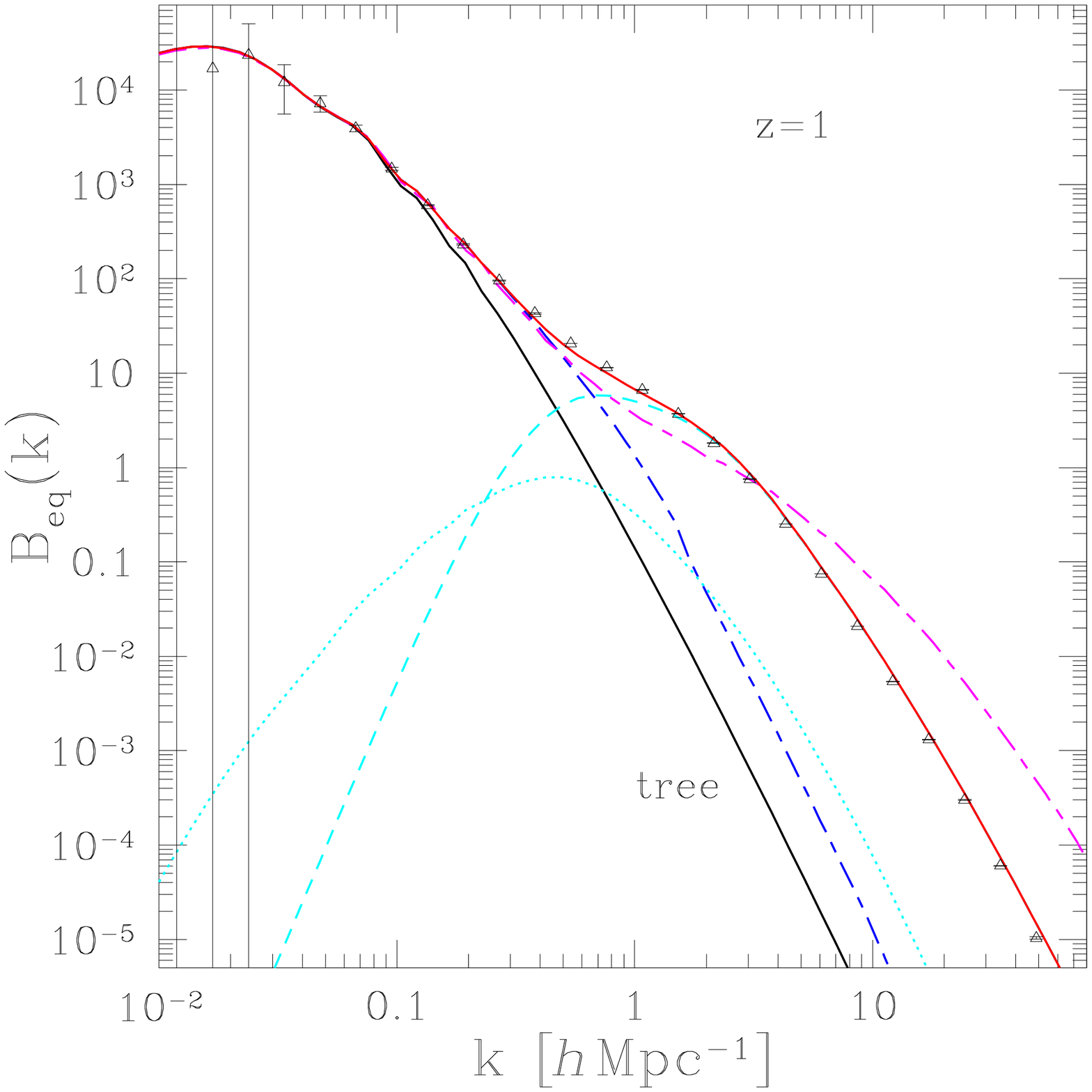}}
\epsfxsize=6.05 cm \epsfysize=5.4 cm {\epsfbox{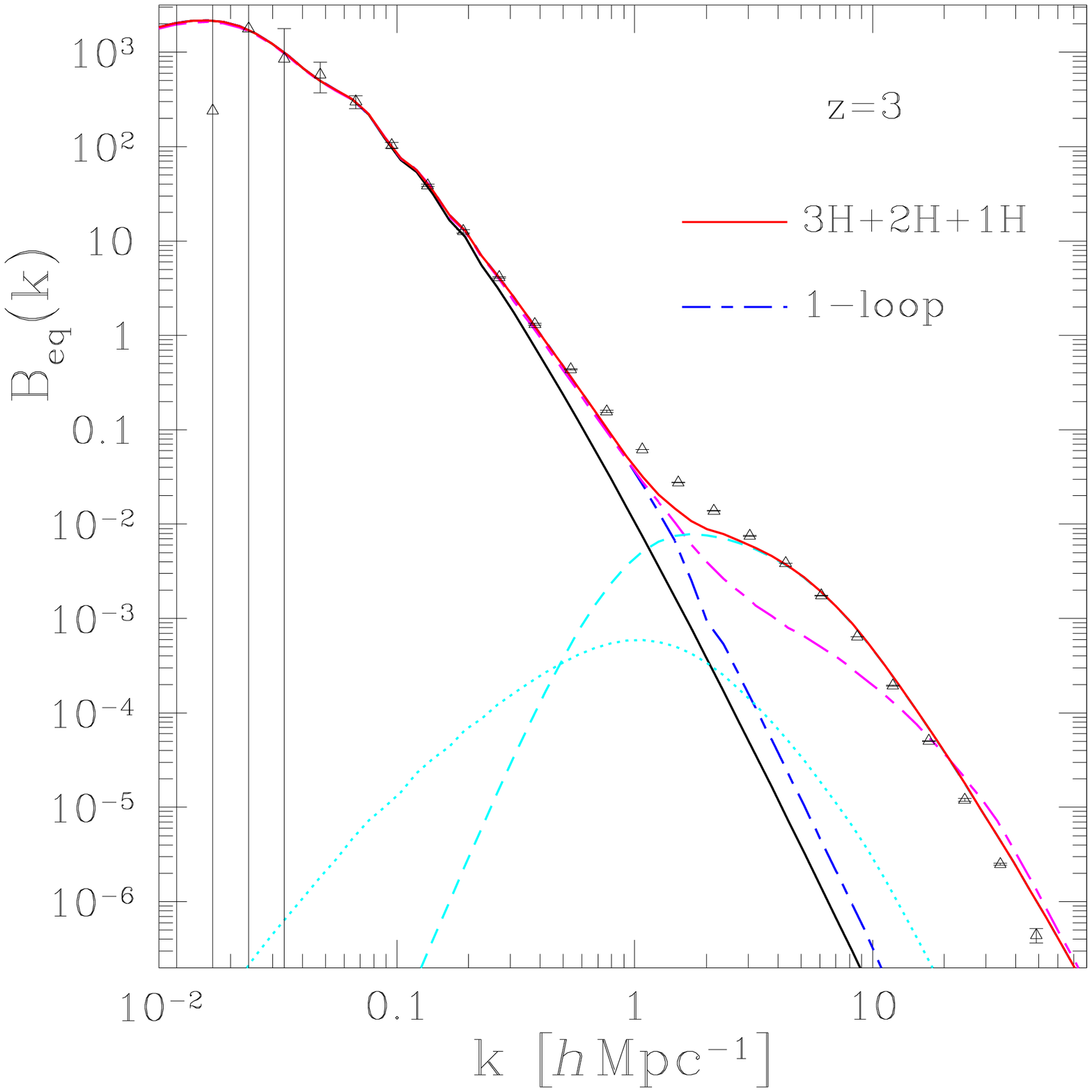}}
\end{center}
\caption{Bispectrum $B_{\rm eq}(k)=B(k,k,k)$ for equilateral configurations,
at redshifts $z=0.35, 1$, and $3$.  The symbols are the results from the numerical
simulations. We show the bispectrum obtained at tree order (black solid line),
standard 1-loop order (blue dot-dashed line), using the full model (red solid line),
as well as the 2-halo (cyan dotted line) and 1-halo (cyan dashed line) contributions.
The magenta dot-dashed line labeled ``Pan'' is the phenomenological model of
\citet{Pan2007}.}
\label{fig_Bk_eq}
\end{figure*}

\begin{figure*}
\begin{center}
\epsfxsize=6.1 cm \epsfysize=5.4 cm {\epsfbox{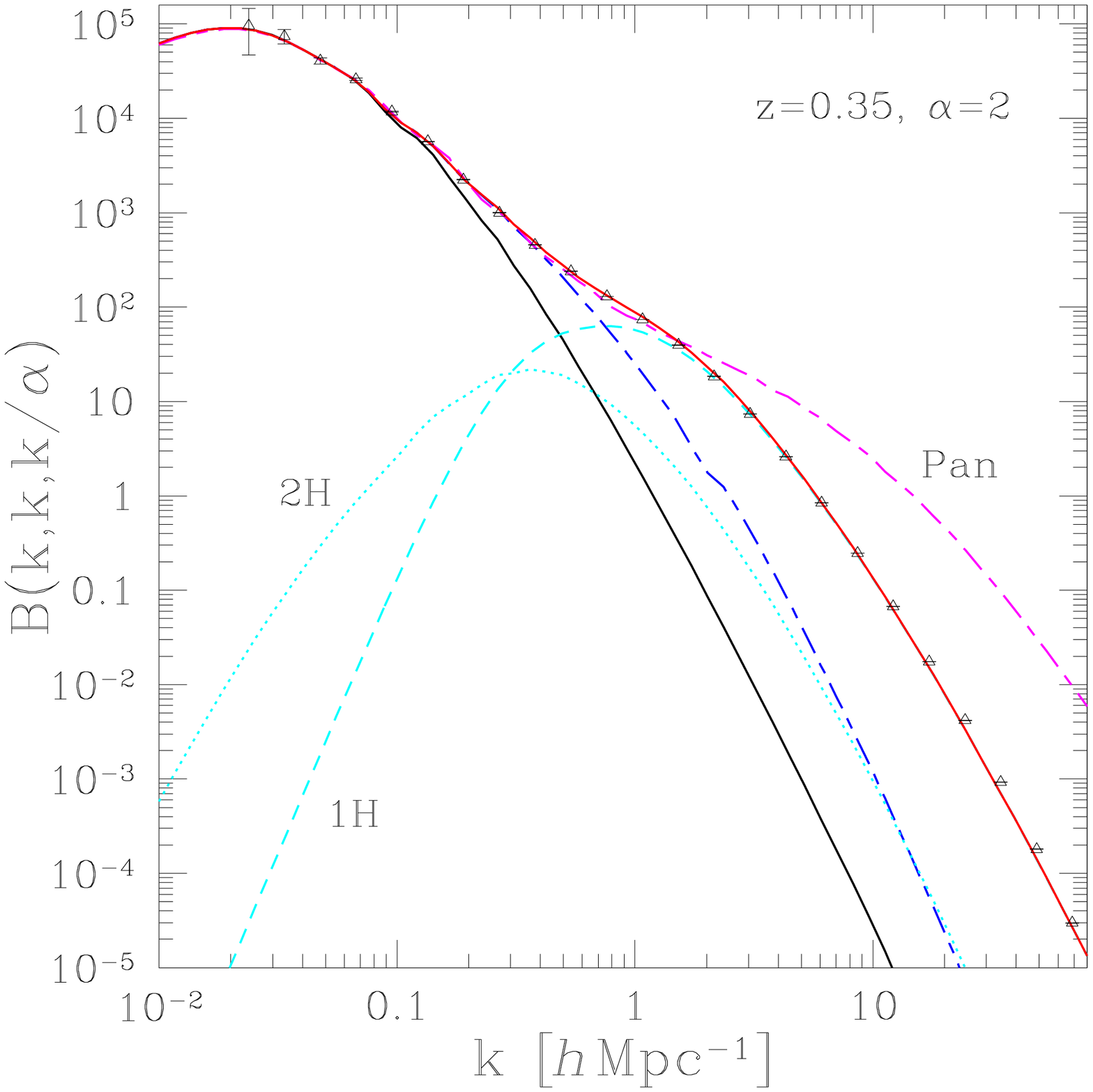}}
\epsfxsize=6.05 cm \epsfysize=5.4 cm {\epsfbox{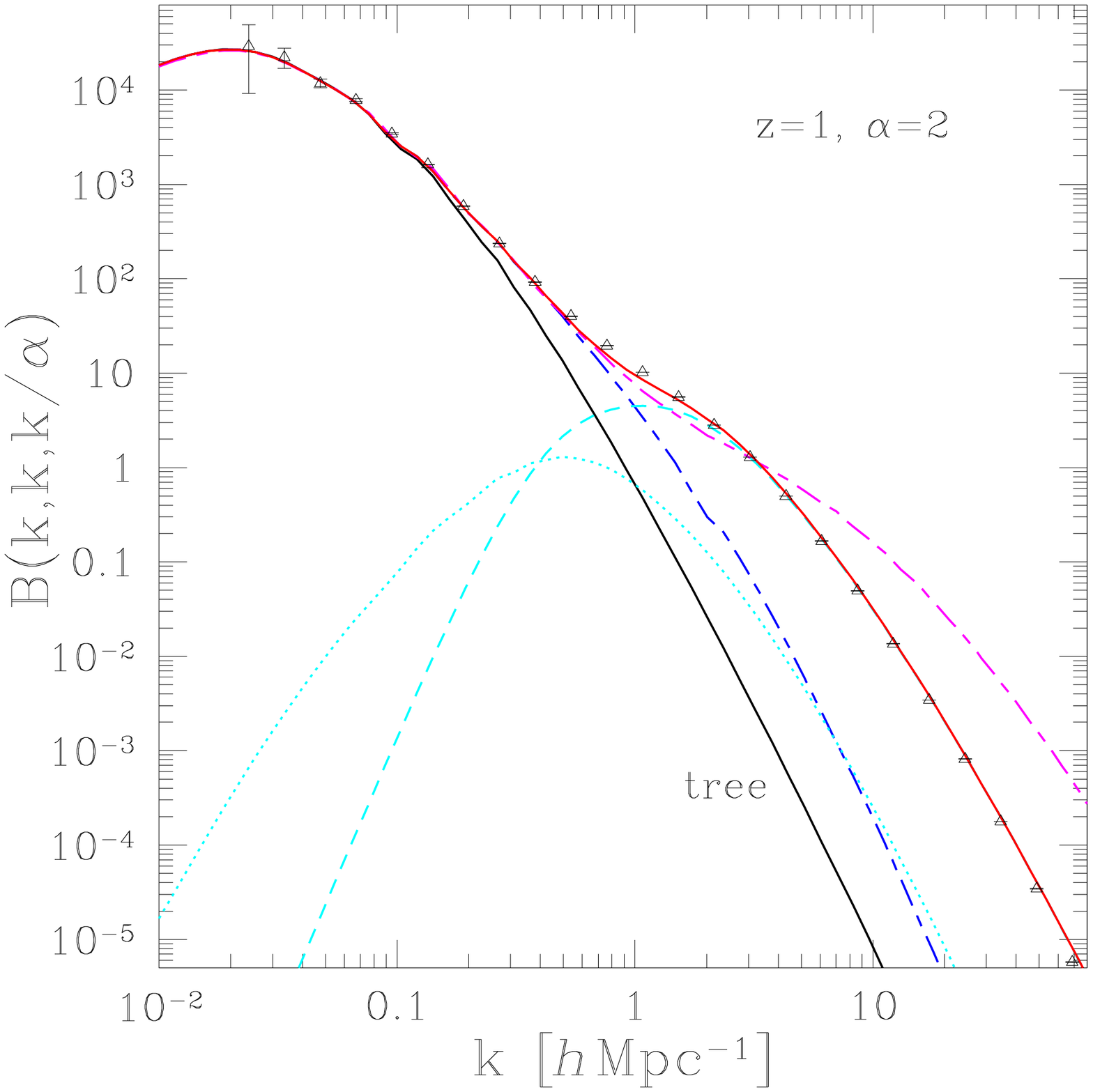}}
\epsfxsize=6.05 cm \epsfysize=5.4 cm {\epsfbox{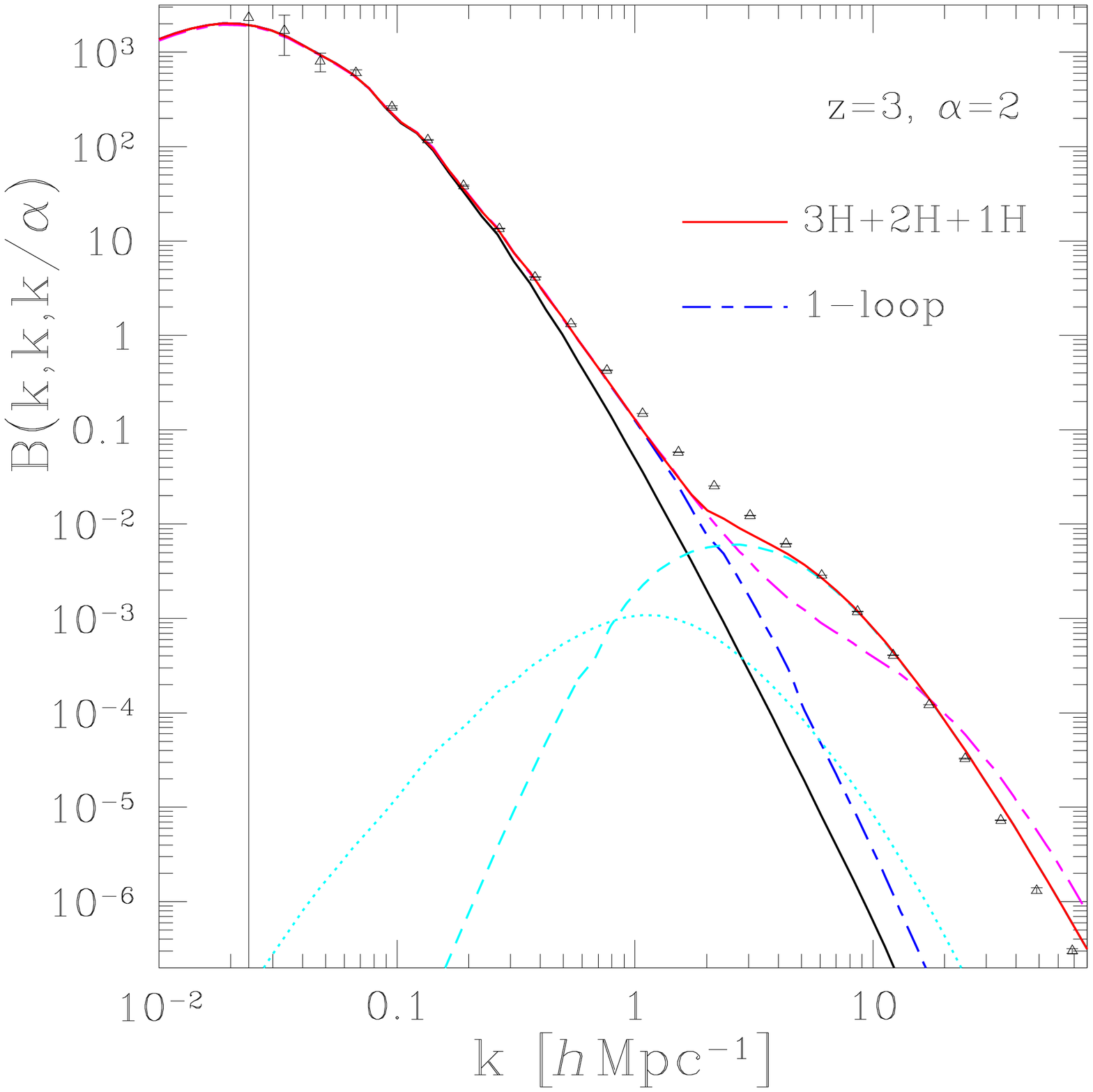}}\\
\epsfxsize=6.1 cm \epsfysize=5.4 cm {\epsfbox{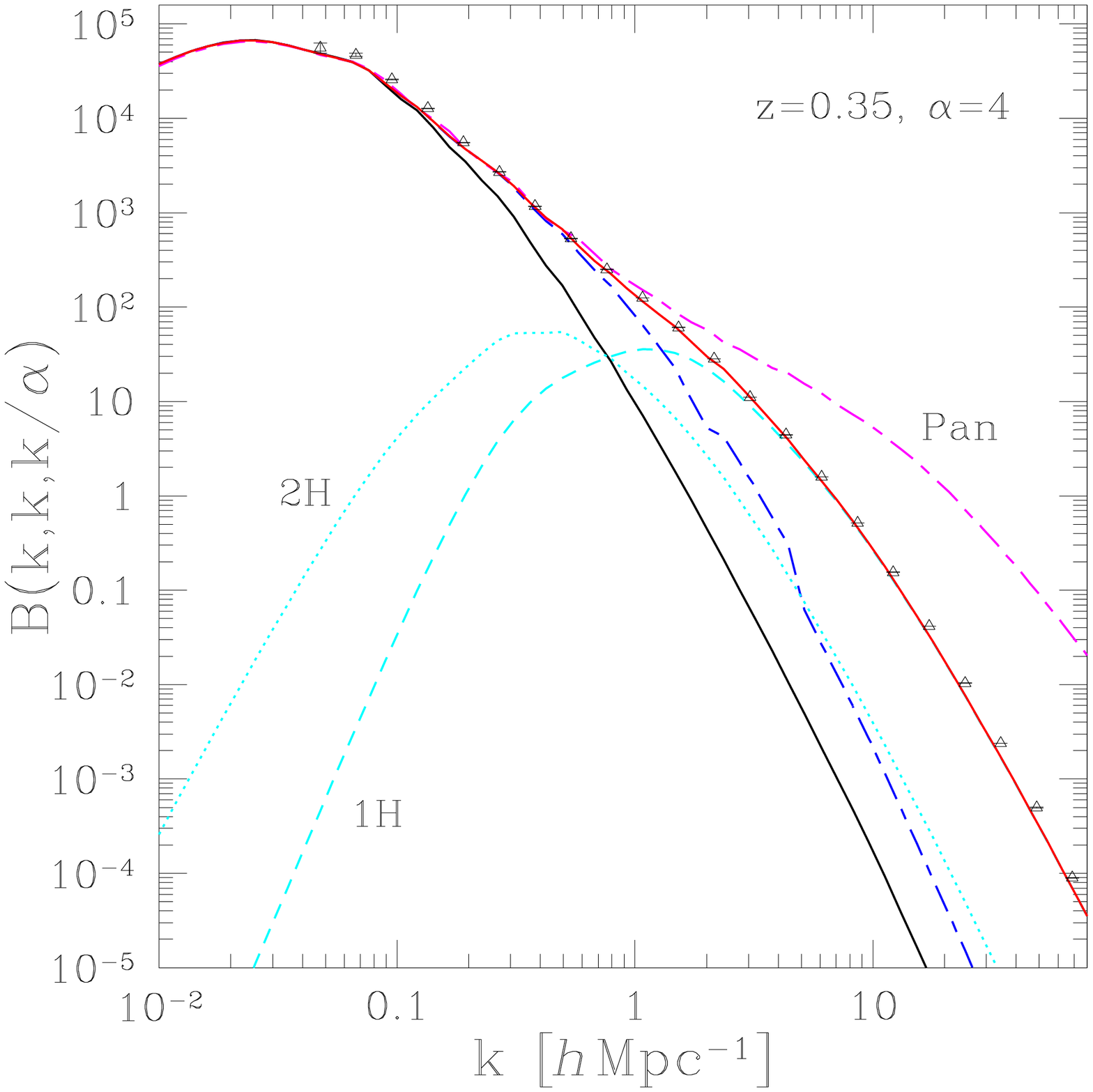}}
\epsfxsize=6.05 cm \epsfysize=5.4 cm {\epsfbox{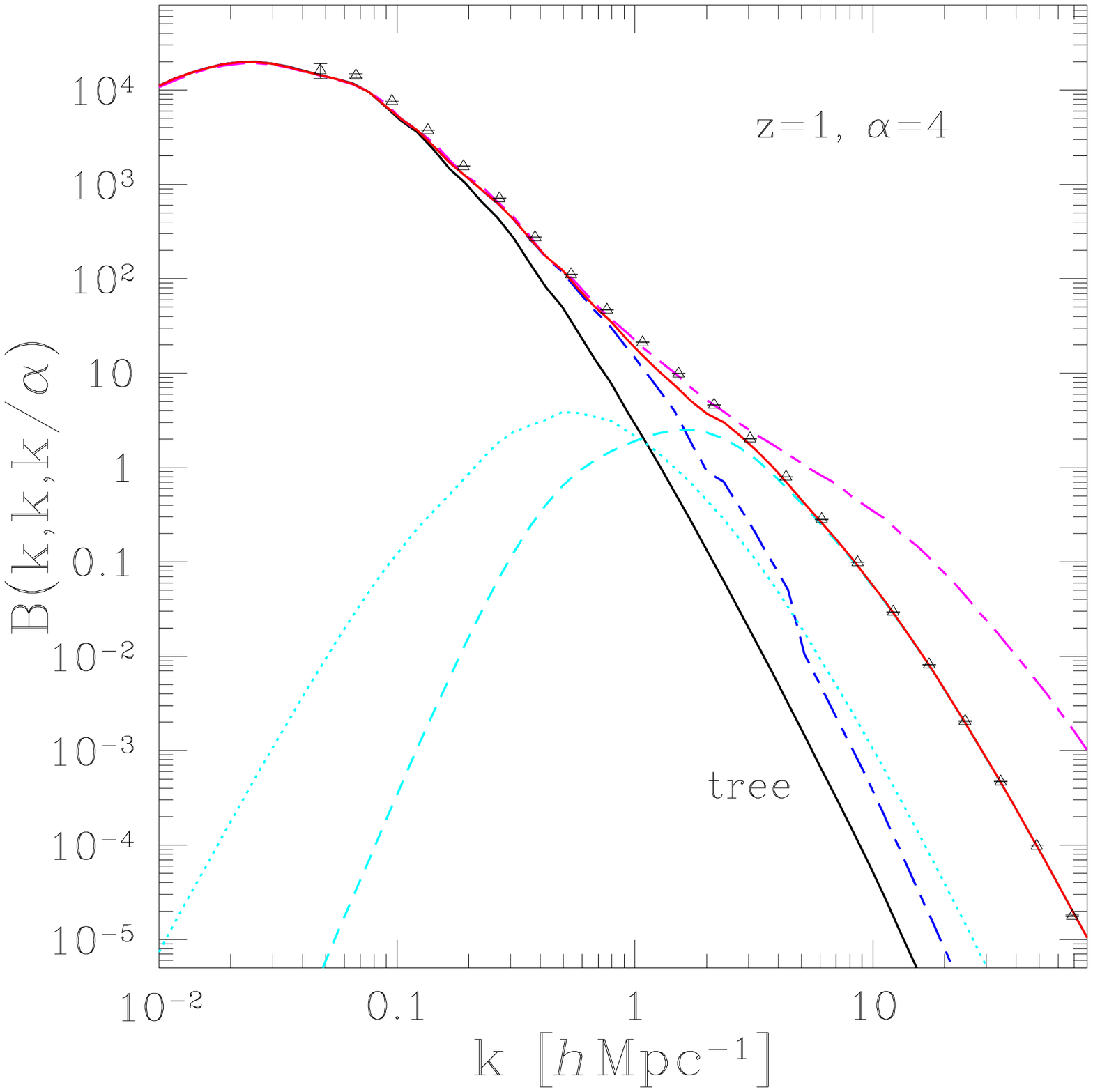}}
\epsfxsize=6.05 cm \epsfysize=5.4 cm {\epsfbox{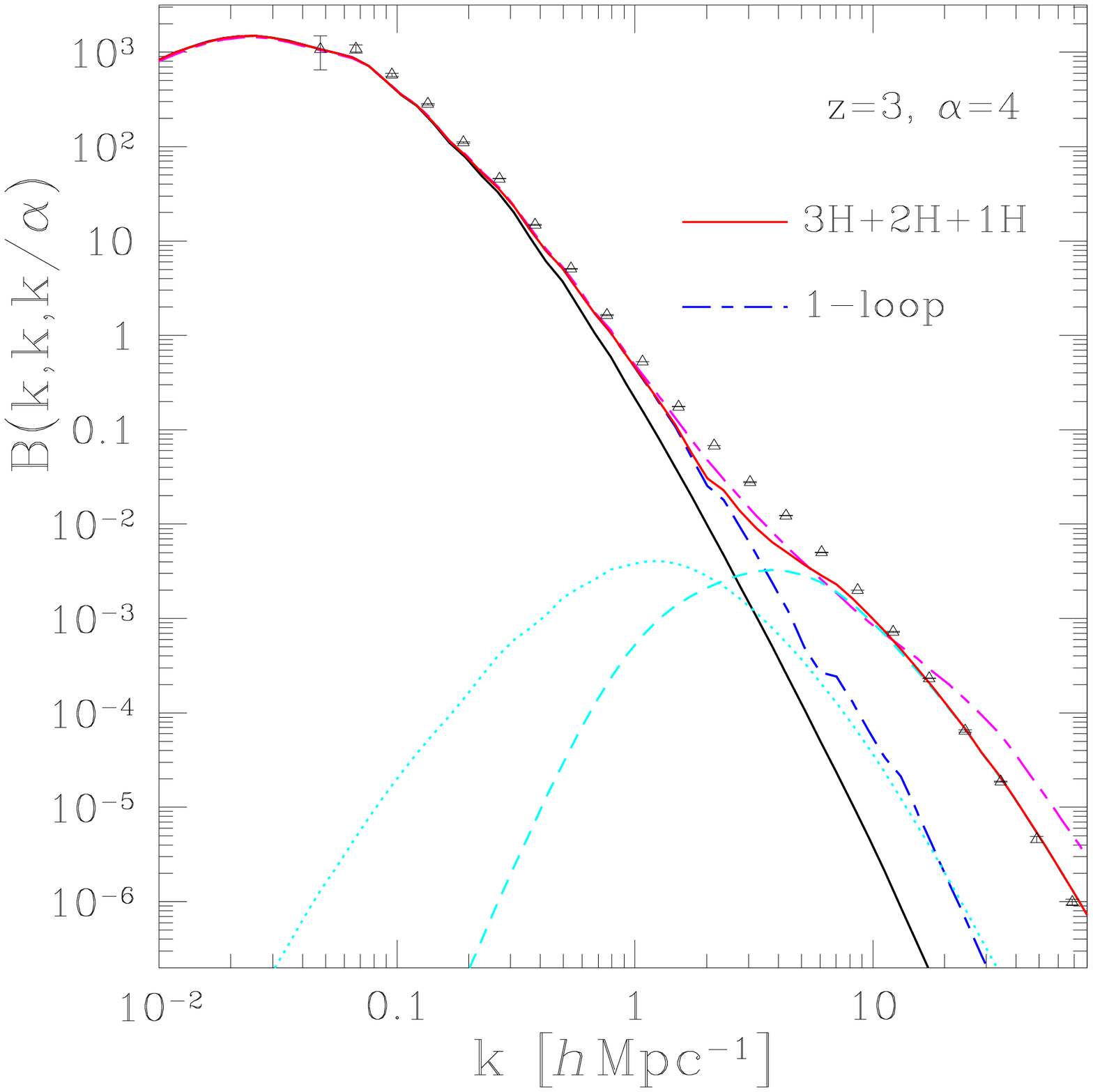}}
\end{center}
\caption{Bispectrum $B(k_1,k_2,k_3)$ for isosceles configurations,
$k_1=k_2=k$ and $k_3=k/\alpha$, at redshifts $z=0.35, 1$, and $3$. 
We consider the cases $\alpha=2$ (upper row) and $\alpha=4$ (lower row).
The symbols are the same as in Fig.~\ref{fig_Bk_eq}.}
\label{fig_Bk_alpha_2_4}
\end{figure*}

We show in Fig.~\ref{fig_Bk_eq} our results for the bispectrum, from linear
to highly nonlinear scales, for equilateral configurations.
Here we take the perturbative 3-halo contribution equal to the standard
1-loop result, which also corresponds to Fig.~\ref{fig_C3-1loop}.
We can see that we obtain a reasonably good agreement with the
numerical simulations over all these scales. This is remarkable since our model
contains no free parameters. Indeed, the 2-halo and 1-halo terms are fully
determined by the halo mass function and density profiles that were used in
\citet{Valageas2010b} for the power spectrum.
This provides a significant improvement over the phenomenological model
presented in \citet{Pan2007}, based on a generalization of the scale
transformation introduced for the two-point correlation function by
\citet{Hamilton1991}, which breaks down on highly nonlinear scales.

However, we can see that at high redshift (right panel at $z=3$) we underestimate
the bispectrum in the transition range between linear and nonlinear scales.
The same behavior appears for the power spectrum, see Fig.~9 in
\citet{Valageas2010b}. This is likely to be due in part to higher order perturbative
terms, which play a greater role at $z=3$ than at $z=0.35$, in agreement with the
detailed study performed in \citet{Valageas2010a} that compares perturbative
and nonperturbative contributions.
On the other hand, it is natural to expect the transition range to be the most
difficult to reproduce by models of the kind studied in this paper.
Indeed, this domain is already beyond perturbation theory but does not yet correspond
to the inner relaxed cores of virialized halos. Therefore, it is at the limit of
validity of the two ingredients (perturbation theory and halo model) used in our
approach.
We will discuss in more detail this transition range and possible improvements on
these scales in Sects.~\ref{depend-halo} and \ref{Improving}.

We clearly see in Fig.~\ref{fig_Bk_eq} the decay on large scales of the
1-halo and 2-halo contributions, in agreement with Eqs.(\ref{lambda-0}) and
(\ref{B-2H-lambda}). As explained in Sect.~\ref{Decomposition}, this is due to the
new counterterms $\tW(k_j q_M)$ of Eqs.(\ref{B-1H-5}) and (\ref{B-2H-4}),
which ensure a physically meaningful behavior.
On the other hand, in agreement with \citet{Sefusatti2010}, we can see that
taking into account the 1-loop perturbative contribution significantly extends
the domain of validity of the 3-halo perturbative term, as compared
with the tree-level contribution.

\subsubsection{Isosceles triangles}
\label{Iso}

\begin{figure*}
\begin{center}
\epsfxsize=6.1 cm \epsfysize=5.4 cm {\epsfbox{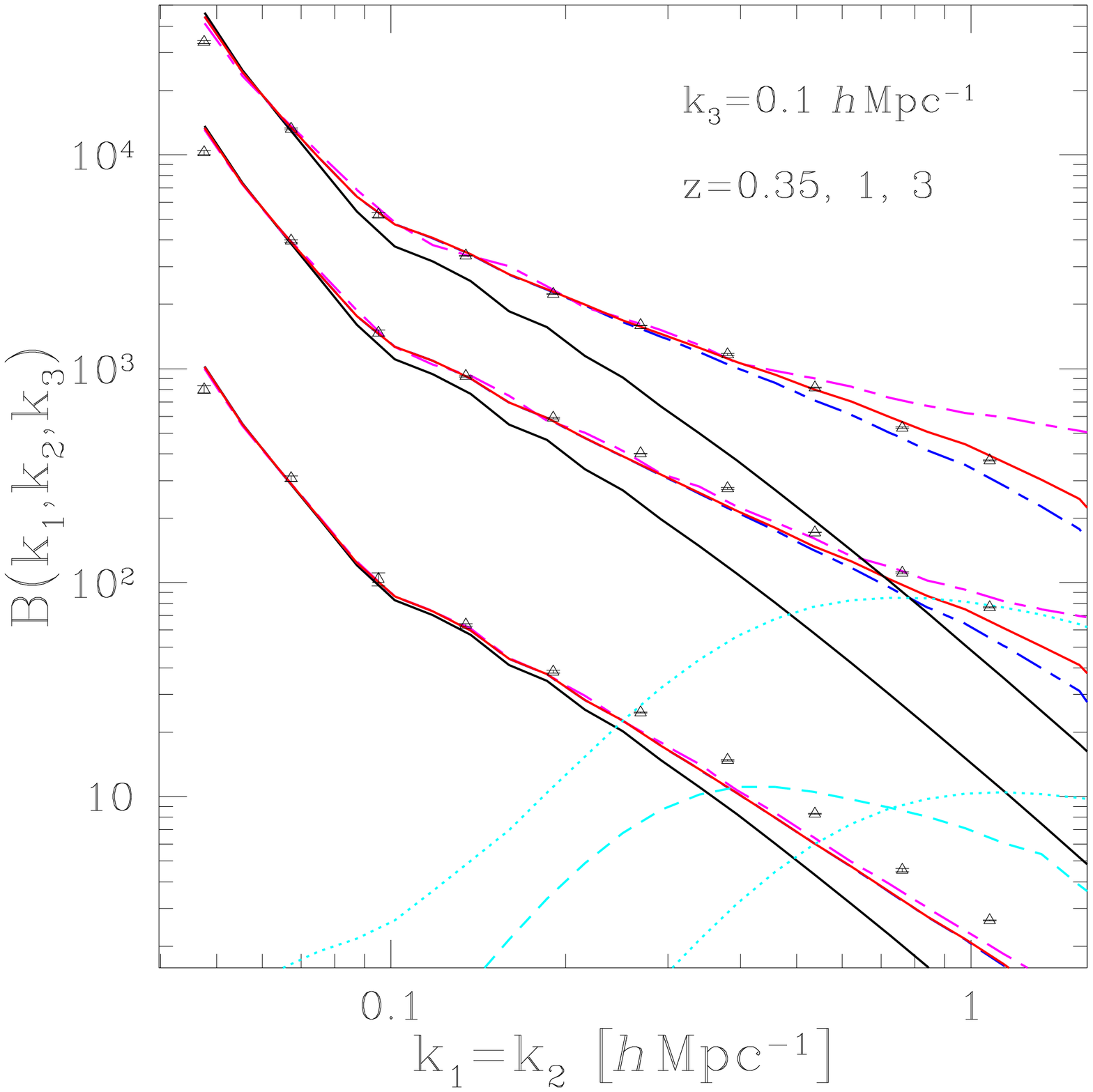}}
\epsfxsize=6.05 cm \epsfysize=5.4 cm {\epsfbox{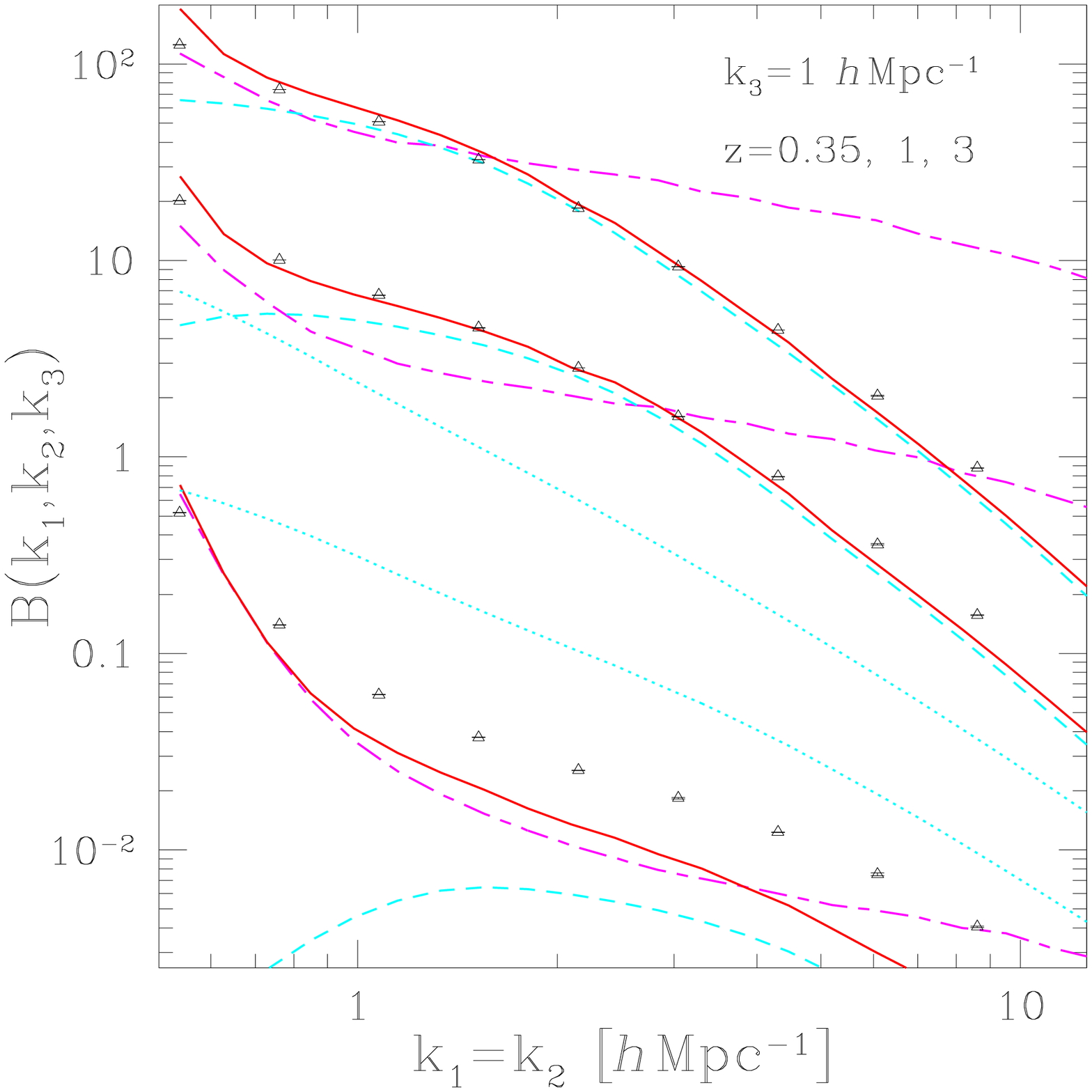}}
\epsfxsize=6.05 cm \epsfysize=5.4 cm {\epsfbox{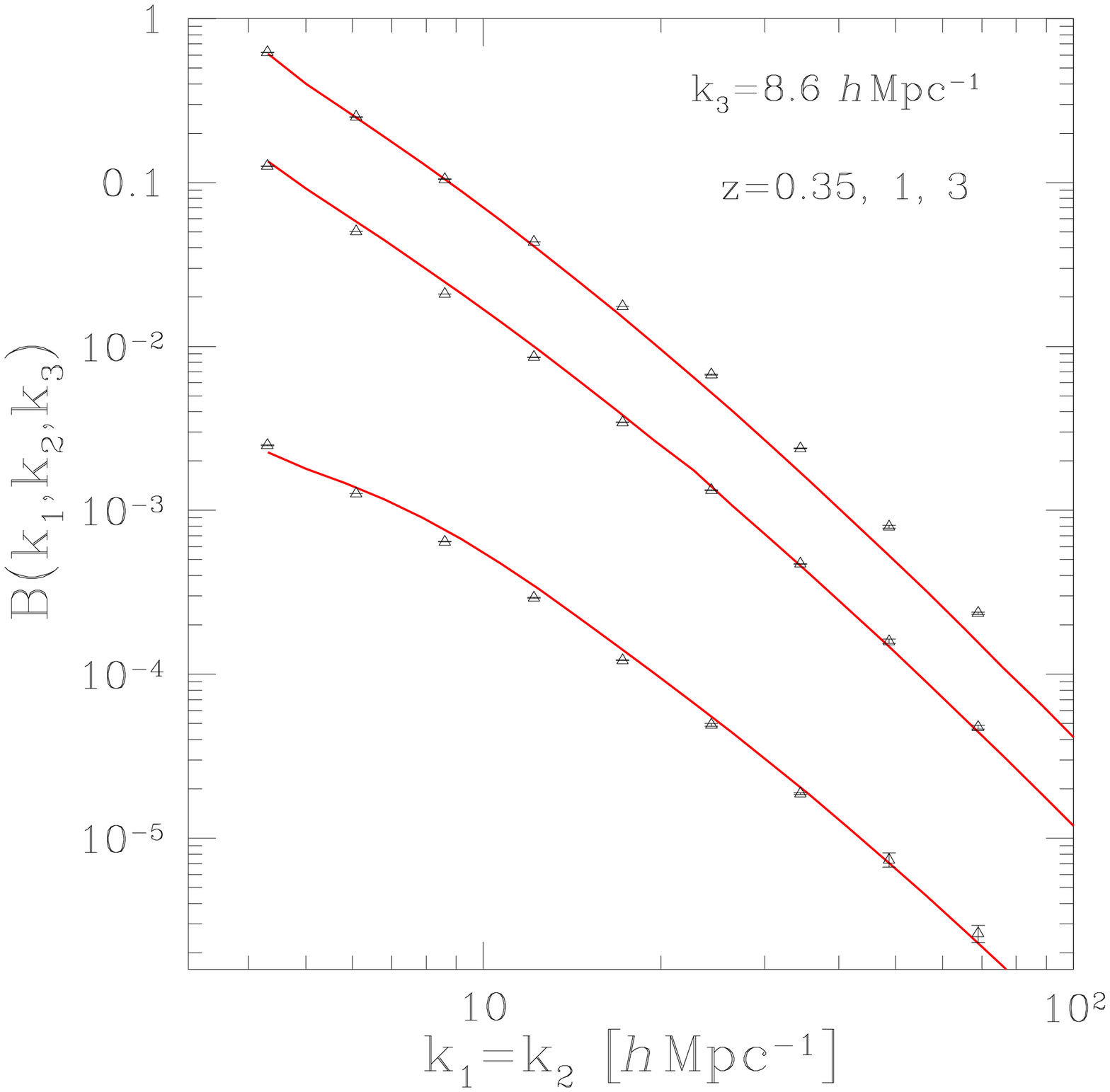}}\\
\epsfxsize=6.1 cm \epsfysize=5.4 cm {\epsfbox{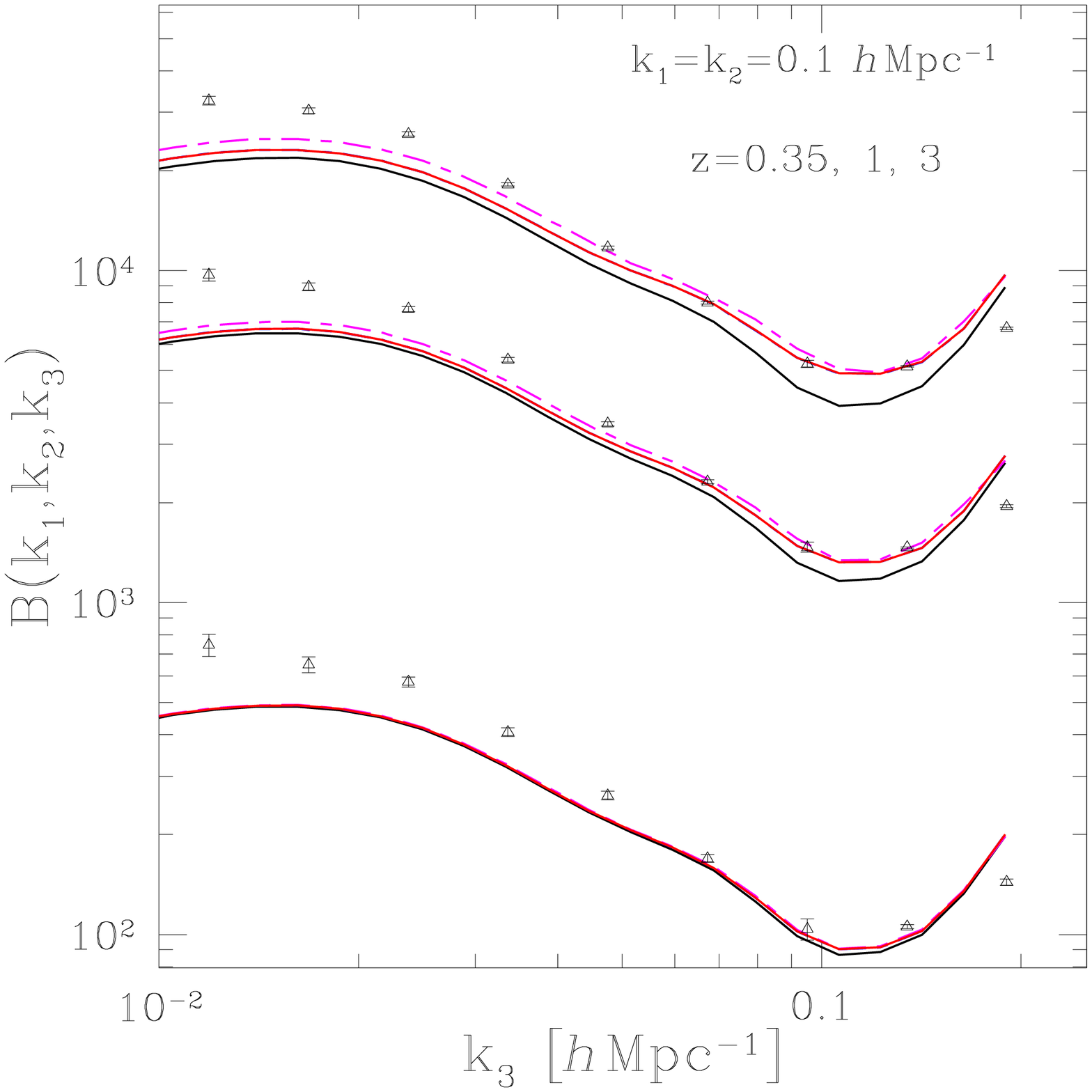}}
\epsfxsize=6.05 cm \epsfysize=5.4 cm {\epsfbox{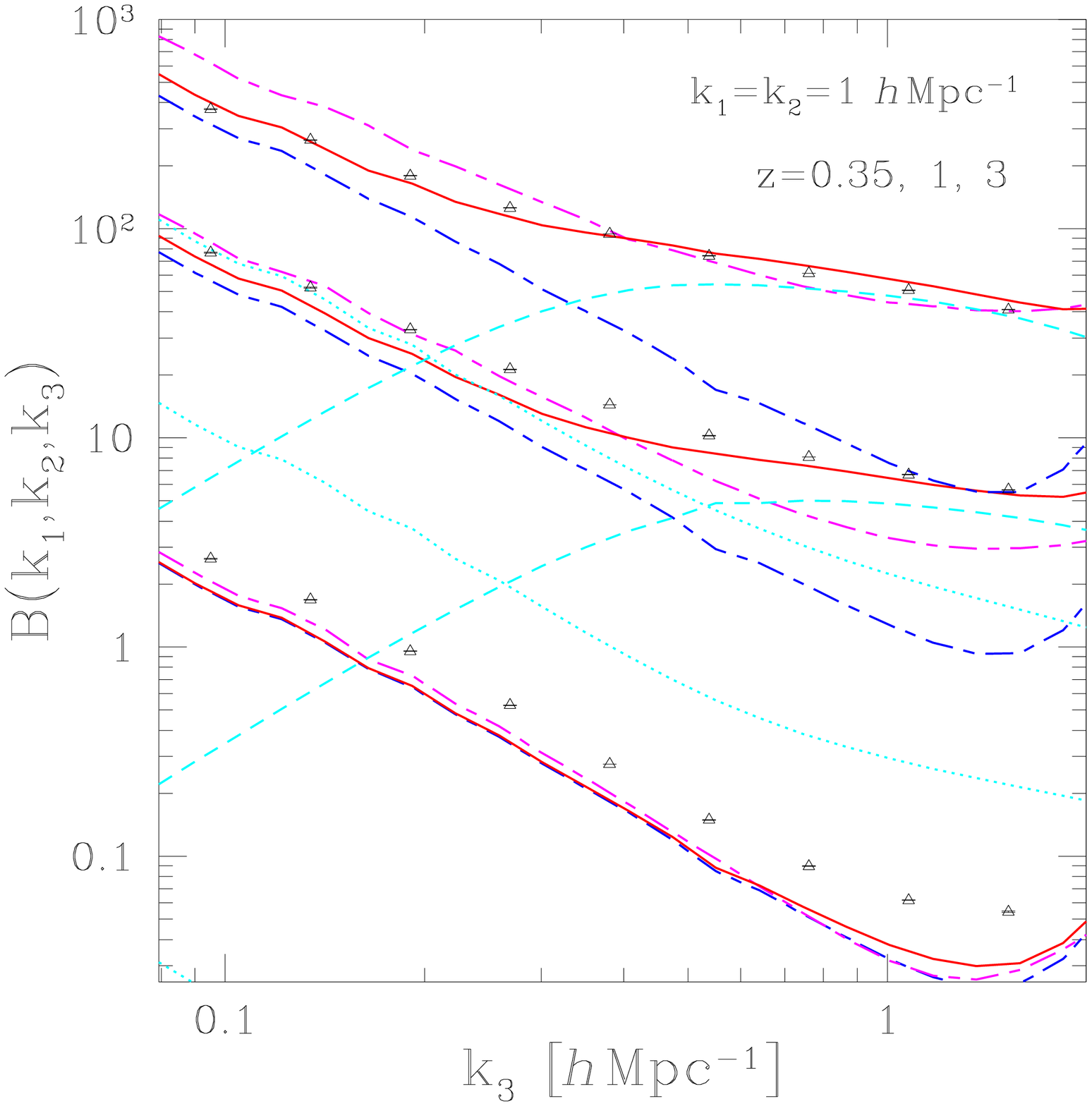}}
\epsfxsize=6.05 cm \epsfysize=5.4 cm {\epsfbox{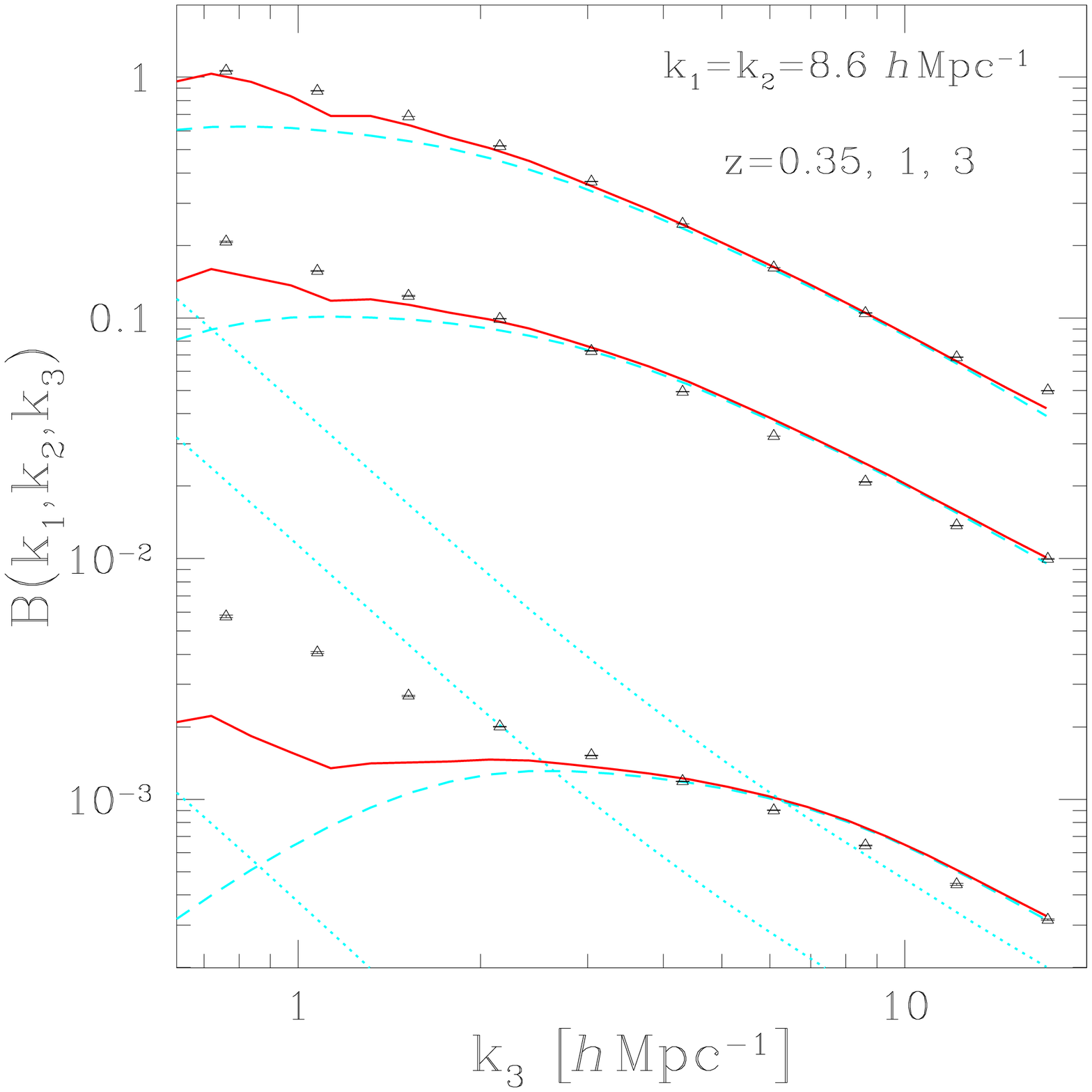}}
\end{center}
\caption{Bispectrum $B(k_1,k_2,k_3)$ for isosceles configurations, 
$k_1=k_2$, at redshifts $z=0.35, 1$, and $3$ (from top to bottom in each panel). 
We show our results as a function of $k_1=k_2$ at fixed $k_3$ (upper row),
and as a function of $k_3$ at fixed $k_1=k_2$ (lower row).
The symbols are the same as in Fig.~\ref{fig_Bk_eq}.}
\label{fig_Bk_k3_k1k2}
\end{figure*}

We show in Fig.~\ref{fig_Bk_alpha_2_4} our results for isosceles triangles
as a function of scale $k$ and for the fixed shapes defined by $k_1=k_2=k$
and $k_3=k/\alpha$, with constant ratios $\alpha$.
We consider the cases $\alpha=2$ and $\alpha=4$.
As expected, as $\alpha$ increases (i.e., the triangles get ``squeezed''),
the 2-halo contribution becomes more important compared with the 1-halo
contribution, because the scale $1/k_3$ grows and it is more likely to have two
distinct halos separated by such a large distance rather than a single halo of
this size. However, for these moderate
values of $\alpha$ the 2-halo contribution always remains subdominant because
the perturbative (i.e., ``3-halo'') contribution is still dominant when the 1-halo term
becomes larger than the 2-halo term.
For more ``squeezed'' shapes, an intermediate regime would appear,
dominated by the 2-halo term.

We can check that our model agrees well with the numerical
simulations and is able to describe both large and small scales.
Again, this extends the analytical predictions beyond the scales
described by the phenomenological model of \citet{Pan2007}.
However, at $z=3$ (right panels) we can see the underestimation on transition
scales already noticed in Fig.~\ref{fig_Bk_eq}. 

Next, we show in Fig.~\ref{fig_Bk_k3_k1k2} the evolution of the bispectrum
with the shape of the triangle formed by $\{\vk_1,\vk_2,\vk_3\}$, for
isosceles configurations $k_1=k_2$. 

First, we consider in the upper row the dependence on $k_1=k_2=k$ at fixed
$k_3$, so that the two equal-length sides run from $k_3/2$ (flat triangle that
reduces to a line) to infinity (squeezed limit). 
In agreement with the behavior observed in Figs.~\ref{fig_Bk_eq} and
\ref{fig_Bk_alpha_2_4}, the bispectrum decreases for higher values of
$k_1=k_2$. The value of the fixed side $k_3$ grows from the left to the
right panel, so that we go from linear to highly nonlinear scales (in each case
we consider the typical range $k_3/2<k_1=k_2<10 k_3$). 
In the left panel, we are dominated by the 3-halo perturbative contribution,
and since $k_3$ remains small, for high values of $k_1=k_2$ the 2-halo
term is larger than the 1-halo term.
In the middle panel, we are already dominated by the 1-halo term (except at low
$k$ at $z=3$ where the 3-halo perturbative contribution is larger), and the
the 2-halo term is subdominant.
In the right panel the bispectrum is completely dominated by the 1-halo
contribution.

\begin{figure*}
\begin{center}
\epsfxsize=6.1 cm \epsfysize=5.4 cm {\epsfbox{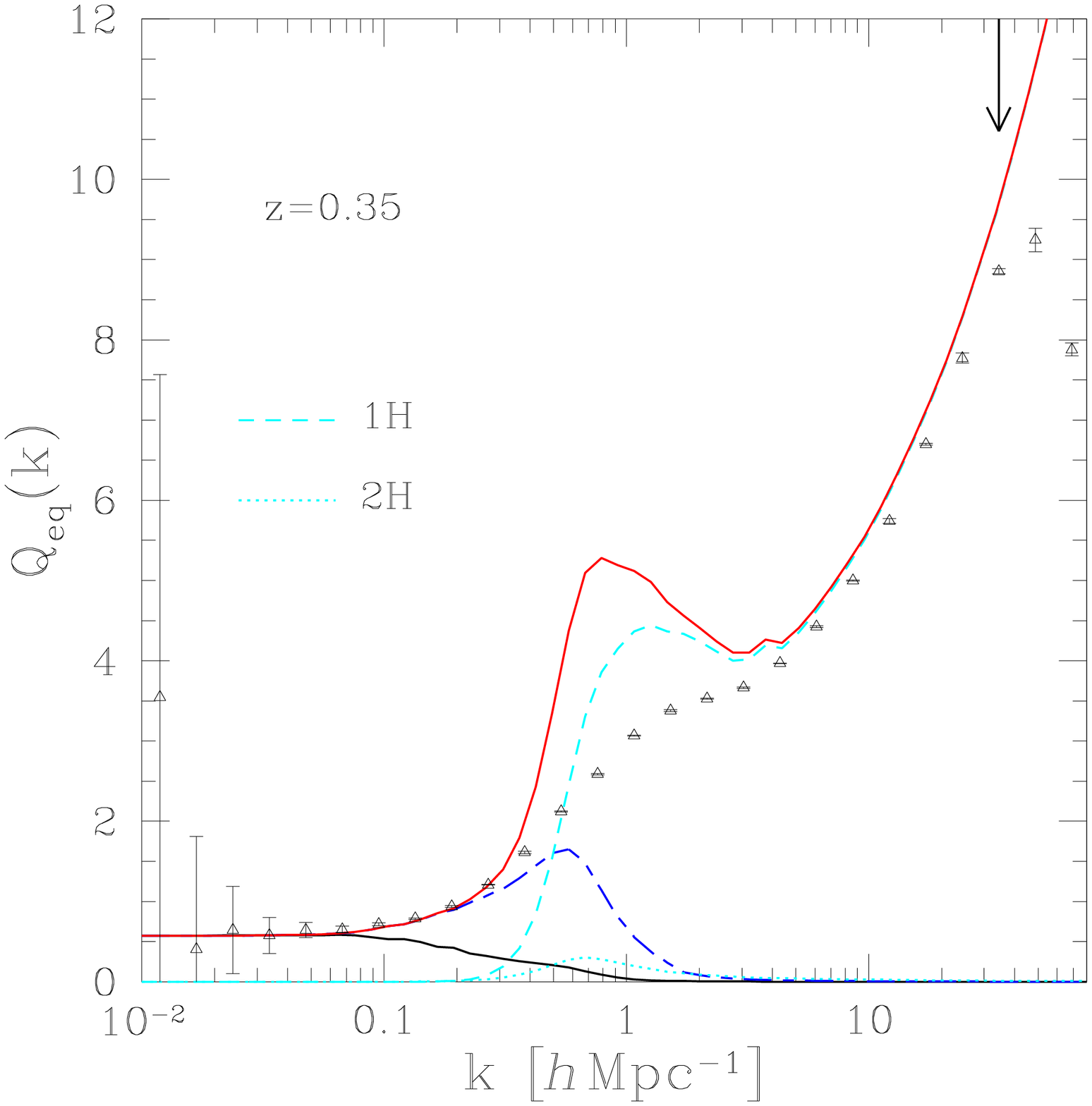}}
\epsfxsize=6.05 cm \epsfysize=5.4 cm {\epsfbox{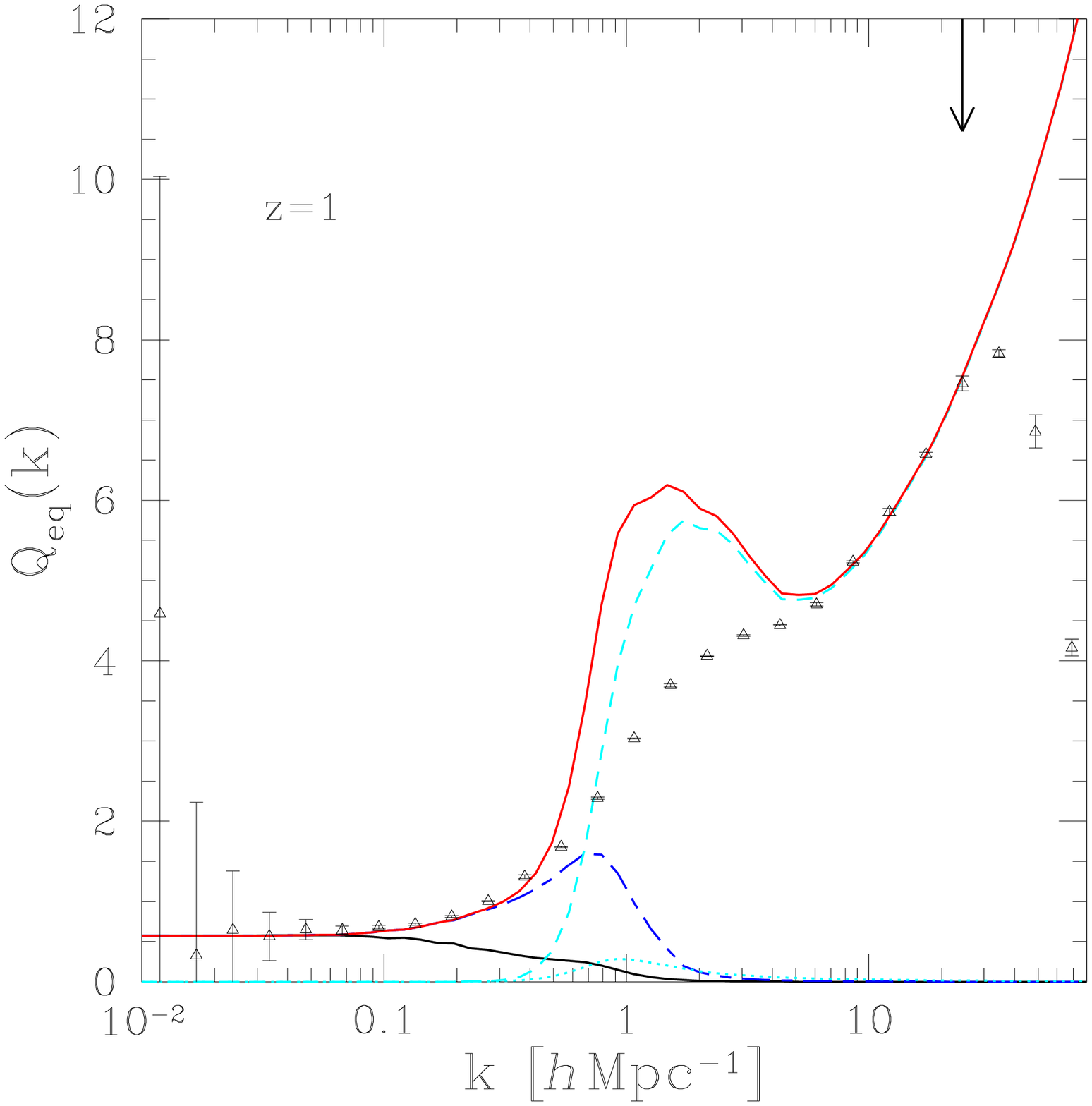}}
\epsfxsize=6.05 cm \epsfysize=5.4 cm {\epsfbox{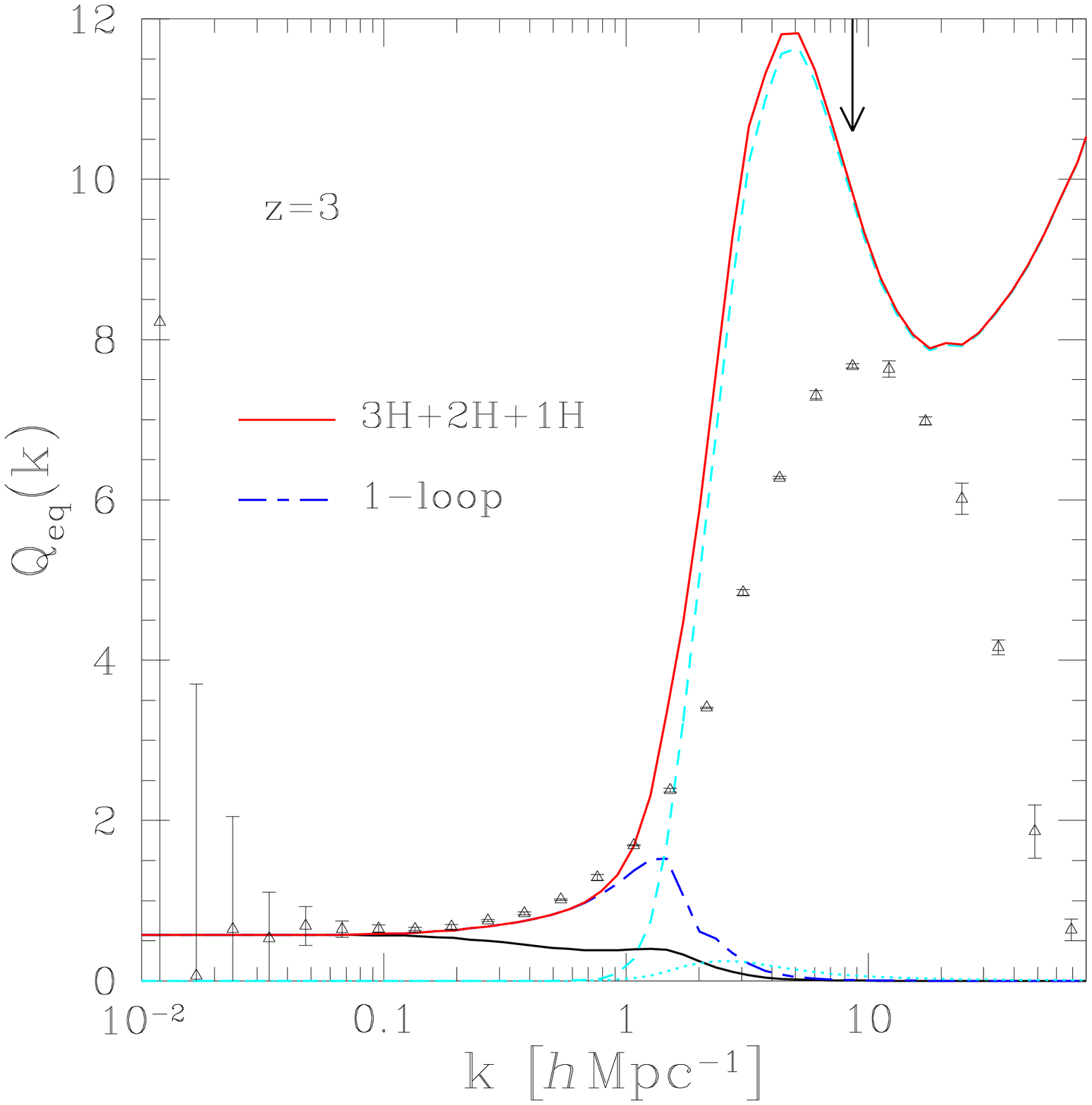}}
\end{center}
\caption{Reduced bispectrum, $Q_{\rm eq}(k)=B_{\rm eq}(k)/[3P(k)^2]$, for
equilateral configurations, at redshifts $z=0.35, 1$, and $3$.  The symbols are
the same as in Fig.~\ref{fig_Bk_eq}. The vertical arrow in the upper right part
shows the wavenumber beyond which the simulation shot noise is greater than
$10\%$.}
\label{fig_Qk_eq}
\end{figure*}

Second, we consider in the lower row the dependence on $k_3$ at fixed
value of the common sides $k_1=k_2=k$, so that the third side runs from
$0$ (squeezed limit) to $2k_1$ (flat triangle that reduces to a line). 
Again, the bispectrum decreases as the dependent wavenumber, here $k_3$,
grows. However, the dependence is much weaker than the one found in the
upper row. In the left panel, we are dominated by the linear-order perturbative
contribution, so that all curves agree with each other. The too high values of
the numerical points at low $k_3$ are a numerical artifact caused by the
finite boxsize and the numerical binning used to compute the bispectrum,
because linear theory must be increasingly accurate on larger scales.
In the middle panel, we can see the crossover between the 3-halo and 1-halo
regimes. The 2-halo term is never dominant on these scales.
In the right panel we are mostly dominated by the 1-halo term.

In all cases we can see that our model agrees well with the
numerical simulations. However, at $z=3$, in the upper middle panel and in the
lower right panel we can see a trace of the underestimation on the transition scales
found in the right panels of Figs.~\ref{fig_Bk_eq} and \ref{fig_Bk_alpha_2_4}.

\subsection{Reduced bispectrum}
\label{Q-comp}

As seen in the previous figures, the bispectrum varies by many orders of
magnitude on the scales of interest. Therefore, it is customary to introduce
the ``reduced bispectrum'' defined as
\beq
Q(k_1,k_2,k_3) = \frac{B(k_1,k_2,k_3)}{P(k_1) P(k_2) + 2 {\rm cyc.}} .
\label{Qdef}
\eeq
Indeed, this ratio goes to a constant on large scales, as seen from
the tree-order expression (\ref{Ba-def}), while it shows a moderate growth
on small scales \citep{Bernardeau2002}.

We compare in Fig.~\ref{fig_Qk_eq} our results for the reduced equilateral
bispectrum, $Q_{\rm eq}(k)=Q(k,k,k)$, with numerical simulations.
Here, in the denominator of Eq.(\ref{Qdef}) we use for the theoretical predictions
the power spectrum obtained with our model described in \citet{Valageas2010b}
(that is, using the same halo model as in this paper and the direct steepest-descent
resummation for the perturbative 2-halo term), while for the numerical curves we
use the power spectrum measured in the simulations.

The vertical arrow in the upper right part of each panel of Fig.~\ref{fig_Qk_eq} 
shows the wavenumber where the shot-noise level of the numerical
simulations becomes larger than $10\%$. 
We can see that this also roughly corresponds
to the scale where the measured reduced bispectrum is maximum.
Therefore, the downturn and the decrease of the reduced bispectrum at higher
$k$ found in the simulations is only a numerical artifact caused by the limited
resolution, and these points must be discarded.
In particular, at lower redshift where the simulations are able to probe deeper
into the nonlinear regime, we clearly see in the left panel at $z=0.35$ the
fast growth of $Q_{\rm eq}(k)$ in the highly nonlinear regime, after a small
plateau on the transition scales. As shown by the two left panels, this
high-$k$ fast increase of $Q_{\rm eq}(k)$ is well reproduced by our model.

We can see that we obtain a reasonably good agreement with the simulations
on both large and small scales. This is consistent with our previous results,
shown in Fig.~\ref{fig_Bk_eq} and in \citet{Valageas2010b}, where we found
that both the bispectrum and the power spectrum are well reproduced
on quasilinear and highly nonlinear scales. 
As expected, we recover the large-scale plateau associated with the tree-level
limit (\ref{Ba-def}), and the early rise owing to higher order perturbative
contributions. 
At very high $k$, in agreement with previous studies \citep{Ma2000b,Cooray2002,Fosalba2005},
the halo model leads to a continuous growth of the reduced bispectrum,
which is a marked difference with the prediction of the stable clustering ansatz
\citet{Peebles1980}. This agrees with our numerical simulations up to the scales
that are well resolved.
As shown in Sect.~2 of \citet{Valageas1999}, the positivity of the matter density
actually implies that the reduced bispectrum, and more generally the
real-space coefficients $S_p=\lag \rho_R^p\rag_c/\lag \rho_R^2\rag_c^{p-1}$,
reach a constant or keep growing on small scales, as the density variance
$\lag \rho_R^2\rag_c$ increases. The limiting behavior of the constant
reduced bispectrum and the constant coefficients $S_p$ is achieved for
the stable clustering ansatz, and more generally for multifractal models
such that $\lag \rho_R^p\rag_c$ are governed by a single fractal
exponent $\alpha$ for the values of $p$ that are considered.
The density field described by a halo model clearly violates these 
conditions\footnote{The halo model can be made to recover the stable-clustering
ansatz predictions if the mass function scales at low mass as $n(M) \dd M \propto
\dd M/M^2$ \citep{Valageas1999,Ma2000b}. This unrealistic formal limit,
where the apparent amount of matter per unit volume is infinite, corresponds to
a multiple counting of ``halos'', which contain an infinite hierarchy of
substructures that are also counted in the mass function, in agreement with
a fractal model.}
because it does not display this scale invariance, with a characteristic
nonlinear mass associated with the falloff of the halo mass function and
reasonably smooth profiles that depend on the mass scale (through their
concentration parameter). This implies that $Q_{\rm eq}(k)$ has to grow
on small scales, as checked in Fig.~\ref{fig_Qk_eq}.

\begin{figure*}
\begin{center}
\epsfxsize=6.1 cm \epsfysize=5.4 cm {\epsfbox{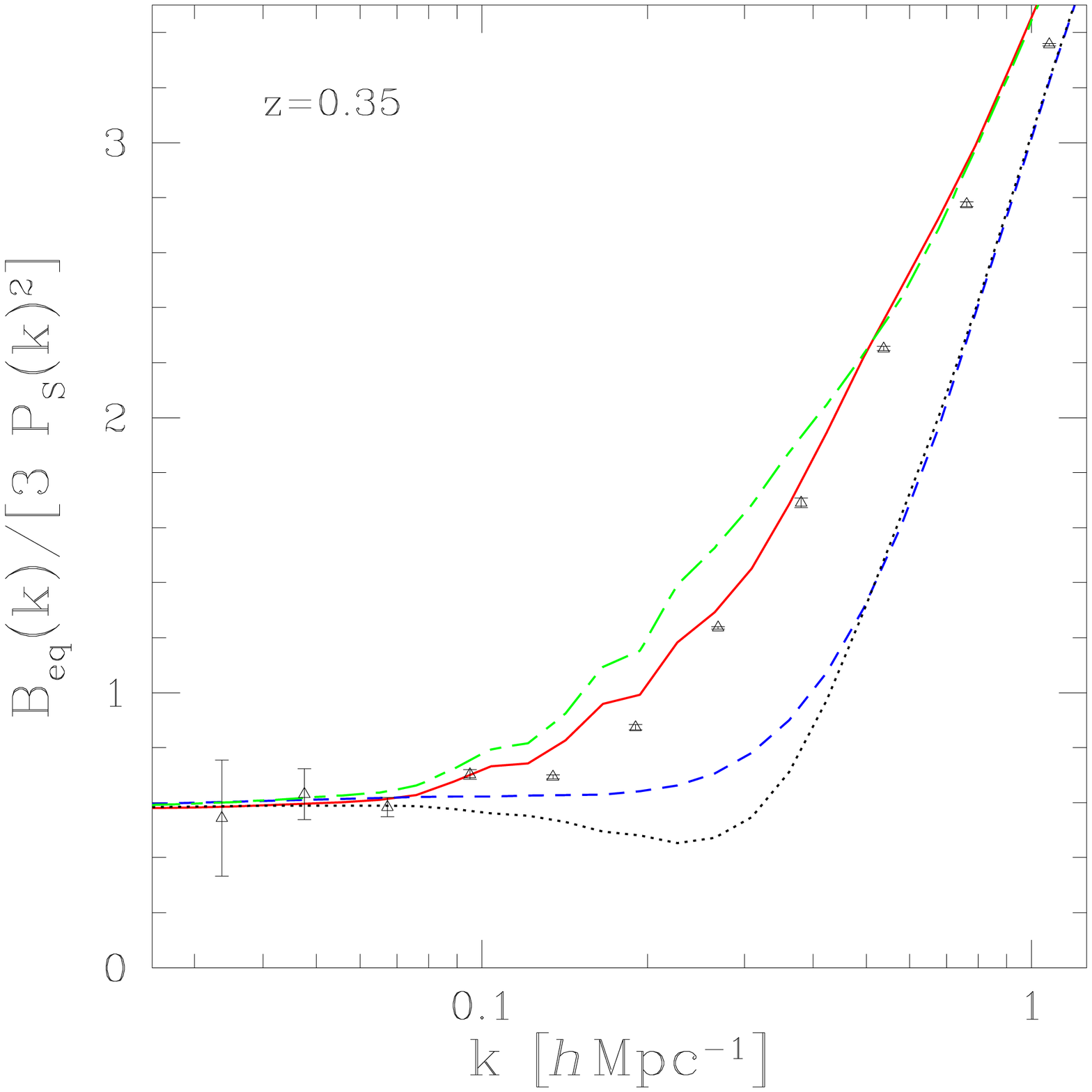}}
\epsfxsize=6.05 cm \epsfysize=5.4 cm {\epsfbox{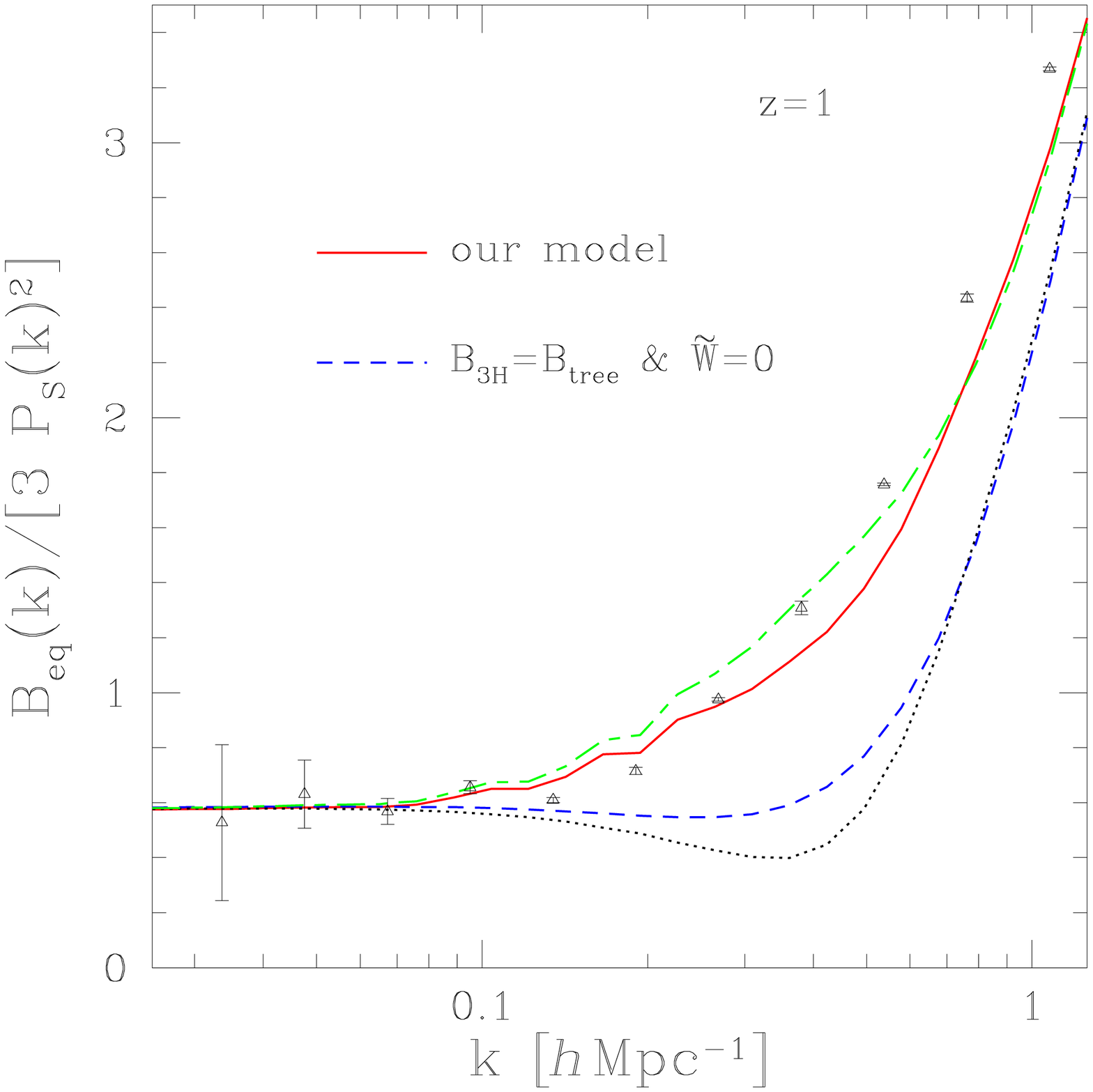}}
\epsfxsize=6.05 cm \epsfysize=5.4 cm {\epsfbox{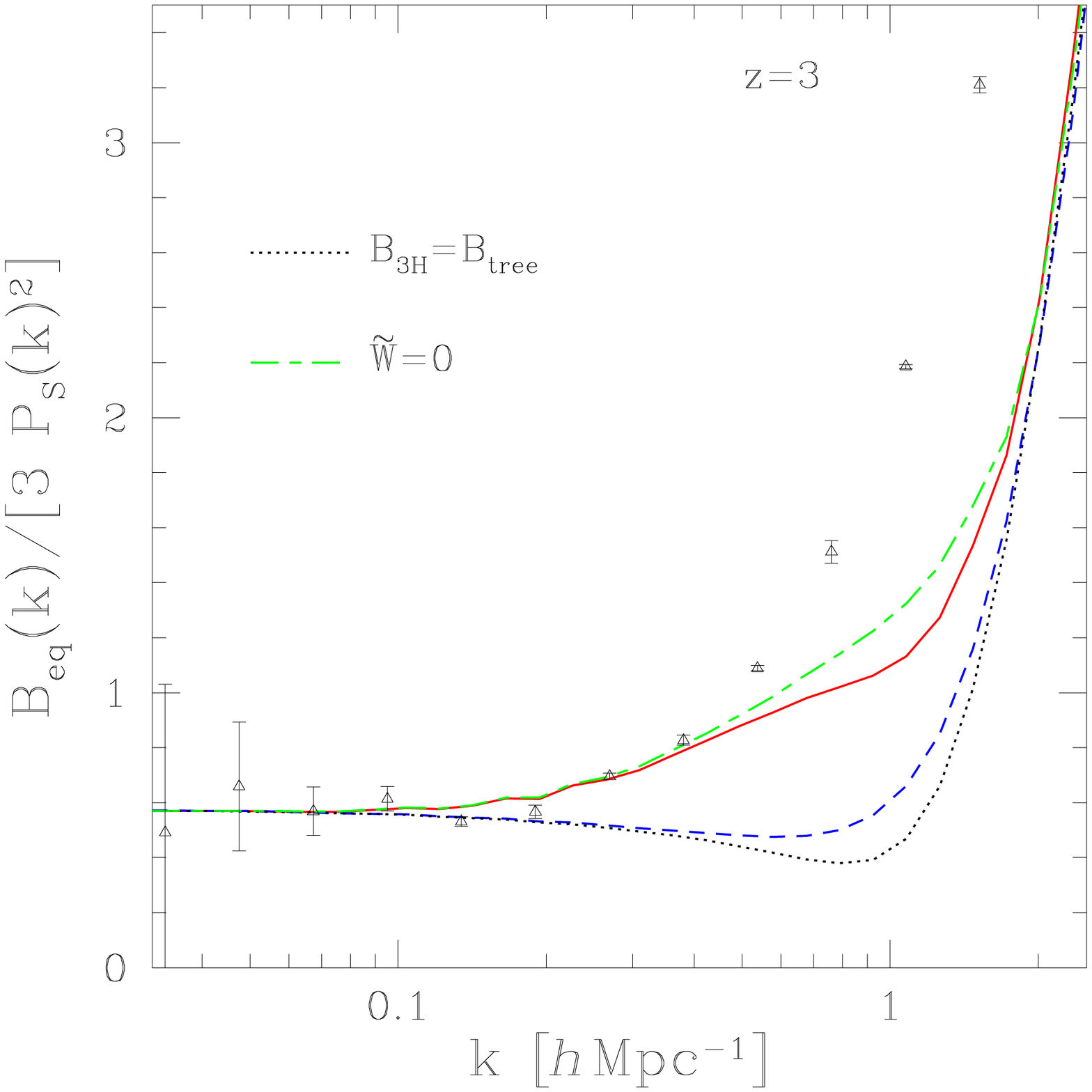}}\\
\epsfxsize=6.1 cm \epsfysize=5.4 cm {\epsfbox{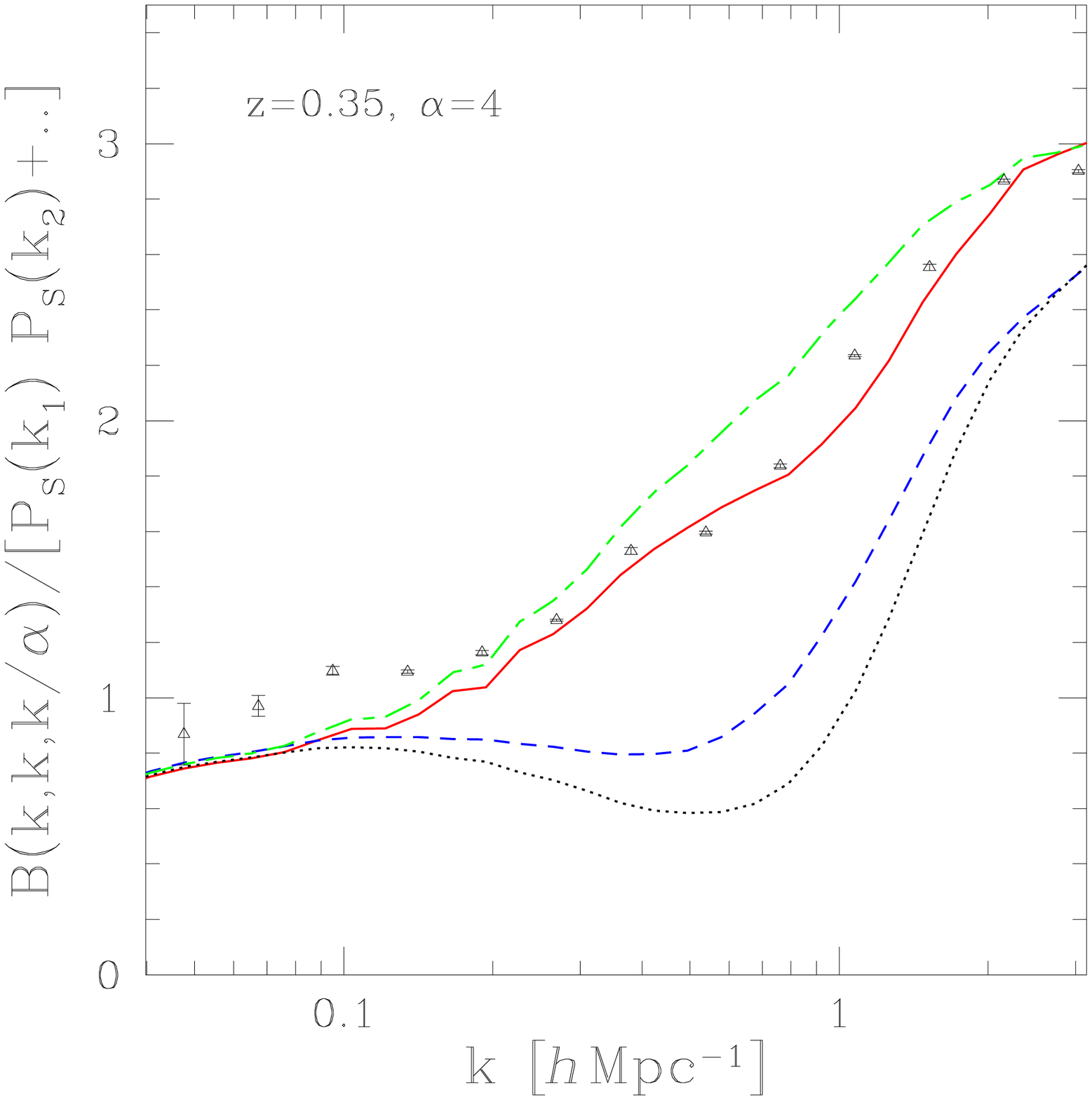}}
\epsfxsize=6.05 cm \epsfysize=5.4 cm {\epsfbox{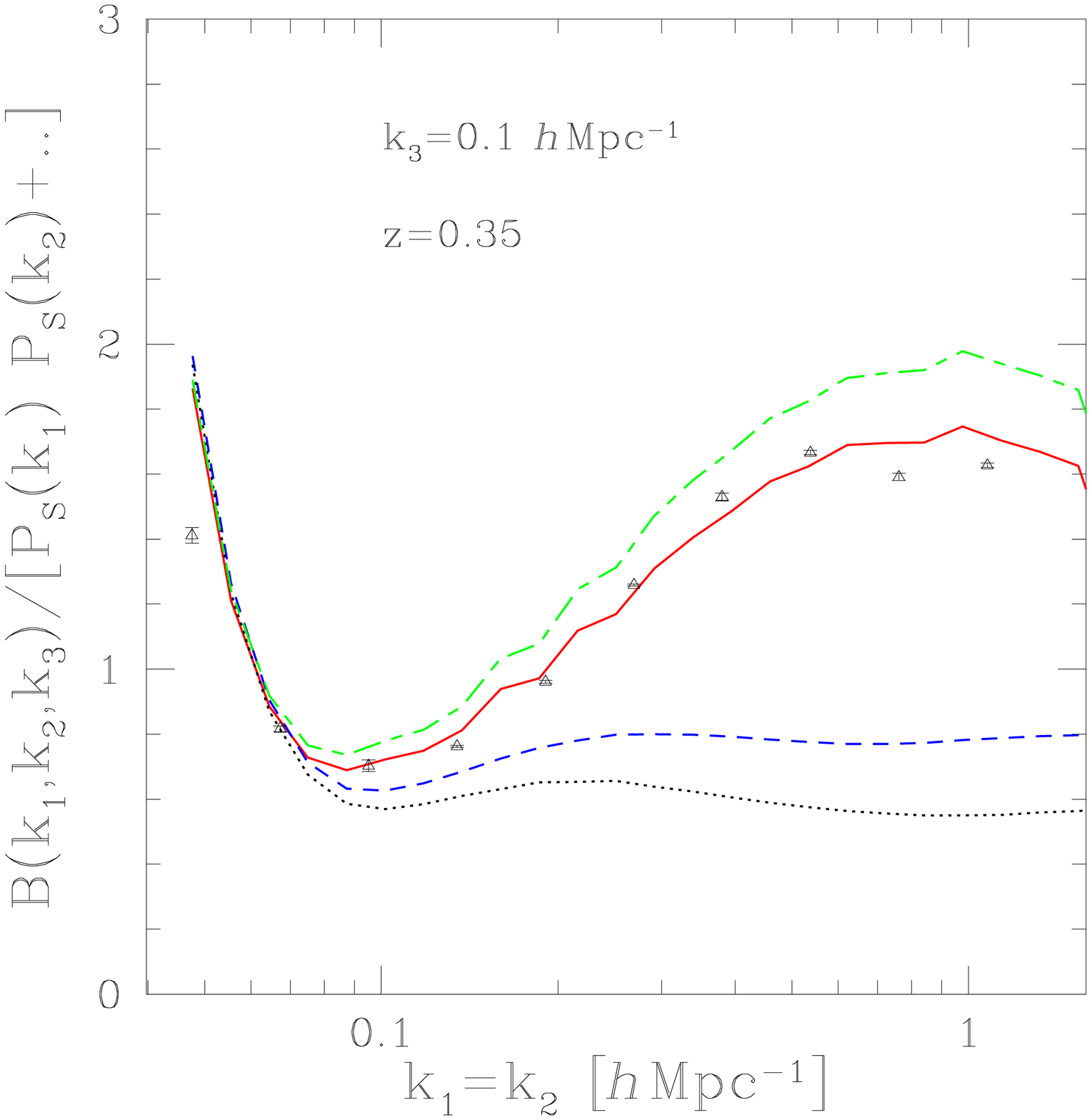}}
\epsfxsize=6.05 cm \epsfysize=5.4 cm {\epsfbox{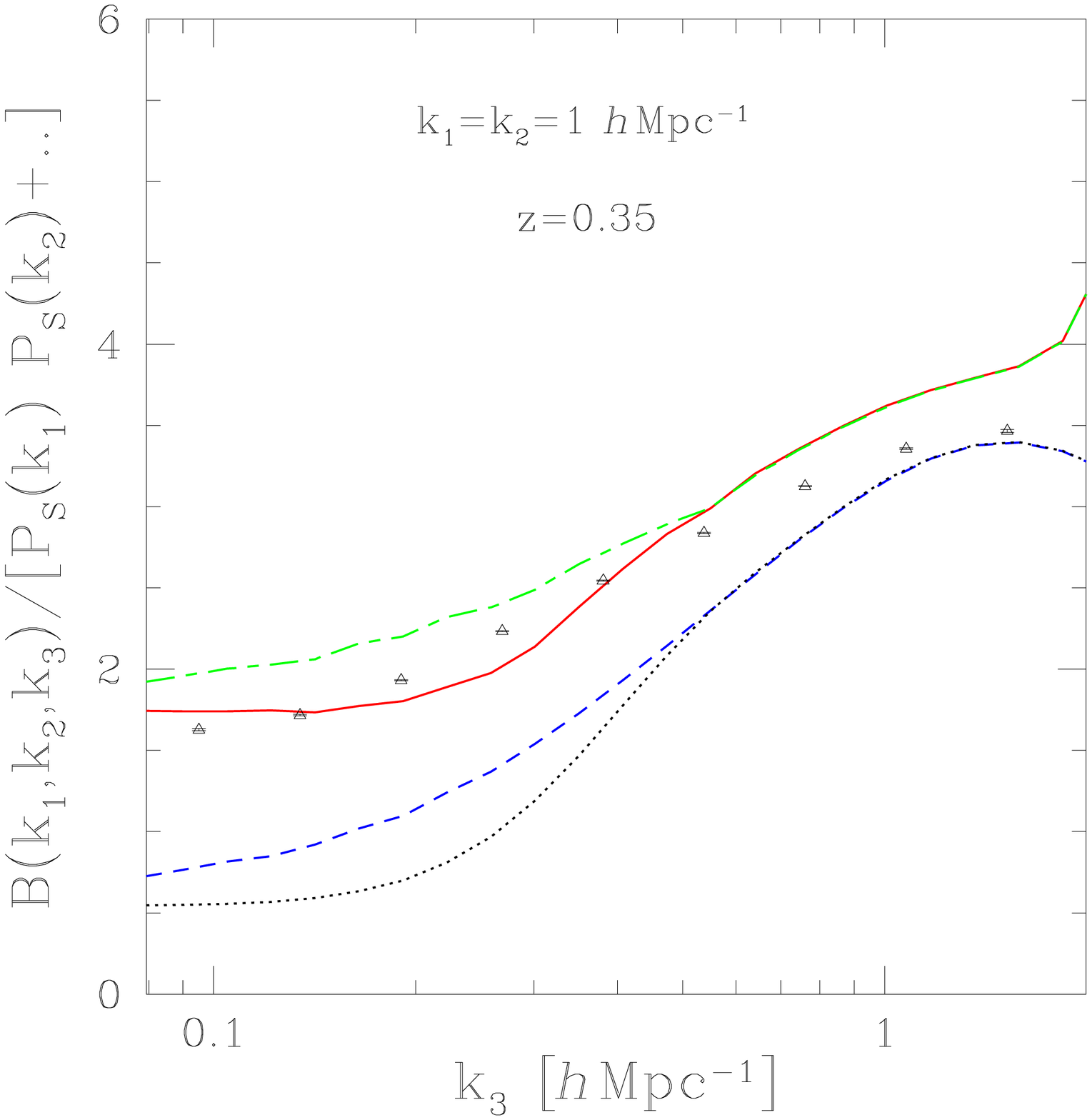}}
\end{center}
\caption{Ratio $B(k_1,k_2,k_3)/[P_{\rm S}(k_1) P_{\rm S}(k_2)+ 2 {\rm cyc.}]$,
where $P_{\rm S}(k)$ is the fit to the power spectrum from \citet{Smith2003}.
The red solid line is our model, as in Figs.~\ref{fig_Bk_eq} to \ref{fig_Qk_eq},
the black dotted line is the approximation $B_{\rm 3H}=B_{\rm tree}$, 
the green dot-dashed line corresponds to no counterterms in the 1-halo
and 2-halo contributions ($\tW=0$), and the blue dashed line makes both
approximations.
{\it Upper row:} equilateral configurations, at redshifts $z=0.35, 1$, and $3$, 
as in Fig.~\ref{fig_Bk_eq}. {\it Lower row:} various isosceles configurations
at redshift $z=0.35$, as in left panels of Figs.~\ref{fig_Bk_alpha_2_4} and
\ref{fig_Bk_k3_k1k2}.}
\label{fig_Qsk_hm}
\end{figure*}

On the transition scales,
we underestimate both the bispectrum and the power spectrum, and it appears
that the latter effect is dominant, which leads to the ``bump'' seen in
Fig.~\ref{fig_Qk_eq}. As seen from the numerical results, this feature is not
physical and only caused by the shortcomings of the theoretical model on this 
range. 
Such a feature was also noticed in previous studies
\citep{Scoccimarro2001,Takada2003,Fosalba2005}, which found that by using a 
larger halo radius, taking into account exclusion constraints in the 2-halo and
3-halo terms, or introducing a high-mass cutoff in the halo mass function, one
might remove this ``bump'' and reach a better agreement with numerical
simulations.
On the other hand, \citet{Guo2009} find that no self-consistent halo model
is able to reproduce well both the matter power spectrum and bispectrum on
these transition scales. 
We will return to this problem and these modifications in
Sect.~\ref{depend-halo} and will
discuss in Sect.~\ref{Improving} an alternative procedure to improve the
model on these scales, using this artificial ``bump'' as a signature of insufficient
accuracy and introducing simple interpolations that resolve most of this problem,
while keeping large and small scales unchanged.

\section{Comparison with previous halo models}
\label{Comparison-wit-previous-models}

We show in Fig.~\ref{fig_Qsk_hm} the impact of two features of our model
that differ from previous implementations of the halo model
\citep{Scoccimarro2001,Fosalba2005,Guo2009}, i) the use of 1-loop perturbation
theory instead of the tree-level contribution (\ref{Ba-def}) for the 3-halo term,
and ii) the counterterms $\tW$ in Eqs.(\ref{B-1H-5}) and (\ref{B-2H-4}) in the
1-halo and 2-halo terms.

To distinguish more clearly between various models, that is, to reduce the size
of the vertical axis, we plot in Fig.~\ref{fig_Qsk_hm} the effective reduced
bispectrum defined as in Eq.(\ref{Qdef}), but where we use the power spectrum
$P_{\rm S}(k)$ from \citet{Smith2003} in the denominator, for the models as
well as for the numerical simulations. Thus, by dividing the bispectra by the
same denominator we avoid introducing additional errors through the
denominator and can make a clear comparison between the various bispectra
and the simulations. 
In other words, Fig.~\ref{fig_Qsk_hm} is not meant to show the actual reduced
bispectrum (\ref{Qdef}), which was displayed in Fig.~\ref{fig_Qk_eq} for
equilateral triangles, but only to show the bispectrum $B$ on a more convenient
scale.
We focus on quasilinear scales where the different models can be distinguished,
since on very large or small scales they converge towards the tree-level
contribution (\ref{Ba-def}) or the 1-halo contribution with a negligible counterterm.

First, we can see in Fig.~\ref{fig_Qsk_hm} that discarding the contribution of
1-loop diagrams (the diagrams (b) to (i) in the representation of
Fig.~\ref{fig_C3-1loop}) leads to a significant underestimate of the bispectrum
on weakly nonlinear scales, $k \sim 0.2 h$Mpc$^{-1}$ at $z \leq 3$,
in agreement with previous works based on perturbation theory alone
\citep{Scoccimarro1998b,Sefusatti2010}.
In particular, at $z=0.35$ the 1-loop contribution is necessary to obtain a good
match to the simulations and appears to be sufficient to make the bridge to
the smaller scales where the 1-halo term is dominant. At $z=3$, the
1-loop contribution also extends up to $k \sim 0.4 h$Mpc$^{-1}$ the good
agreement with simulations, while using the tree-level contribution alone
yields significant discrepancies as soon as $k > 0.2 h$Mpc$^{-1}$. However, it is
no longer sufficient to make the bridge with the highly nonlinear scales dominated
by the 1-halo term, because we now underestimate the bispectrum on the transition
scales, $k \sim 1 h$Mpc$^{-1}$. This behavior was already noticed in 
Figs.~\ref{fig_Bk_eq} and \ref{fig_Bk_alpha_2_4}.

Second, setting the counterterms to zero in the 2-halo and 1-halo contributions
(while keeping the 1-loop contribution in the 3-halo term) obviously increases
the bispectrum and worsens the agreement with simulations. This is most
clearly seen on the quasilinear scales, which are already well described by
standard perturbation theory so that the extra power associated with the
unphysical constant asymptotes at low $k$ of the uncorrected 2-halo and
1-halo contributions spoil the good agreement with simulations.

Making simultaneously both approximations, discarding 1-loop diagrams
and halo counterterms, as in usual implementations of the halo model,
also yields significantly worse results. Although both effects partly compensate,
neglecting 1-loop contributions dominates so that the resulting bispectrum is
significantly too low on weakly nonlinear scales.
Therefore, in addition to a more satisfactory physical behavior (systematic
agreement on large scales, up to higher order, and decay of 2-halo and 1-halo
contributions), our approach gives a better accuracy on weakly nonlinear scales.
This shows the importance of including both 1-loop contributions and
the 2-halo and 1-halo counterterms.

\section{Dependence on various ingredients of the model}
\label{ingredients}

We now investigate the impact of various ingredients of the model
on the predictions obtained for the bispectrum.

\subsection{Dependence on the perturbative scheme}
\label{Pert-num}

\begin{figure*}
\begin{center}
\epsfxsize=6.1 cm \epsfysize=5.4 cm {\epsfbox{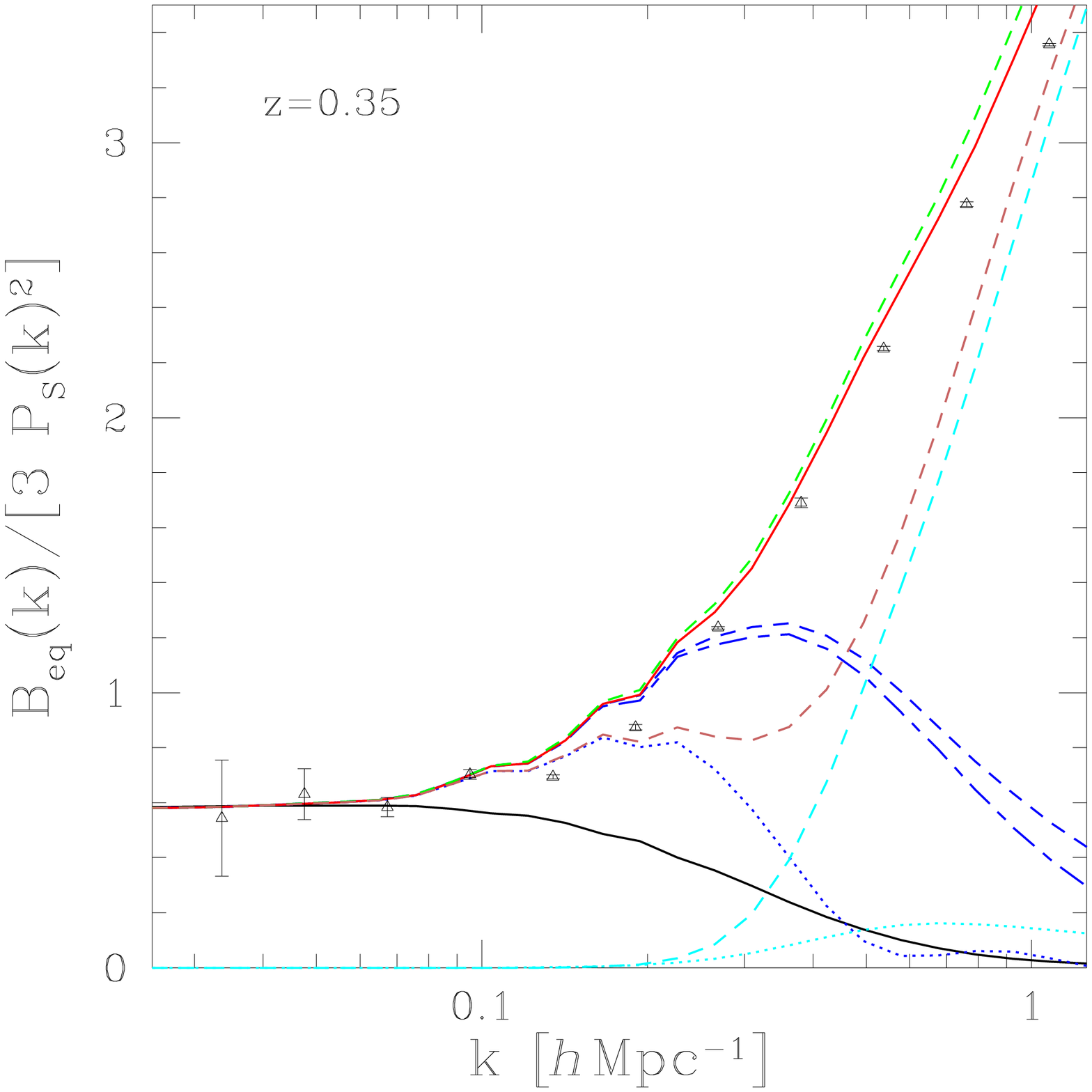}}
\epsfxsize=6.05 cm \epsfysize=5.4 cm {\epsfbox{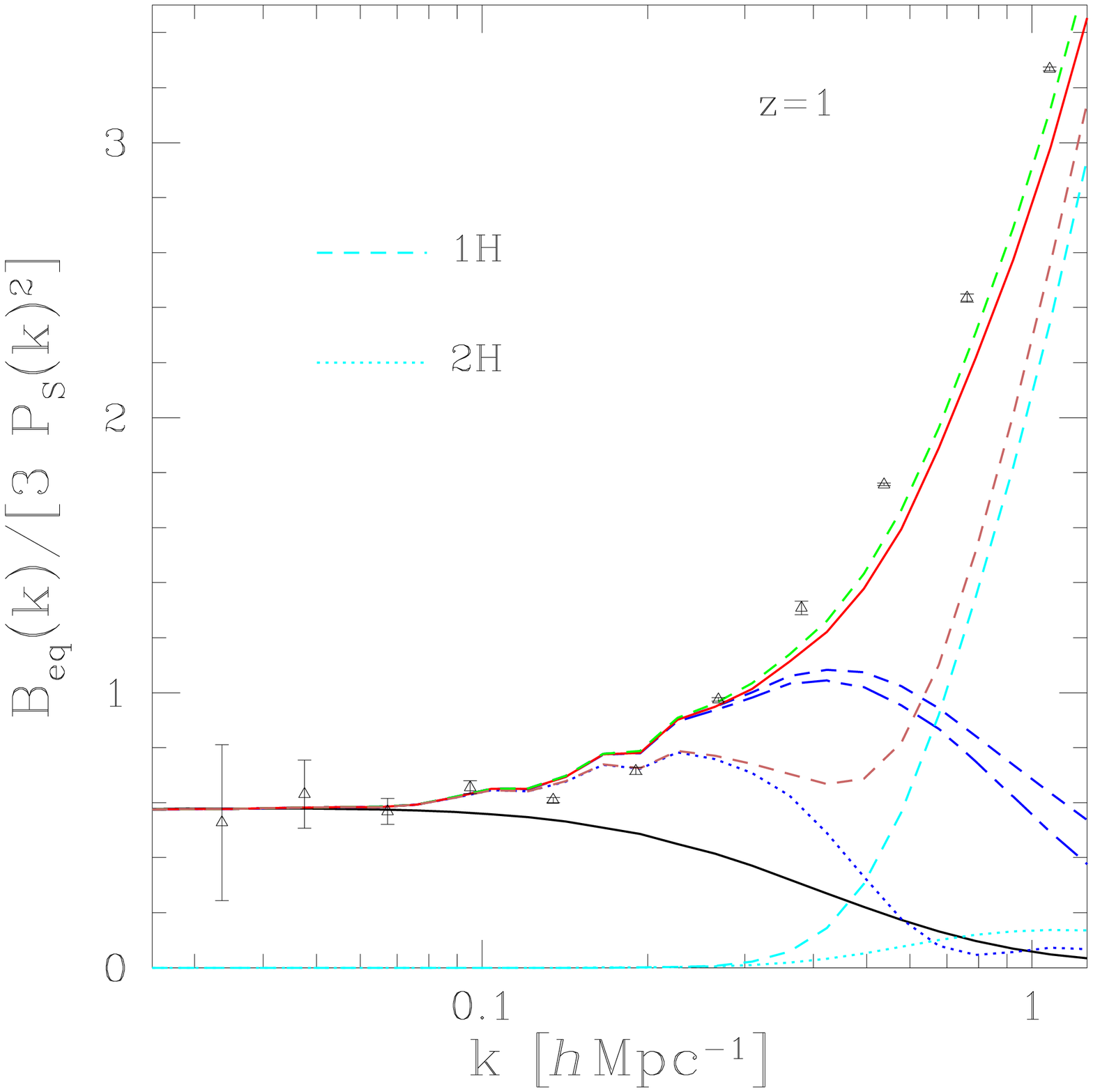}}
\epsfxsize=6.05 cm \epsfysize=5.4 cm {\epsfbox{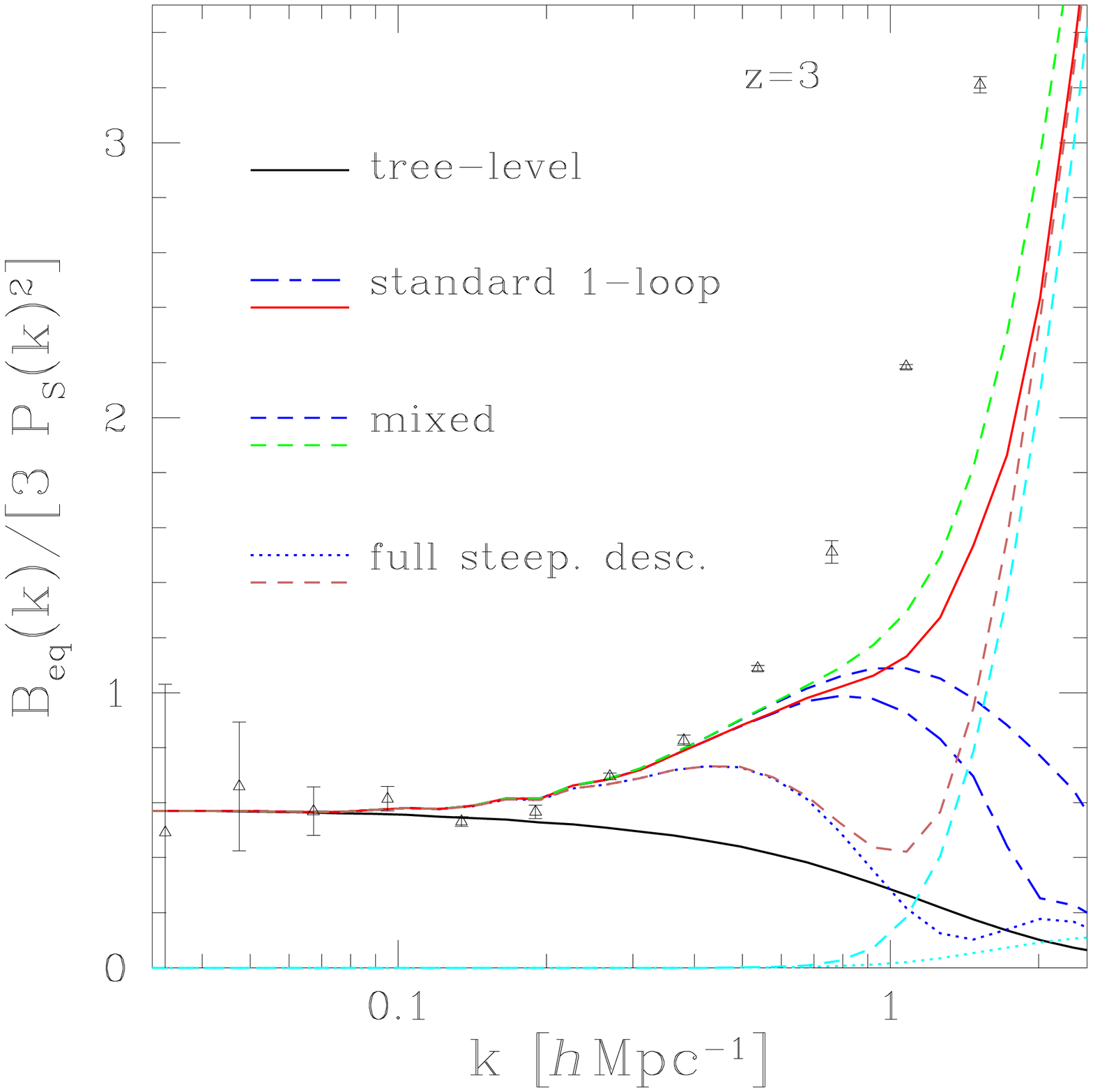}}\\
\epsfxsize=6.1 cm \epsfysize=5.4 cm {\epsfbox{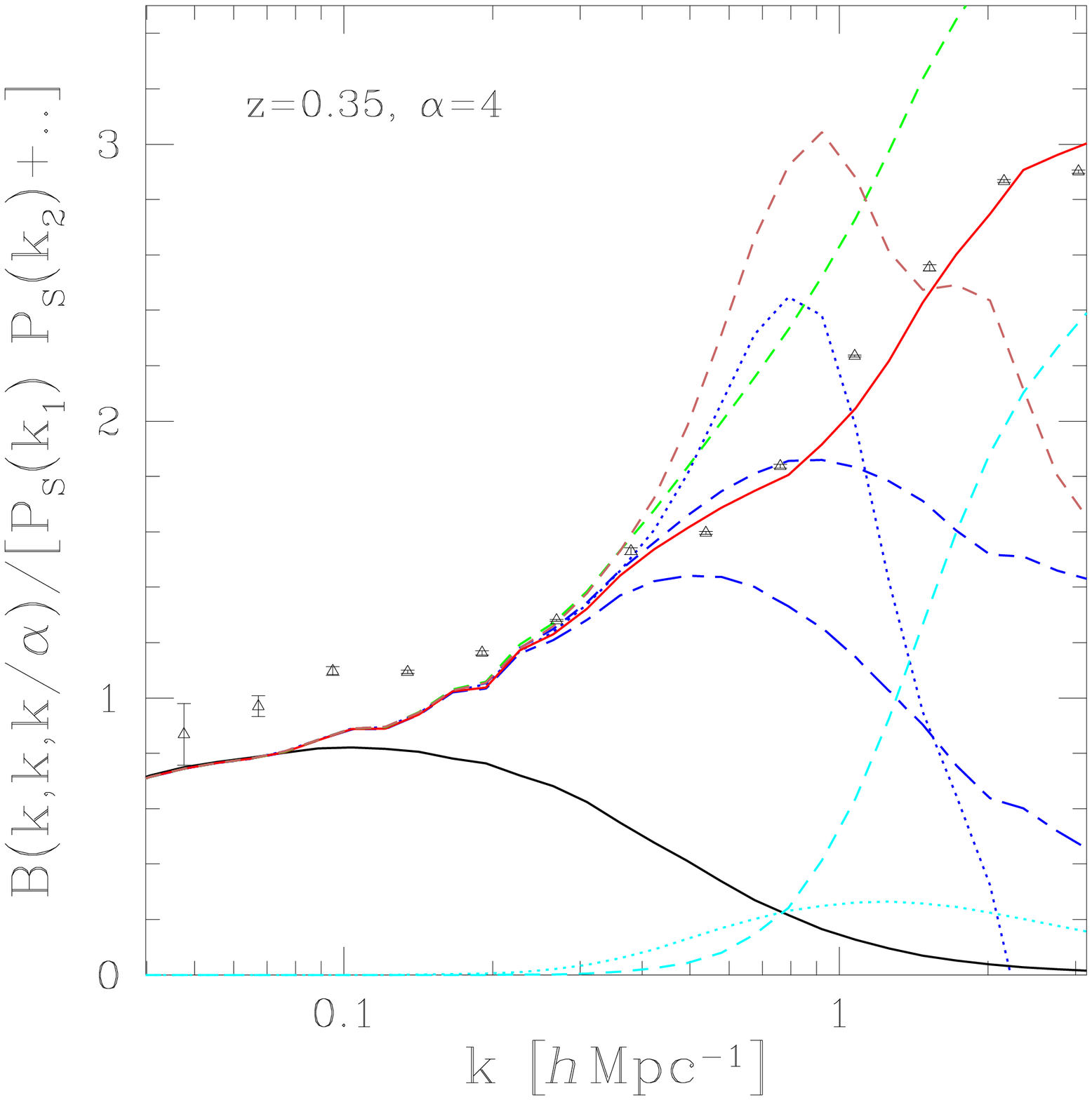}}
\epsfxsize=6.05 cm \epsfysize=5.4 cm {\epsfbox{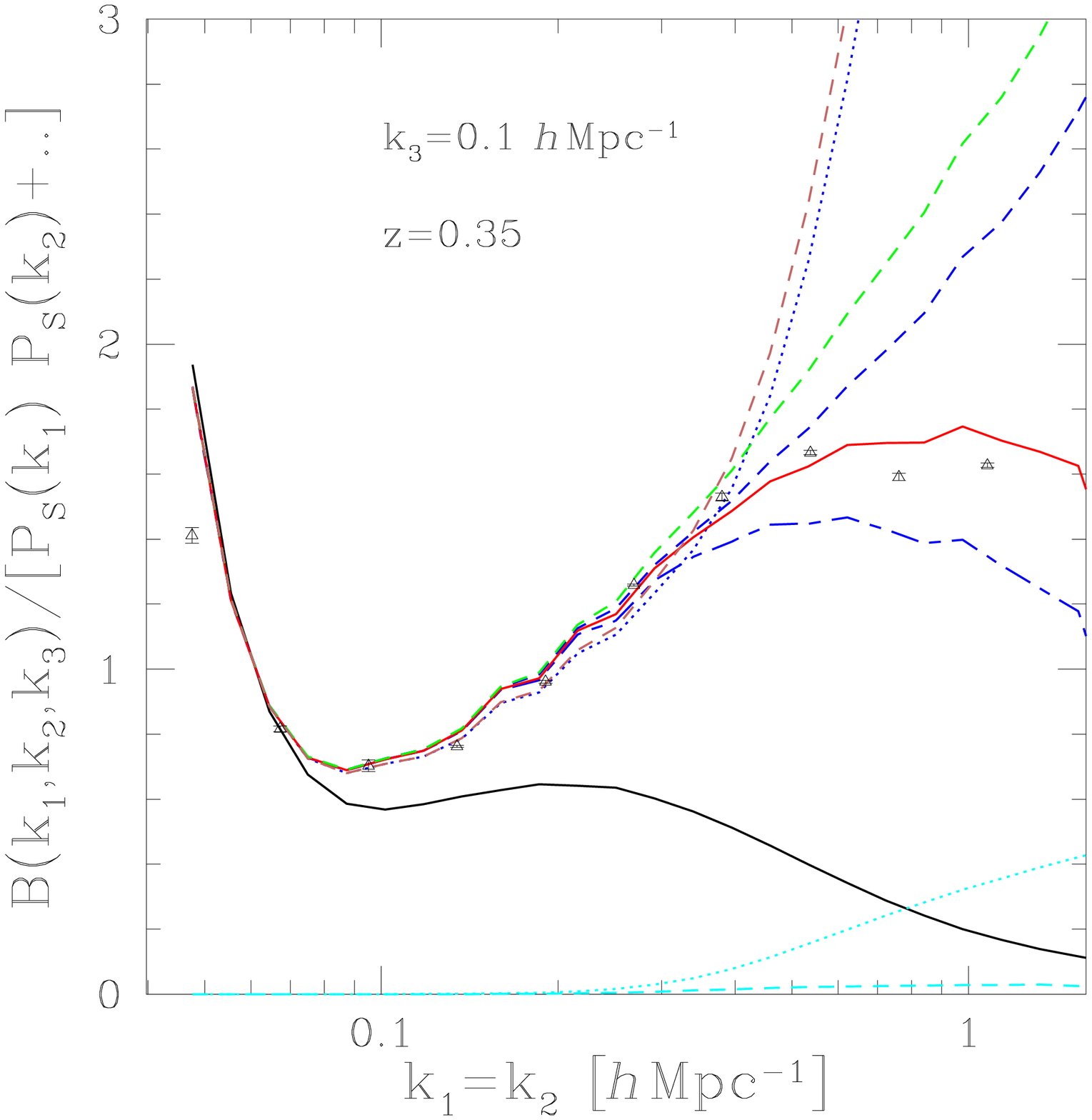}}
\epsfxsize=6.05 cm \epsfysize=5.4 cm {\epsfbox{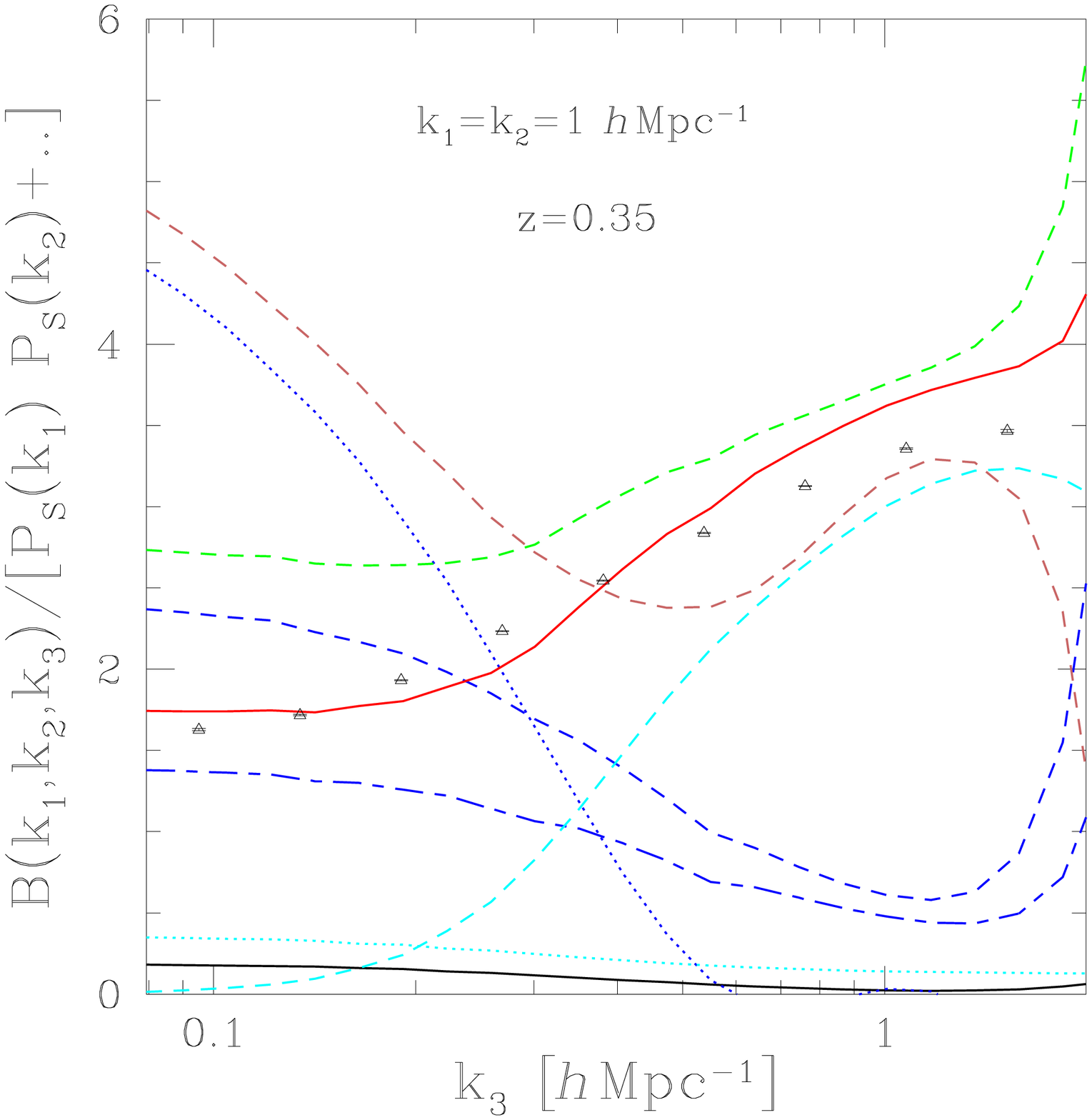}}
\end{center}
\caption{Ratio $B(k_1,k_2,k_3)/[P_{\rm S}(k_1) P_{\rm S}(k_2)+ 2 {\rm cyc.}]$,
as in Fig.~\ref{fig_Qsk_hm}. We show the results obtained by using for the
3-halo term each of the three 1-loop perturbative expansions given in
Figs.~\ref{fig_C3-1loop}, \ref{fig_C3-sd}, and \ref{fig_C3-mixed}. 
In each case, we show both the 3-halo contribution alone (which
decays at high $k$, upper line in front of the labels) and the full result
(i.e., the sum of the 3-halo, 2-halo, and 1-halo terms, lower line in front of the
labels). We also plot the tree-level perturbative result (black solid line).
{\it Upper row:} equilateral configurations, at redshifts $z=0.35, 1$, and $3$, 
as in Fig.~\ref{fig_Bk_eq}. {\it Lower row:} various isosceles configurations
at redshift $z=0.35$, as in left panels of Figs.~\ref{fig_Bk_alpha_2_4} and
\ref{fig_Bk_k3_k1k2}.}
\label{fig_Qsk}
\end{figure*}

We investigate in this section the dependence of our results on the
perturbative scheme used for the 3-halo contribution. 
As in Fig.~\ref{fig_Qsk_hm}, the bispectra obtained from the analytical models
and the numerical simulations are divided by the same power spectra, from
\citet{Smith2003}, to make a clear comparison.

We compare in Fig.~\ref{fig_Qsk} the results obtained using for the perturbative
3-halo contribution either the standard 1-loop result of Fig.~\ref{fig_C3-1loop},
the full steepest-descent resummation of Fig.~\ref{fig_C3-sd}, or the
mixed case of Fig.~\ref{fig_C3-mixed}. For each choice we show both
the 3-halo term itself (blue curves that decrease at high $k$) and the full
prediction, where we sum the 3-halo, 2-halo, and 1-halo contributions.
(The 2-halo and 1-halo contributions are the same for all three choices.)

Let us first consider the upper row, which corresponds to equilateral triangles.
(At $z=3$ we can see the underestimate of the bispectrum in the transition
range that we had already noticed in Fig.~\ref{fig_Bk_eq}.)
The standard and mixed cases give very close results. At $z=3$, the mixed
case, giving a slightly greater 3-halo contribution, yields a slightly better
agreement with numerical simulations around $1 h$Mpc$^{-1}$.
However, this is not conclusive, especially since the 1-halo term is already
non-negligible on these scales and it is not sufficient to significantly improve
the match to simulations.
The complete steepest-descent resummation of Fig.~\ref{fig_C3-sd} yields
a 3-halo contribution that decays significantly faster at high $k$.
At $z=0.35$ and $z=1$, two numerical points in the range
$0.1<k<0.2 h$Mpc$^{-1}$
suggest that this may be a true improvement on these quasilinear scales.
This improvement might be important for the analysis of baryon acoustic oscillations; 
it has been shown that resummation schemes give better predictions for the 
power spectrum on this scale (e.g., \citet{Crocce2008,Nishimichi2009,Valageas2010b},
and we could expect 
the same thing for the bispectrum. However, beyond $0.2 h$Mpc$^{-1}$ the faster 
decay significantly worsens the agreement with the simulations.

The upper row of Fig.~\ref{fig_Qsk} does not allow us to clearly discriminate
between the various schemes because they mainly differ in the transition regime,
where the 1-halo contribution is already important and all models tend to
underestimate the bispectrum. However, looking now at the lower row, associated
with isosceles triangles, and especially at the two lower right panels that show the
evolution of the bispectrum with the shape of the triangle, we can see that
the standard 1-loop perturbation theory yields a significantly better agreement
with simulations than the two other perturbative expansions.
This is also the simplest scheme for practical purposes.

Thus, from the analysis of both equilateral and isosceles triangles, we can conclude
that using  the standard 1-loop perturbation theory for the perturbative 3-halo
term is the most efficient and accurate scheme among the three approaches
studied in this paper.
This is quite different from the power spectrum, studied in \citet{Valageas2010b},
where the 1-loop steepest-descent resummation is clearly better than standard
perturbation theory. There, it is more accurate on weakly nonlinear scales but it
is also the only choice (with other resummation schemes) that allows a combined
model for both large and small scales. Indeed, the standard 1-loop perturbation theory
yields a 2-halo contribution that keeps growing on small scales (for the
logarithmic power of Eq.(\ref{Deltak-def}) below) and that worsens the agreement
with simulations (even though it is smaller than the 1-halo term).

We do not have this small-scale problem for the bispectrum, as seen in
Figs.~\ref{fig_Bk_eq}, \ref{fig_Bk_alpha_2_4}, and \ref{fig_Qk_eq}. Indeed,
the 1-loop standard perturbation theory gives a contribution to the bispectrum
that quickly decreases at high $k$ and does not spoil the good agreement with
simulations shown by the 1-halo term in this regime.
This is best seen in Fig.~\ref{fig_Qk_eq}, which shows that the 1-loop standard
perturbation theory contribution to the reduced equilateral bispectrum
$Q_{\rm eq}(k)$ itself also decreases at high $k$, in spite of the denominator
$3P(k)^2$.
Therefore, contrary to the power spectrum, it is now possible to use the 1-loop
standard perturbation theory to build a combined model that covers all scales.
Moreover, as shown by Fig.~\ref{fig_Qsk}, it happens that this is also more
accurate than the two other perturbative schemes investigated in this paper,
which we presented in Figs.~\ref{fig_C3-sd} and \ref{fig_C3-mixed}.

It is possible that other perturbative schemes that we did not study here provide more
accurate results. On the other hand, higher orders of standard perturbation theory
may yield contributions that are increasingly large on small scales, so that
beyond a certain order they lead to a 3-halo term that is unphysically large
and spoils the good agreement with simulations on small scales.
Then, one would need to use other perturbative approaches, such as the
resummation schemes investigated here, or to add some ad-hoc cutoff on small
scales.

For completeness, we mention a few different approaches. 
A first method, introduced in \citet{Crocce2006a,Crocce2006b} from a
diagrammatic approach, reorganizes the standard perturbation theory
and performs partial resummations by introducing both correlation functions
and propagators (or response functions), in a fashion somewhat similar to the
approach described in Sect.~\ref{Perturbative-contribution}. However, a key step in
this method is the matching between the low-order low-$k$ behavior and
the high-$k$ asymptote of the propagator. This ensures a good behavior of this
quantity in the nonlinear regime, which has been checked against numerical
simulations \citep{Crocce2006b,Bernardeau2008}, and expresses a well-understood
``sweeping effect'' associated with the random transport of density structures by
large-scale velocity flows 
\citep{Valageas2007b,BernardeauVal2008,BernardeauVal2010a}.
Another advantage of this approach is that its extensions to high-order quantities 
\citep{Bernardeau2008}, to non-Gaussian initial conditions \citep{Bernardeau2010},
and to higher perturbative orders \citep{Anselmi2011}, have already been
studied.

A second approach, developed in \citet{Valageas2007a,Valageas2007b} as
a ``large-$N$ 2PI'' expansion, and in \citet{Taruya2008,Taruya2009} as a
``closure approximation'', is quite similar to the one described in 
Sect.~\ref{Perturbative-contribution}. Although in principle it may give more
accurate results, its full implementation is more complex because self-energy
terms depend on the nonlinear quantities that are looked for. This leads to
coupled nonlinear equations, which depend on scales and times, and to heavier
numerical computations \citep{Valageas2007a}. Because for practical purposes one
requires fast algorithms this presents a significant disadvantage, unless one
introduces further simplifications \citep{Taruya2009}.

A third method, introduced in \citet{Pietroni2008}, directly obtains the hierarchy
of equations obeyed by the many-body correlation functions from the
equations of motion, in a fashion similar to the standard BBGKY hierarchy. 
Then, truncation at a given order (at the trispectrum in \citet{Pietroni2008})
defines an approximation for lower-order correlations. This method has already
been used to study the effect of massive neutrinos \citep{Lesgourgues2009}
and of primordial non-Gaussianities \citep{Bartolo2010}. An advantage of this
approach is that it does not involve propagators, but only the usual many-body
correlations also encountered in standard perturbation theory, and it is written
in terms of single-time quantities. This is a great simplification that should
allow efficient numerical computations.

A fourth approach, developed in \citet{Matsubara2008,Matsubara2008a},
is based on a Lagrangian framework. This would be well-suited to the approach
described in this paper and presents the advantage that it provides direct
extensions to redshift-space statistics. Unfortunately, as noticed in
\citet{Valageas2010b}, in its present form this Lagrangian perturbation theory
does not fare as well as its Eulerian counterparts when tested against numerical
simulations. However, this approach remains interesting, especially in view of
applications to redshift space.  

In any case, it would not be too difficult to combine any of these methods with
halo models by using the framework described in this paper, which is more general
than the use of the standard perturbation theory or the two resummation schemes
described in Sect.~\ref{Perturbative-contribution}.

\subsection{Dependence on halo properties}
\label{depend-halo}

\begin{figure}
\begin{center}
\epsfxsize=7.8 cm \epsfysize=5.4 cm {\epsfbox{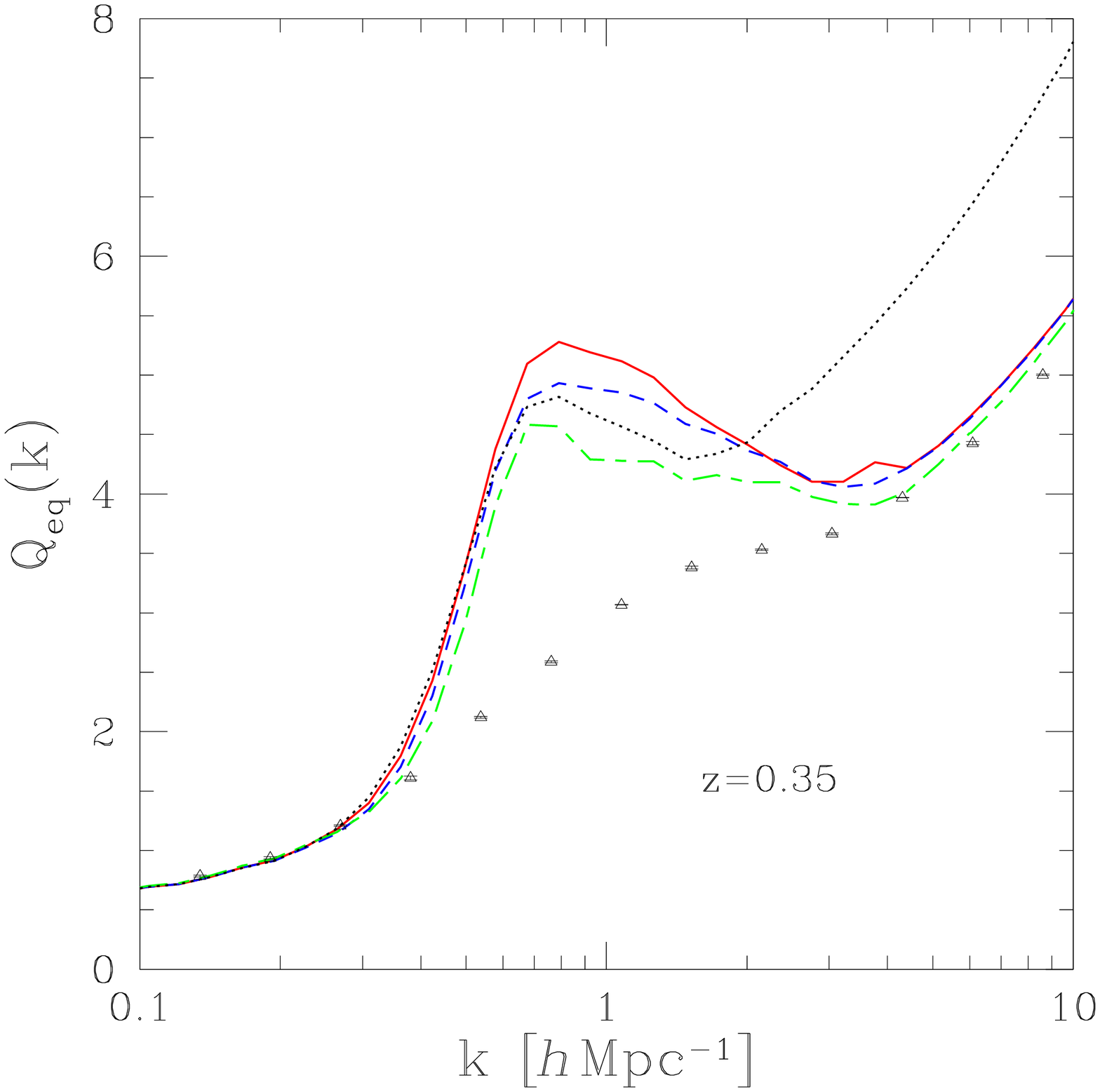}}\\
\epsfxsize=7.8 cm \epsfysize=5.4 cm {\epsfbox{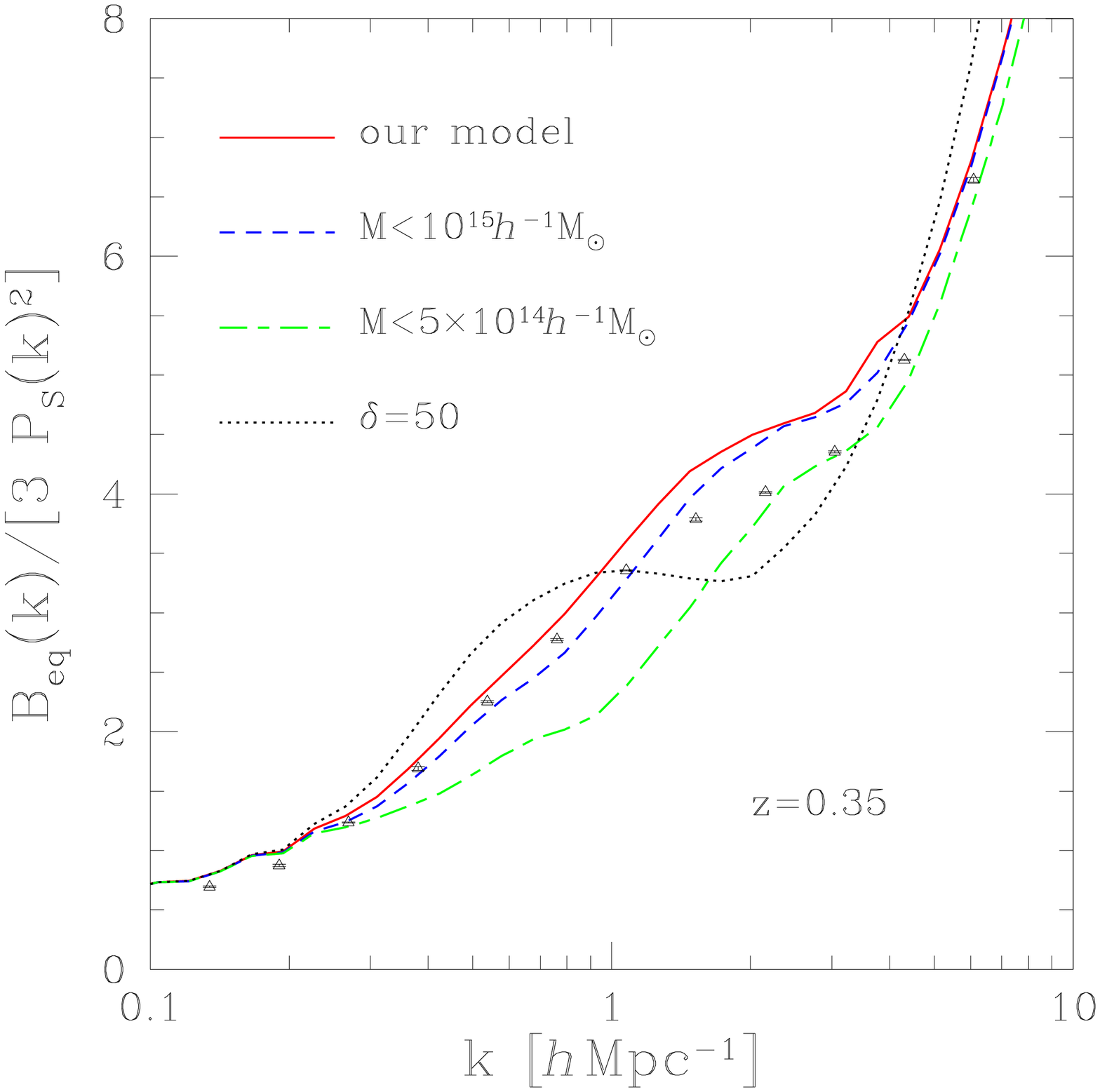}}
\end{center}
\caption{Reduced bispectrum (upper panel, as in Fig.~\ref{fig_Qk_eq}) and the
bispectrum scaled by $3P_{\rm S}(k)^2$ (lower panel, as in the upper row of
Fig.~\ref{fig_Qsk}) obtained for modified halo properties on equilateral
triangles at $z=0.35$. We plot our fiducial model (red solid line, as in previous
figures),
the impact of a high mass cutoff ($M<10^{15}h^{-1}M_{\odot}$, blue dashed
line; $M<5\times 10^{14}h^{-1}M_{\odot}$, green dot-dashed line), and the
effect of defining halos by the lower nonlinear density threshold $\delta=50$
(black dotted line).}
\label{fig_Qk_M}
\end{figure}

We now investigate the dependence of our results on the details of the halo
model. As we recalled in Sect.~\ref{Q-comp}, the ``bump'' shown by the
reduced bispectrum in Fig.~\ref{fig_Qk_eq} was also found in previous
works, which noticed that this feature may be cured to some extent by
tuning halo parameters \citep{Scoccimarro2001,Takada2003,Fosalba2005}.
Therefore, we show in Fig.~\ref{fig_Qk_M} the reduced bispectrum (upper
panel, as in Fig.~\ref{fig_Qk_eq}) and the scaled bispectrum (lower panel, as in
Fig.~\ref{fig_Qsk}) for several modifications of halo parameters at redshift
$z=0.35$.

\begin{figure}
\begin{center}
\epsfxsize=7.8 cm \epsfysize=5.4 cm {\epsfbox{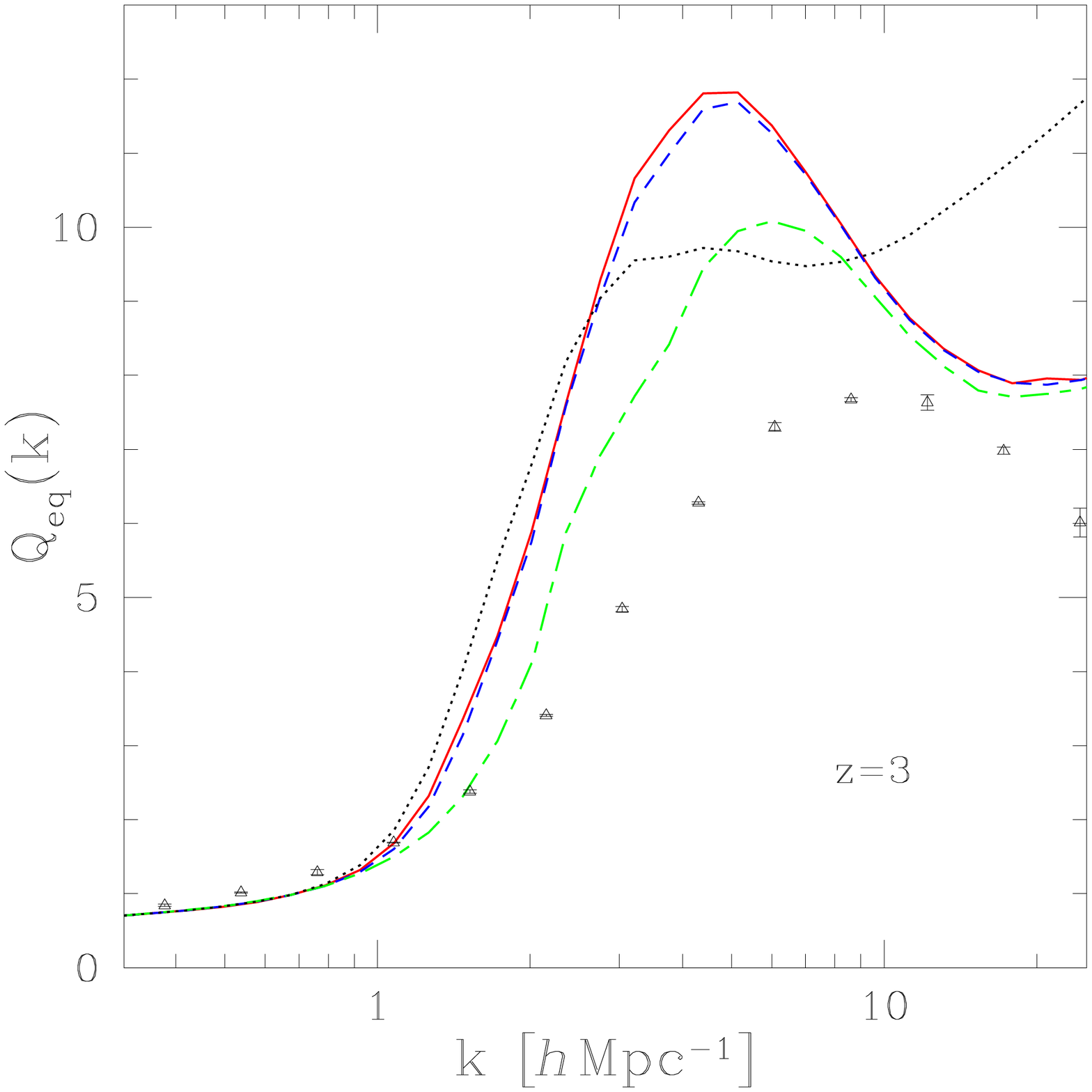}}\\
\epsfxsize=7.8 cm \epsfysize=5.4 cm {\epsfbox{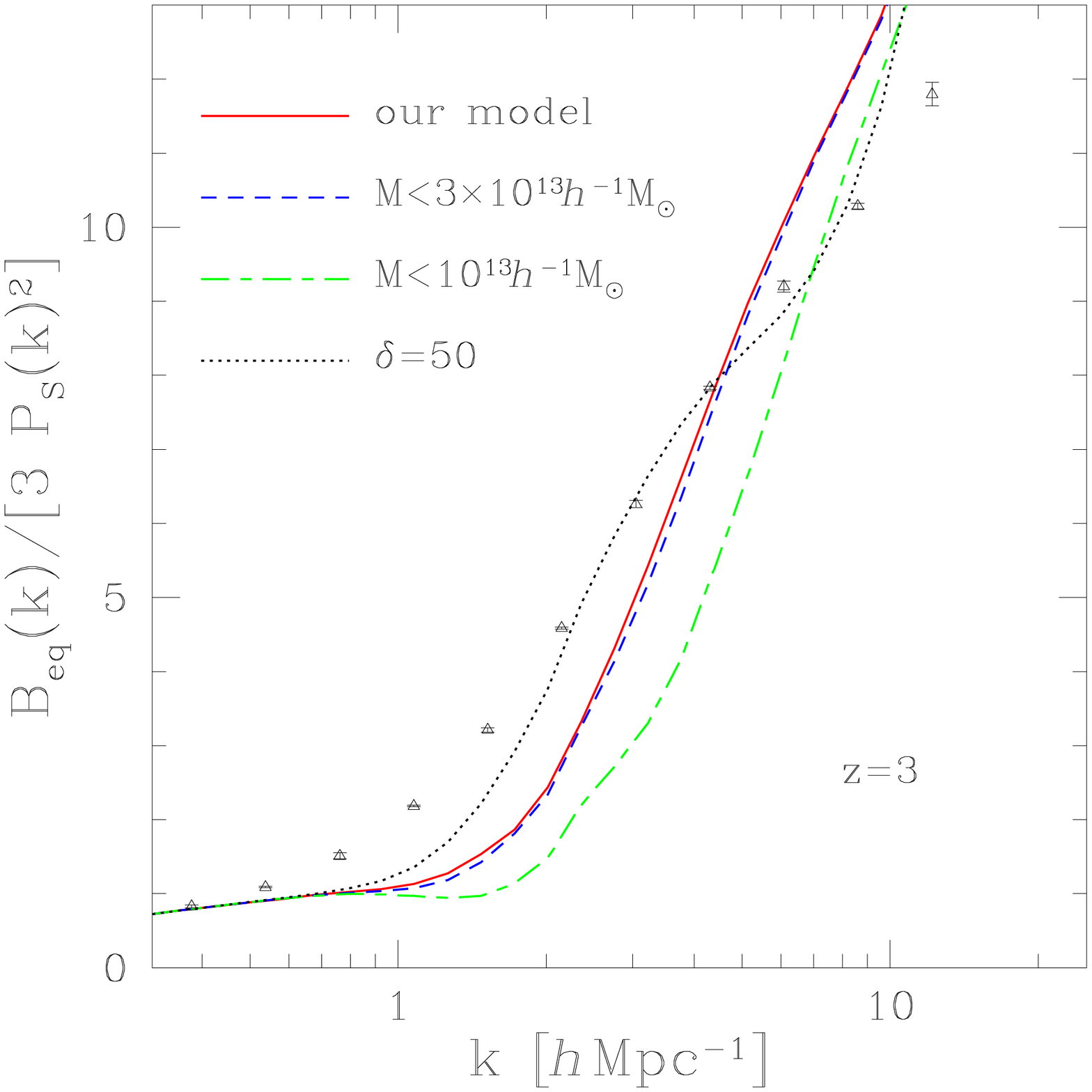}}
\end{center}
\caption{Reduced bispectrum (upper panel) and the
bispectrum scaled by $3P_{\rm S}(k)^2$ (lower panel) obtained for modified
halo properties on equilateral triangles, as in Fig.~\ref{fig_Qk_M} but at $z=3$.}
\label{fig_Qk_M_z3}
\end{figure}

Following \citet{Scoccimarro2001,Wang2004,Fosalba2005}, we investigate
the impact of truncating the halo mass function below $M<10^{15}$ or
$M<5\times 10^{14}h^{-1}M_{\odot}$. In agreement with these studies,
removing large halos decreases both the matter power spectrum and bispectrum
on mildly nonlinear scales (since smaller scales are associated with small halos),
and the net effect on the reduced bispectrum is also a small decrease of the
artificial ``bump''. However, the upper panel shows that even truncating at
$M<5\times 10^{14}h^{-1}M_{\odot}$ is not sufficient to erase the ``bump'' and
to provide a good agreement with simulations for the reduced bispectrum.
Two differences with previous studies are that the 3-halo term includes
1-loop perturbative contributions, which are still important on these scales
as shown in Figs.~\ref{fig_Qk_eq} and \ref{fig_Qsk}, and that the 1-halo and
2-halo contributions contain the counterterms $\tW$, so that the
relative importance of the 2-halo and 1-halo contributions is somewhat
smaller than in previous models on the weakly nonlinear scales associated
with the early rise of the ``bump''.
On the other hand, because both the power spectrum and the bispectrum
simultaneously decrease when we add such a high-mass cutoff and these effects
partly compensate in the reduced bispectrum, before we obtain a significant
improvement for the latter quantity both the power spectrum and the bispectrum
have already been decreased by a large amount that disagrees with numerical
results, as shown in the lower panel for the bispectrum.
Moreover, in our large simulations we find halos up to
$3\times 10^{15}h^{-1}M_{\odot}$, so that the cutoff at
$M<5\times 10^{14}h^{-1}M_{\odot}$ is already too low to be fully justified by
the finite box size of the simulations.

Then, following \citet{Fosalba2005}, we investigate the impact of larger halo
radii. More precisely, we consider the impact of defining halos by a nonlinear
density threshold $\delta=50$ instead of $\delta=200$, following the
procedure described in Sect.~6.1 of \citet{Valageas2010b}. 
This also involves a slightly different halo mass - concentration parameter relation,
so as to keep a satisfactory match to numerical results for the power spectrum.
As seen in Fig.~\ref{fig_Qk_M}, this modification does not bring a significant
improvement either.

We find similar results at higher redshifts, as seen in Fig.~\ref{fig_Qk_M_z3}
at $z=3$ (where we consider the smaller mass thresholds
$M<3\times 10^{13}$ and $M<10^{13}h^{-1}M_{\odot}$).

Therefore, we reach the same conclusion as \citet{Guo2009}, that tuning these
halo parameters cannot simultaneously provide accurate results for the
power spectrum and bispectrum, especially compared with large
simulations where massive halos are present, which removes the justification
for introducing severe high-mass cutoffs.

The discrepancies found on these transition scales are not so surprising
because one expects the transition range to be the most difficult
to describe in a systematic fashion, since both the perturbative expansion
(that neglects shell crossing) and the halo model (which assumes spherical and
relaxed objects in our implementation) may break down in this intermediate
regime. Another possibility is that we are missing important high-order
perturbative terms that have not been included in the perturbative expansions
used here, which are only complete up to one-loop order.
Indeed, for the power spectrum perturbative terms up to order $\sim 9$ (at $z=0$)
or $\sim 66$ (at $z=3$) are likely to be relevant \citep{Valageas2010a}.
On the halo-model side, because in reality the density field on these transition scales
shows a crossover from relaxed inner halo shells to outer infalling shells
and filaments that cannot any longer be described in terms of well-defined
halos, one cannot expect a systematic convergence to the numerical results
by adding such simple modifications to halo properties. 
Then, one is probably
limited to some extent by the intrinsic limitations of the halo model itself.

\section{Improving the predictions for $P(k)$ and $B_{\rm eq}(k)$ using the reduced bispectrum $Q_{\rm eq}(k)$}
\label{Improving}

\begin{figure*}
\begin{center}
\epsfxsize=6.1 cm \epsfysize=5.4 cm {\epsfbox{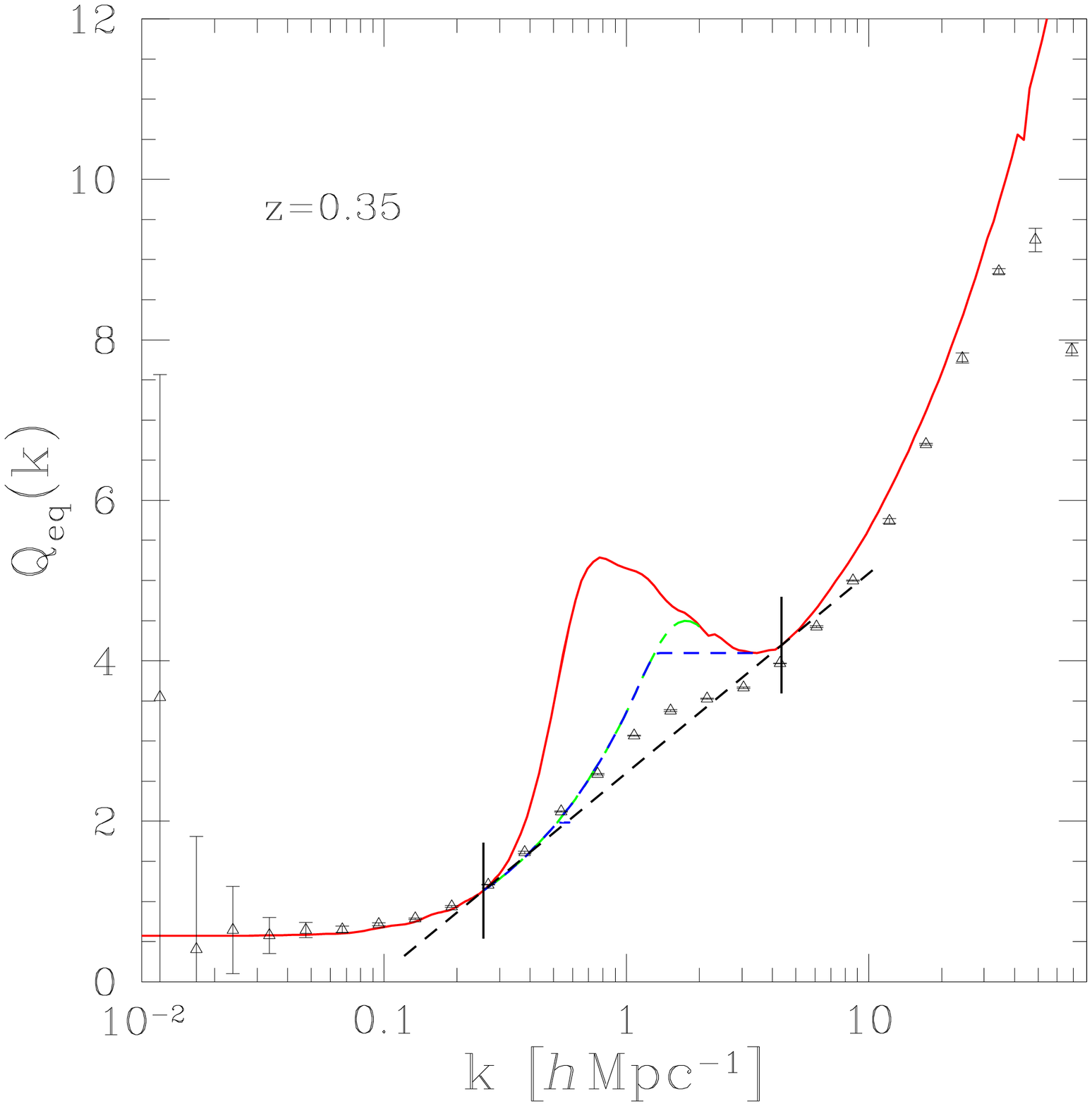}}
\epsfxsize=6.05 cm \epsfysize=5.4 cm {\epsfbox{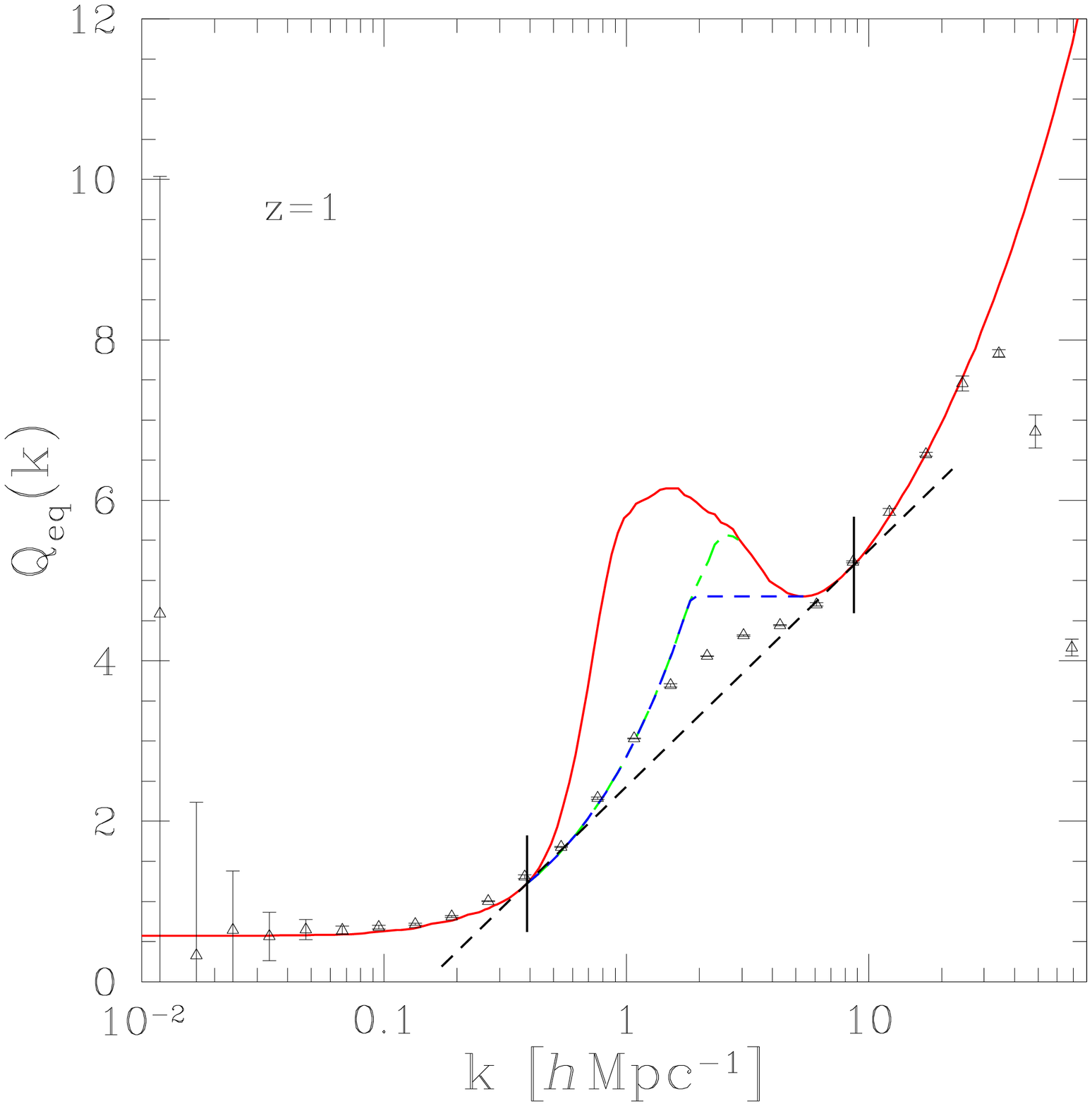}}
\epsfxsize=6.05 cm \epsfysize=5.4 cm {\epsfbox{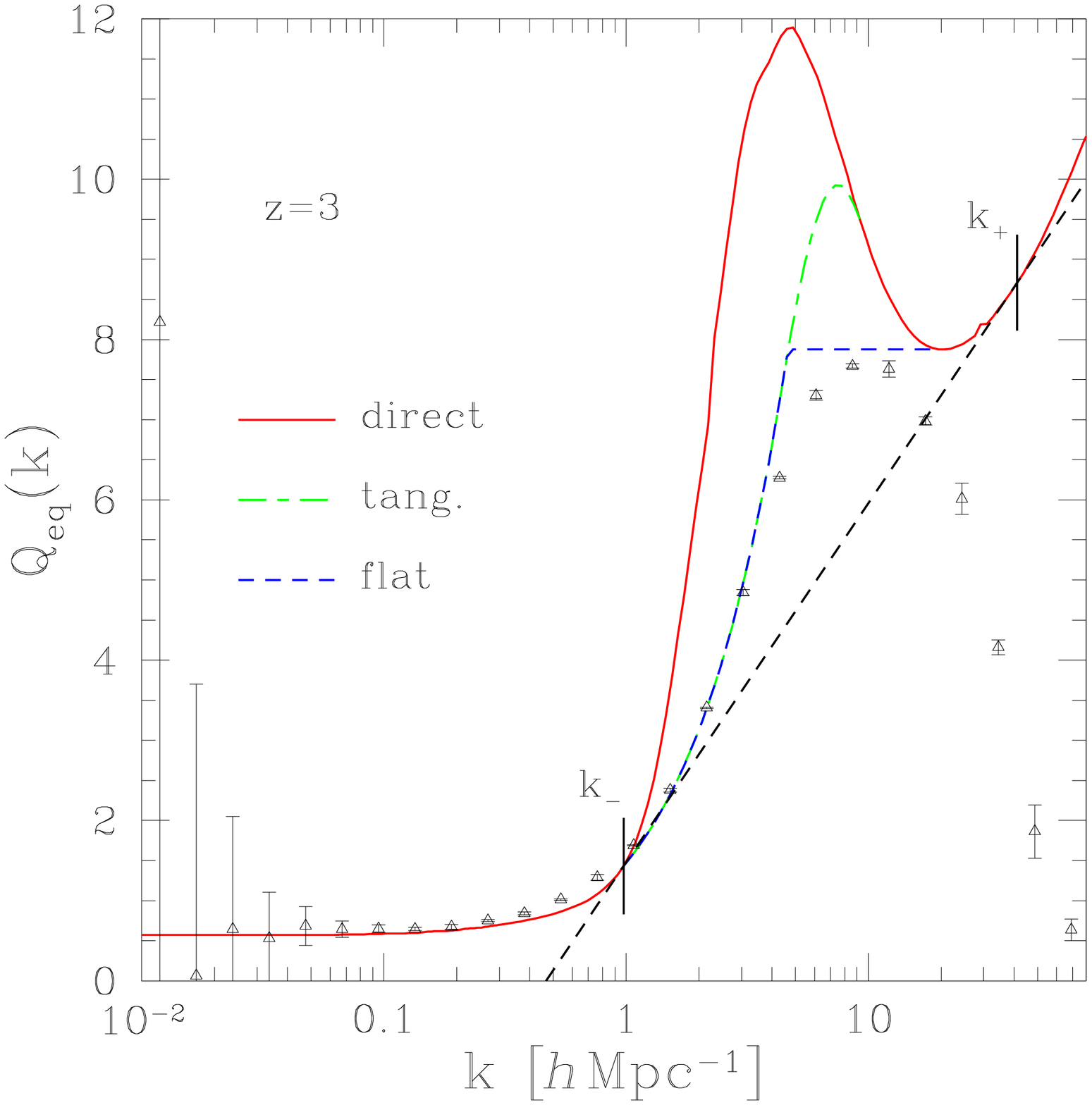}}
\end{center}
\caption{Reduced bispectrum, $Q_{\rm eq}(k)=B_{\rm eq}(k)/[3P(k)^2]$, for
equilateral configurations, at redshifts $z=0.35, 1$, and $3$. The points are the
results from the numerical simulations and the solid line is our ``direct'' model,
as in Fig.~\ref{fig_Qk_eq}. 
The black dashed line is the lower tangent to the ``direct'' curve $Q_{\rm eq}(k)$,
on the transition region. It defines the two contact points of abscissa $k_-$
and $k_+$, shown by the two small vertical lines, between which we introduce
the two modifications ``tang.'' (green dot-dashed line) and ``flat'' (blue
dashed line), explained in the main text.}
\label{fig_Qk_eq_mod}
\end{figure*}

The artificial ``bump'' found in Fig.~\ref{fig_Qk_eq} for the reduced equilateral
bispectrum $Q_{\rm eq}(k)$, which is mostly caused by the underestimation of the power 
$P(k)$, suggests a simple trick to improve the predictions for $P(k)$ and $B_{\rm eq}(k)$.
The idea is to use the unphysical ``bump'' shown by the predicted reduced
bispectrum $Q_{\rm eq}(k)$ to automatically detect the range $[k_-,k_+]$ where
the model is not sufficiently accurate. The procedure that we investigate in this
paper is to draw in the $(\log k, Q_{\rm eq})$ plane the lower tangent line
to the predicted curve that was plotted in Fig.~\ref{fig_Qk_eq}.
We show this construction in Fig.~\ref{fig_Qk_eq_mod}, where we again plot the
prediction of our model, described in the previous sections and labeled ``direct''
in this figure, as well as the lower tangent on the region where the ``bump''
appears.
The two contact points between the model curve and its tangent line define the
two wavenumbers, $k_-$ and $k_+$, between which the artificial ``bump''
arises and the model needs to be improved.

\begin{figure}
\begin{center}
\epsfxsize=8.2 cm \epsfysize=6.2 cm {\epsfbox{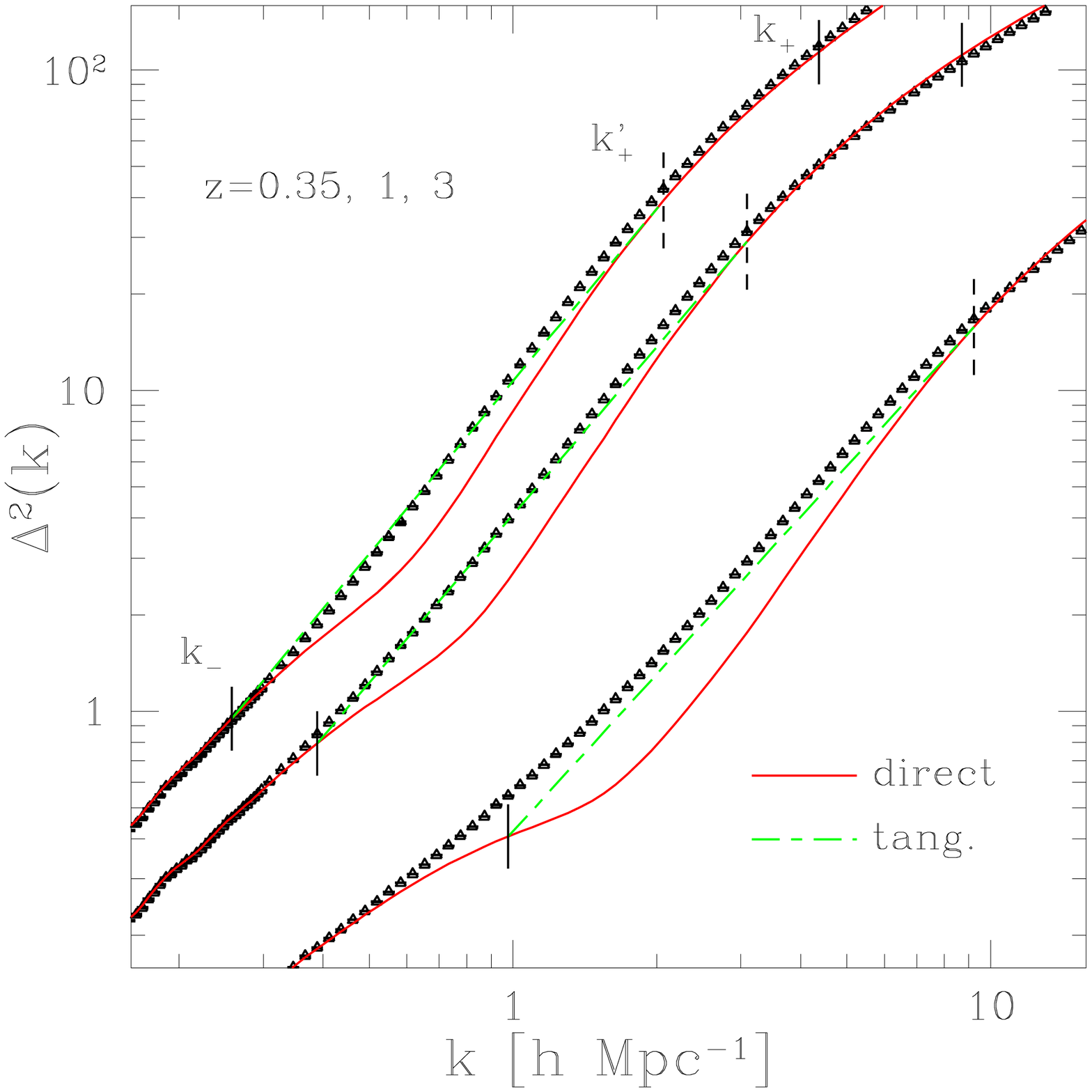}}
\end{center}
\caption{Power per logarithmic interval of $k$, as defined in
Eq.(\ref{Deltak-def}), at redshifts $z=0.35$, $1$, and $3$ (from top to bottom).
We show the ``direct''
model (red solid line), studied in \citet{Valageas2010b}, and the geometrical
modification ``tang.'' (green dot-dashed line), given by the upper tangent
that runs through the point of abscissa $k_-$. The small dashed vertical line
shows the location $k_+'$ of the contact point of this tangent line with the
``direct'' curve.}
\label{fig_lDk}
\end{figure}

First, we modify the density power spectrum as shown in Fig.~\ref{fig_lDk},
where we plot the power per logarithmic interval of $k$, defined as
\beq
\Delta^2(k) = 4\pi k^3 P(k) .
\label{Deltak-def}
\eeq
The modified power ``tang.'' is obtained from the ``direct'' prediction
described in \citet{Valageas2010b} by drawing in the $(\log k, \log \Delta^2)$ 
plane the upper tangent line on the interval $[k_-,k_+]$ that runs through the
left point $(\log k_-, \log \Delta^2(k_-))$. For the cases shown in Fig.~\ref{fig_lDk}
the right contact point has an abscissa $k_+'<k_+$ (by construction we have
$k_-<k_+' \leq k_+$), and the power spectrum is only modified on the interval
$[k_-,k_+']$. 
As seen in Fig.~\ref{fig_lDk}, in this fashion we correct most of the artificial
``dip'' shown by our model without modifying the large-scale and small-scale
regimes where the ``direct'' predictions are 
satisfactory\footnote{
Although we show in Fig.~\ref{fig_lDk} this geometrical construction for the
logarithmic power $\Delta^2(k)$, applying the
same construction to the power $P(k)$, that is, in the $(\log k, \log P)$ plane,
yields the same results (since the abscissa $k_+'$ of the contact point with the
upper tangent line is the same).}.

\begin{figure}
\begin{center}
\epsfxsize=8.2 cm \epsfysize=6.2 cm {\epsfbox{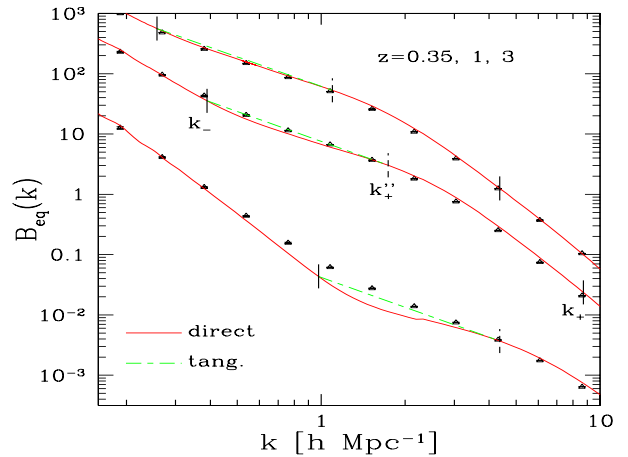}}
\end{center}
\caption{Equilateral bispectrum at redshifts $z=0.35$, $1$, and $3$
(from top to bottom). We show the ``direct'' model (red solid line), already shown
in Fig.~\ref{fig_Bk_eq},
and the geometrical modification ``tang.'' (green dot-dashed line) given by the
upper tangent that runs through the point of abscissa $k_-$. The small dashed
vertical line shows the location $k_+''$ of the contact point of this tangent line
with the ``direct'' curve.}
\label{fig_lBk_eq}
\end{figure}

Second, for equilateral triangles we modify the bispectrum in the same manner.
Thus, as shown in Fig.~\ref{fig_lBk_eq}, we again draw in the
$(\log k,\log B_{\rm eq})$ plane the upper tangent line, on the interval $[k_-,k_+]$,
that runs through the left point $(\log k_-,\log B_{\rm eq}(k_-))$.
This yields another right contact point at $k_+''$, which in the cases shown in
Fig.~\ref{fig_lBk_eq} is located to the left of $k_+$ as there is again a knee
in the shape of the bispectrum.
This ensures that we follow the ``direct'' prediction beyond the
knee, where it shows a good match to the numerical simulations, while
correcting most of the artificial ``dip'' shown by the model.

As explained above, the interest of this procedure, based on the reduced
equilateral bispectrum $Q_{\rm eq}(k)$, is to automatically define the
wavenumbers $k_-$ and $k_+$. One could imagine defining 
these boundaries in a simpler manner from the power spectrum itself, for
instance by $\Delta^2(k_{\pm})=\Delta^2_{\pm}$, with fixed values
$\Delta^2_{\pm}$ that would mark the transition regime (such as
$\Delta^2_-=1$ and $\Delta^2_+=30$). However, owing to the curvature of
the linear CDM power spectrum, the interval $[k_-,k_+]$ where the model
departs from the numerical simulations is not given by these redshift-independent
conditions, as can be seen in Fig.~\ref{fig_lDk}.
In particular, at higher redshift (i.e., lower value of the effective slope $n$ of the
linear power spectrum on the relevant scales) the threshold $\Delta^2_-$
is seen to decrease. This is partly captured by our procedure, described in
Fig.~\ref{fig_Qk_eq_mod}, which is sensitive to the shape of the initial
conditions, through the shape of the predicted reduced bispectrum
$Q_{\rm eq}(k)$.

Finally, going back to Fig.~\ref{fig_Qk_eq_mod}, from the improved model
``tang.'' constructed for the power spectrum and the bispectrum, shown
in Figs.~\ref{fig_lDk} and \ref{fig_lBk_eq}, we can build a new prediction
``tang.'' for the reduced bispectrum from Eq.(\ref{Qdef}).
For equilateral triangles this reads as
$Q_{\rm eq}^{\rm tang.}(k) = B_{\rm eq}^{\rm tang.}(k)
/[3P^{\rm tang.}(k)^2]$.
The result, plotted in Fig.~\ref{fig_Qk_eq_mod}, shows a significant
improvement over the ``direct'' prediction, which was already
plotted in Fig.~\ref{fig_Qk_eq}. However, we can see that although it is
much smaller, the artificial ``bump'' has not been fully removed by the
modifications to the power spectrum and the bispectrum.
If one is interested in the reduced bispectrum $Q_{\rm eq}(k)$, one can
introduce a last improvement by replacing this small ``bump'' by a flat
plateau. This ``flat'' model is obtained from the ``tang.'' curve by
running down the $Q_{\rm eq}^{\rm tang.}(k)$ curve over the interval
$[k_-,k_+]$, starting from the right boundary $k_+$, and imposing a monotonic
decrease as $k$ decreases. Thus, $Q_{\rm eq}^{\rm flat}$ and 
$Q_{\rm eq}^{\rm tang.}$ are identical over most of $[k_-,k_+]$ (and on all
larger and smaller scales), except under the remaining ``bump'' of
$Q_{\rm eq}^{\rm tang.}$, where $Q_{\rm eq}^{\rm flat}$ is constant and equal
to the local minimum of $Q_{\rm eq}^{\rm tang.}$ to the right of this ``bump''.

The geometrical improvements described in Figs.~\ref{fig_Qk_eq_mod},
\ref{fig_lDk}, and \ref{fig_lBk_eq} are not as elegant as one would wish for.
Indeed, by combining perturbative expansions with halo models, as in this
article and in \citet{Valageas2010b}, one would hope to build a model
that shows a good match to numerical simulations on all scales, without further
modifications. Unfortunately, as noticed above, at the current stage
there remain some discrepancies on the transition scales. 
As discussed in Sect.~\ref{depend-halo}, this is not surprising because transition
scales may be at the limit of validity of both perturbation theory and halo models.
Indeed, shell crossing (which is beyond the reach of the perturbative approaches
studied here, based on the fluid approximation) is already important, and
these scales do not correspond to relaxed halo inner shells, but rather to
outer infalling clumps and to filaments (which cannot be described as isolated
objects with a well-defined profile). 
In particular, we have seen in Fig.~\ref{fig_Qk_M} that tuning halo parameters
does not easily provide a significantly better agreement with simulations.
Moreover, this would require some ad-hoc modifications
and new free parameters, which make the model less predictive.
Then, the geometrical modifications introduced above can be seen as a simple
procedure to obtain a good matching between the weakly and highly
nonlinear regimes, and are no less satisfactory than more ``algebraic'' matchings.
They offer the advantage to bypass all these complicating matters and the
free parameters they would involve, but they share with approaches based on
modifications to the halo parameters the lack of
a systematic method toward increasingly high accuracy.

In addition to the automatic definition of the interval $[k_-,k_+]$, associated
with the transition range, an important property of this procedure is that the
power spectrum and the equilateral bispectrum are not modified outside of
this interval. Therefore, we keep the good match obtained on larger and smaller
scales. In particular, large scales are still obtained by systematic perturbative
expansions and small scales by phenomenological halo models, so that
the final result is still a combination of these two approaches and keeps their
distinct benefits.

We describe in App.~\ref{correlation-function} the impact on the real-space
two-point correlation function of this modification to the power spectrum.
In particular, we check that we keep an accuracy on the order of $1\%$
on weakly nonlinear scales, $x>10 h^{-1}$Mpc, while reaching an accuracy
on the order of $10\%$ on nonlinear scales.


\section{Typical accuracy of combined models}
\label{Typical-accuracy}

\begin{figure*}
\begin{center}
\epsfxsize=6.1 cm \epsfysize=5.4 cm {\epsfbox{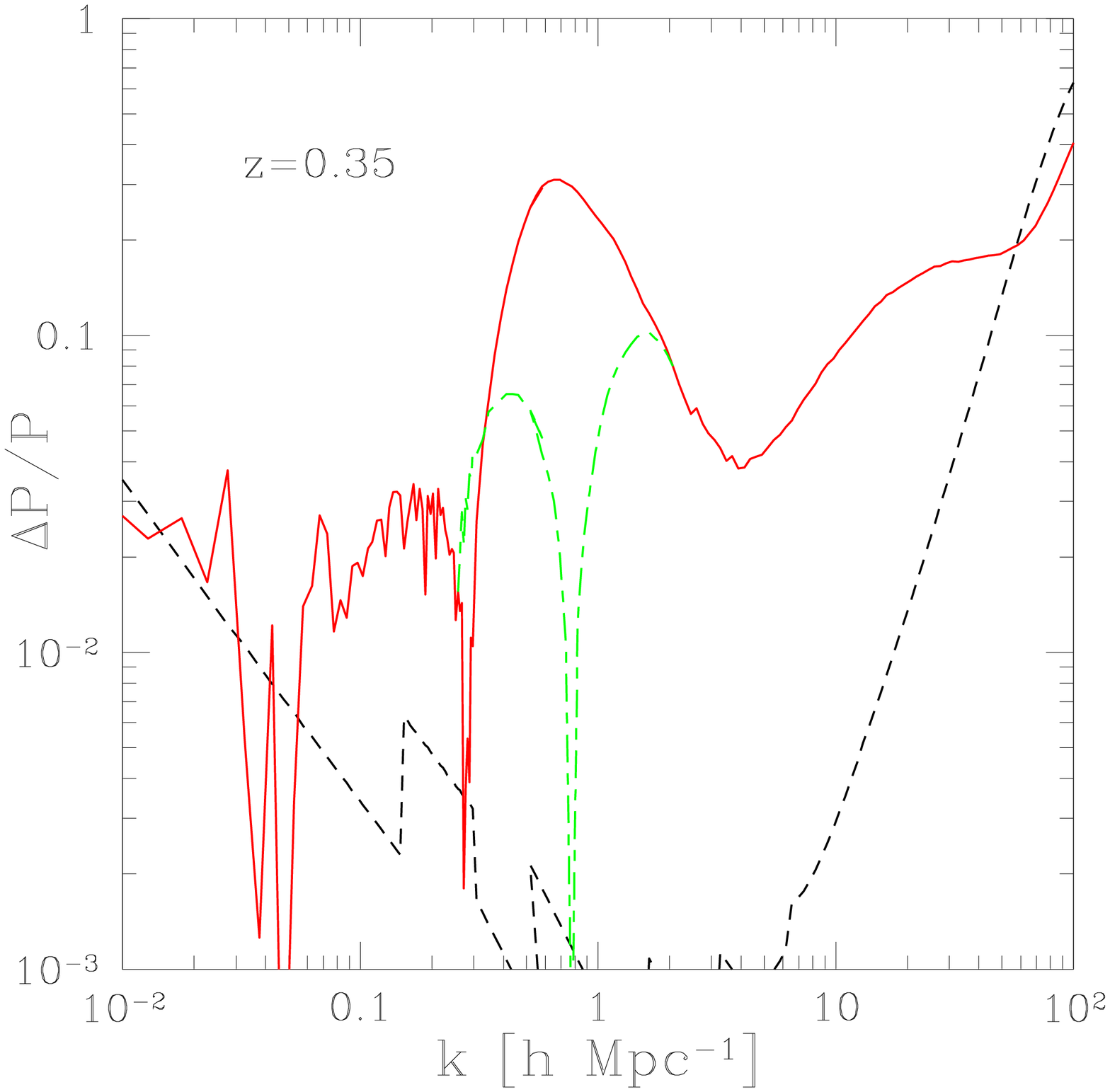}}
\epsfxsize=6.05 cm \epsfysize=5.4 cm {\epsfbox{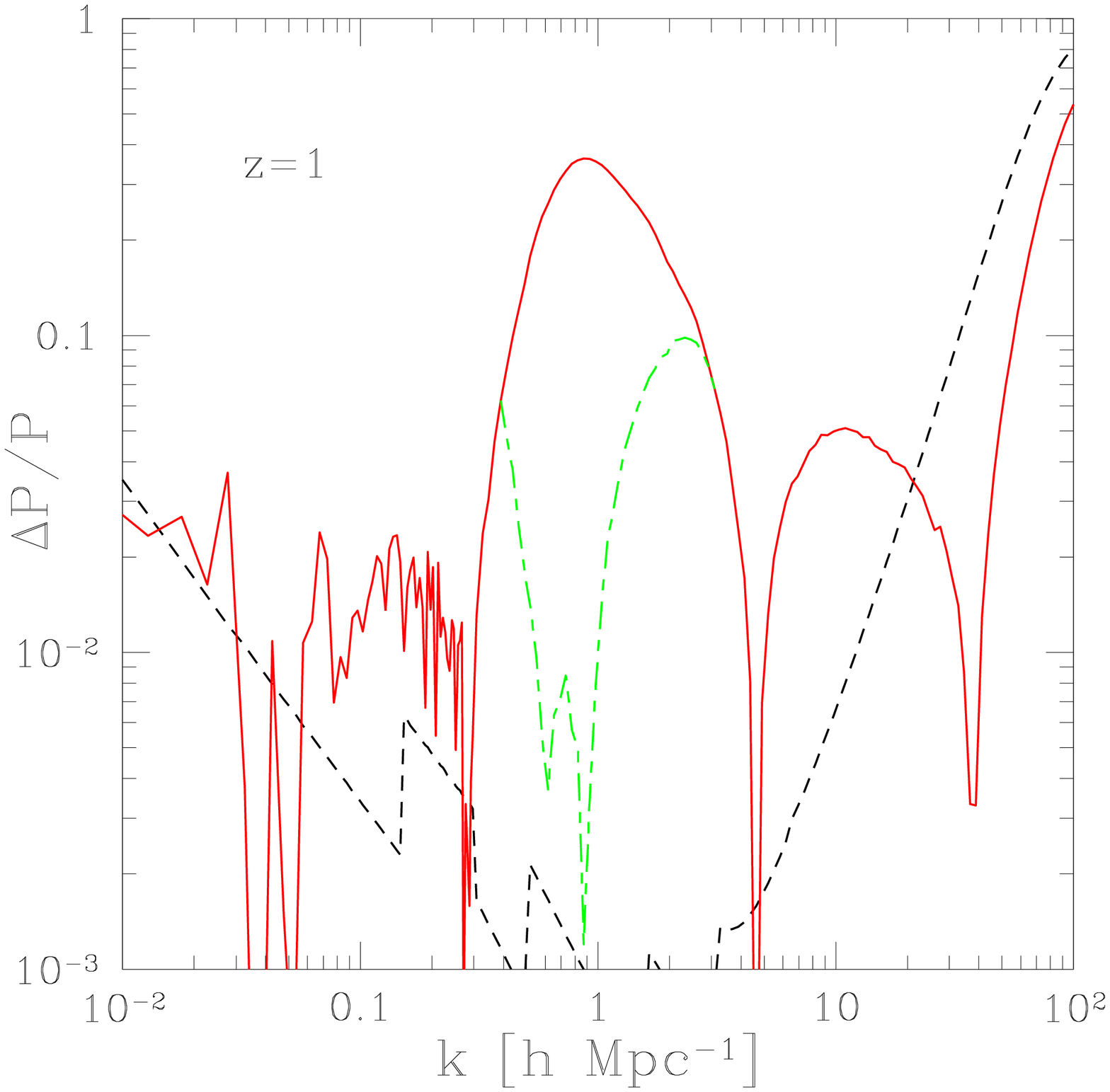}}
\epsfxsize=6.05 cm \epsfysize=5.4 cm {\epsfbox{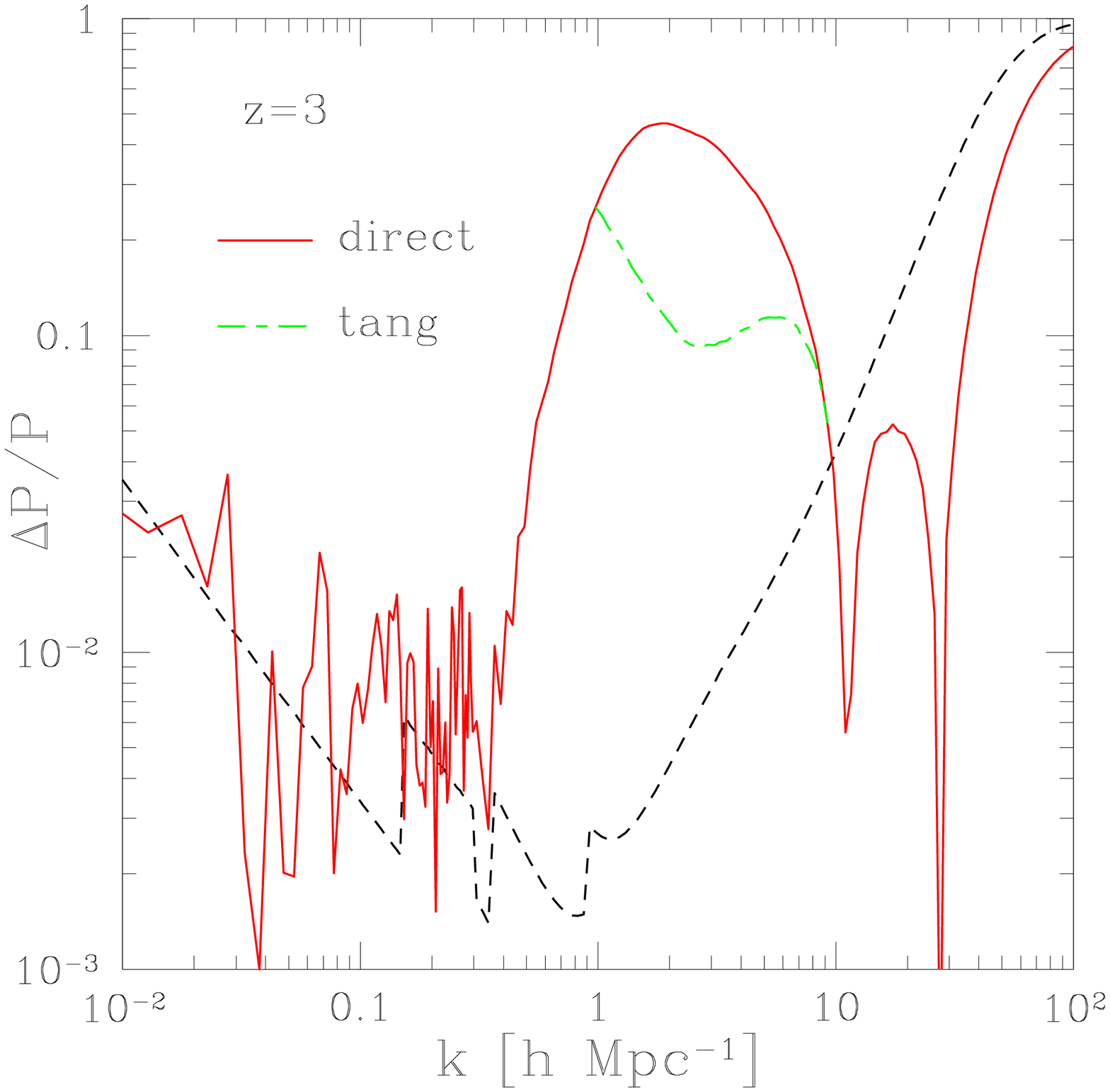}}
\end{center}
\caption{Accuracy of our models and our numerical simulations at redshifts
$z=0.35$, $1$, and $3$ for the power spectrum.
The red solid line ``direct'' is the relative difference (\ref{Delta-P}) between the
simulations and our model as described in \citet{Valageas2010b}.
The green dashed line ``tang.'' corresponds to the geometrical modification 
shown in Fig.~\ref{fig_lDk}.
The black dashed line is the error level of the simulations, including both the
relative statistical error (which grows at low $k$) and the relative shot-noise
(which grows at high $k$).}
\label{fig_dP}
\end{figure*}

We now consider the accuracy of the combined model described in the
previous sections and in \citet{Valageas2010b}.
We first plot in Fig.~\ref{fig_dP} the relative difference between our model and
the numerical simulations for the power spectrum,
\beq
\frac{\Delta P}{P}(k) = \frac{|P_{\rm model}(k)-P_{\rm N-body}(k)|}
{P_{\rm N-body}(k)} .
\label{Delta-P}
\eeq
The ``direct'' curve corresponds to the model described in \citet{Valageas2010b},
without the geometrical modifications introduced in the previous
Sect.~\ref{Improving}, and was already displayed in Fig.~22 of
\citet{Valageas2010b}. We can see that the modification $P^{\rm tang}(k)$,
shown in Fig.~\ref{fig_lDk} in terms of the logarithmic power $\Delta^2(k)$,
provides a significant improvement on the transition scales, especially for
$z\leq 1$. 
Thus, we obtain an accuracy on the order of $10\%$ on the transition scales,
and $1\%$ on weakly nonlinear scales. On very large and very small scales
the curves in Fig.~\ref{fig_dP} are dominated by the error bars of the
numerical simulations.

\begin{figure*}
\begin{center}
\epsfxsize=6.1 cm \epsfysize=5.4 cm {\epsfbox{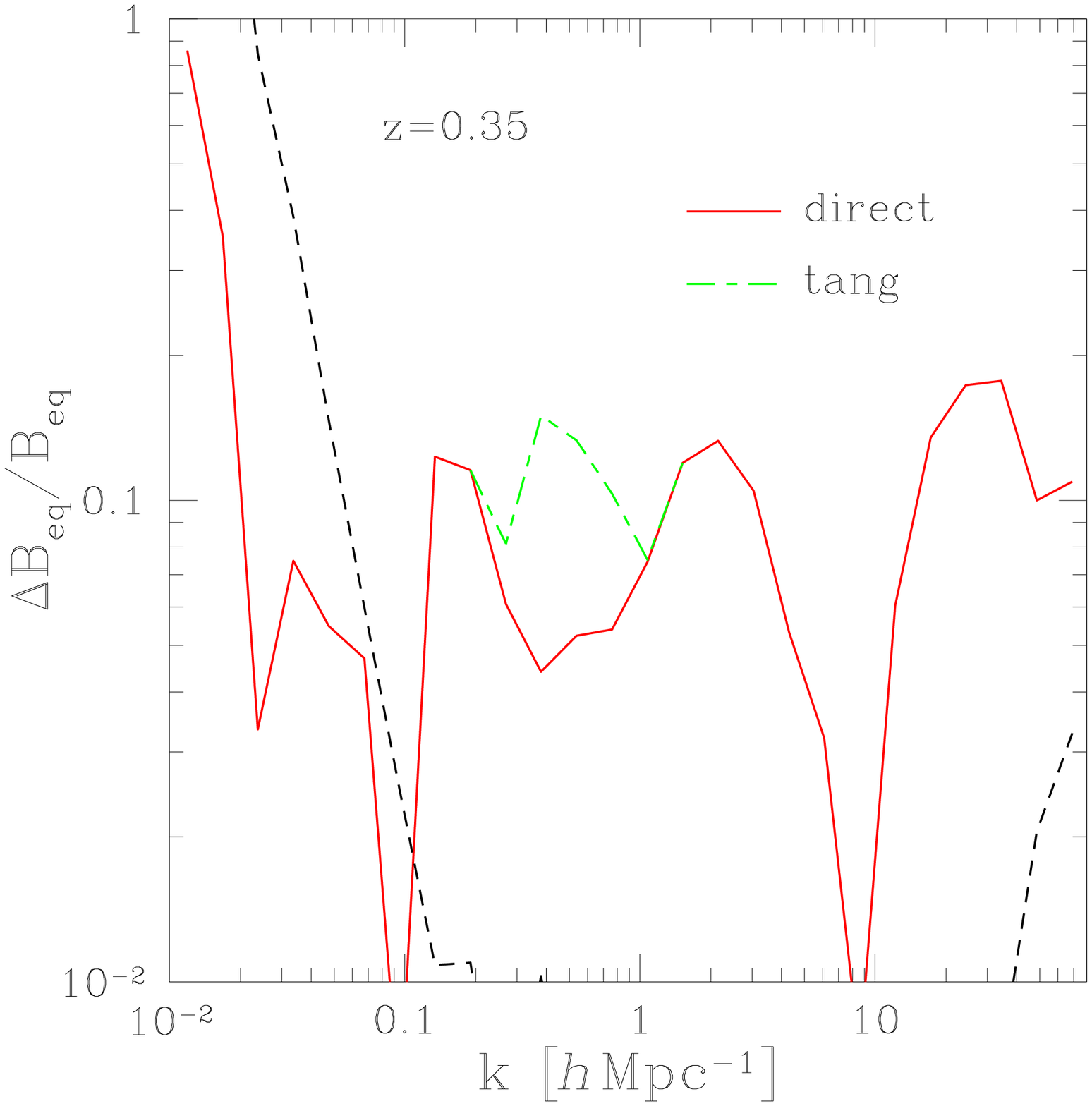}}
\epsfxsize=6.05 cm \epsfysize=5.4 cm {\epsfbox{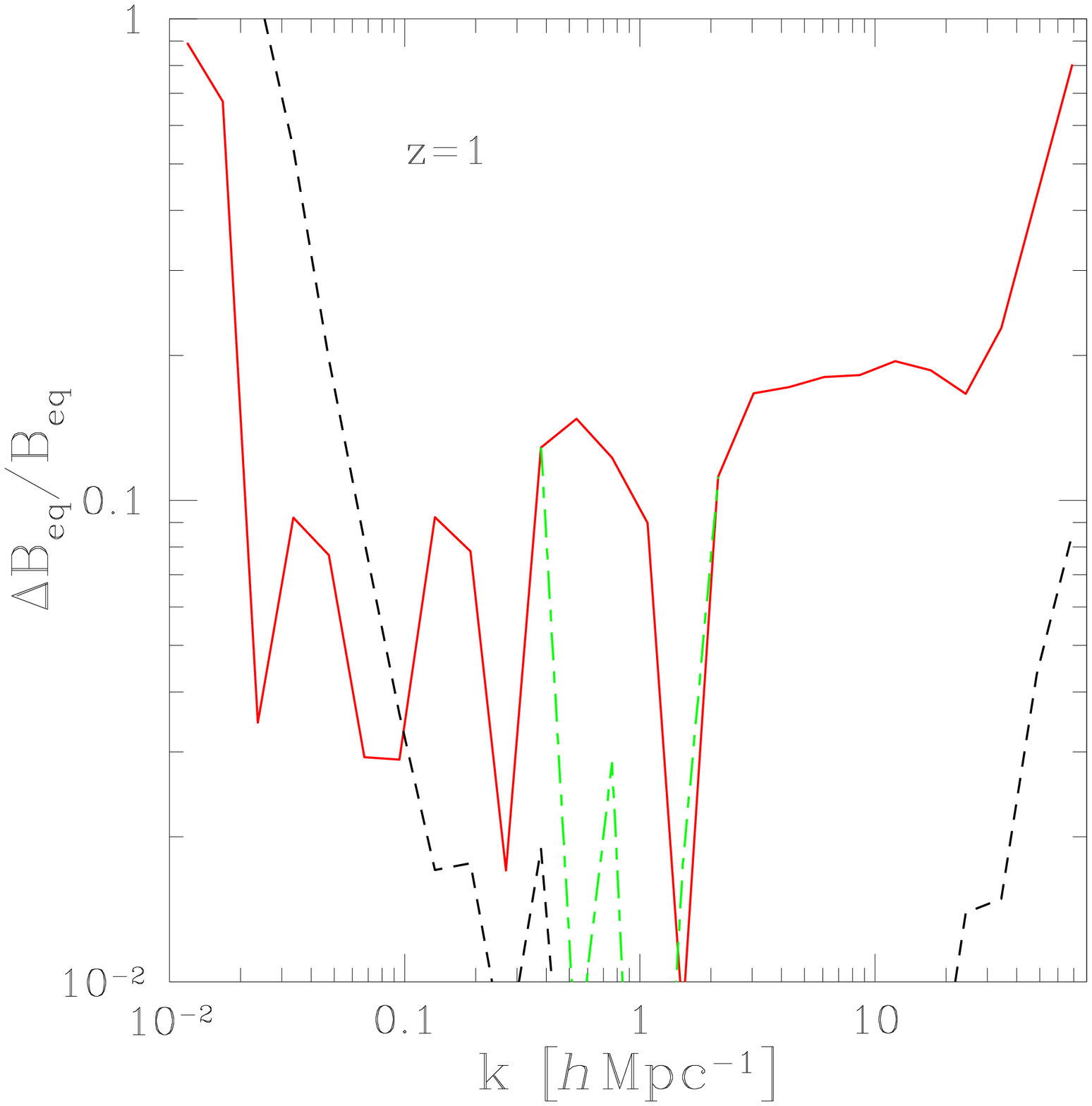}}
\epsfxsize=6.05 cm \epsfysize=5.4 cm {\epsfbox{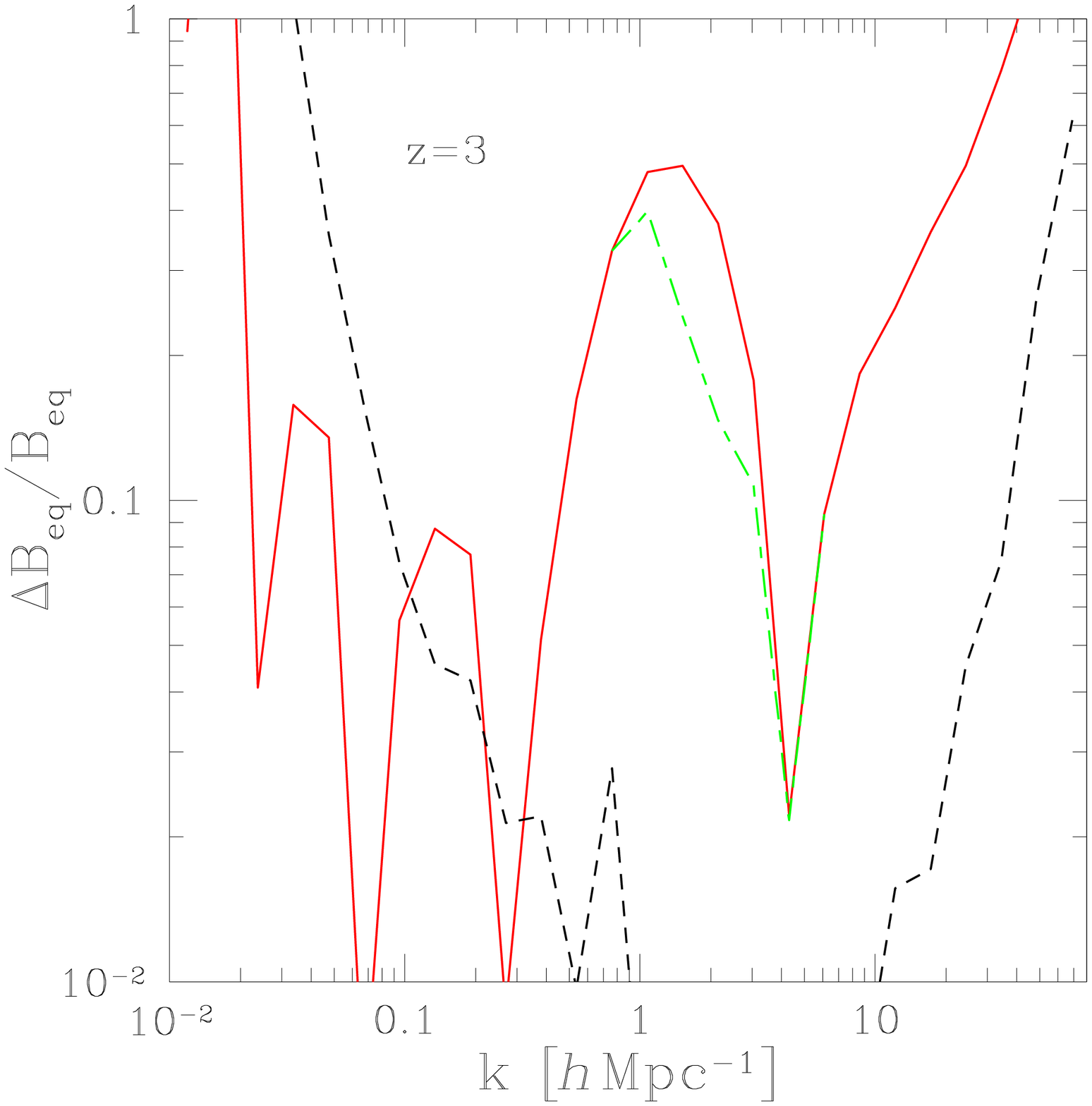}}
\end{center}
\caption{Accuracy of our model and our numerical simulations at redshifts
$z=0.35$, $1$, and $3$ for the bispectrum on equilateral configurations.
We plot our ``direct'' model (red solid line) and its geometrical modification
``tang.'' (green dot-dashed line), shown in Fig.~\ref{fig_lBk_eq}. The black dashed
line shows the statistical and shot-noise errors of the simulations.}
\label{fig_dBk_eq}
\end{figure*}

\begin{figure*}
\begin{center}
\epsfxsize=6.1 cm \epsfysize=5.4 cm {\epsfbox{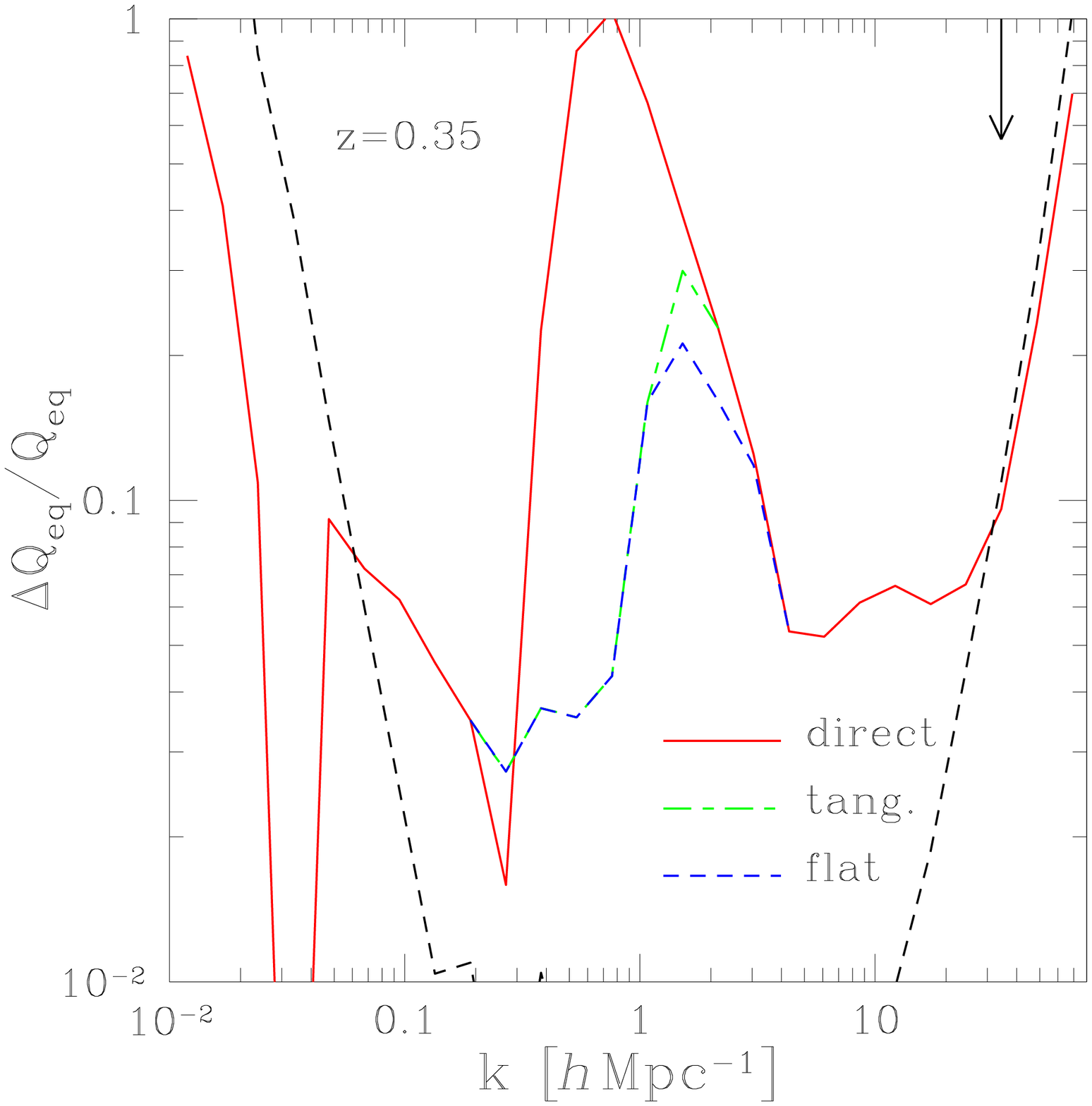}}
\epsfxsize=6.05 cm \epsfysize=5.4 cm {\epsfbox{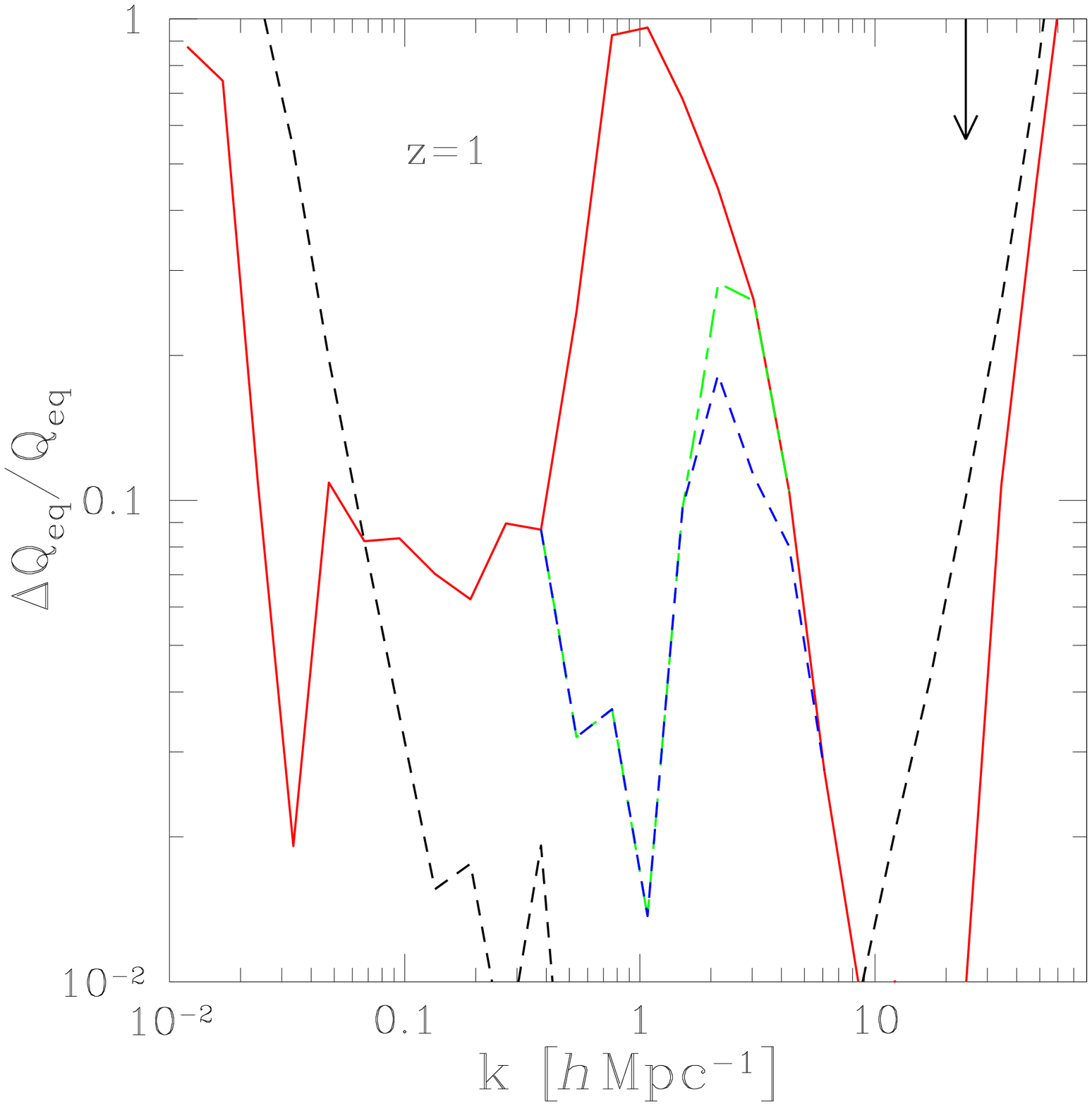}}
\epsfxsize=6.05 cm \epsfysize=5.4 cm {\epsfbox{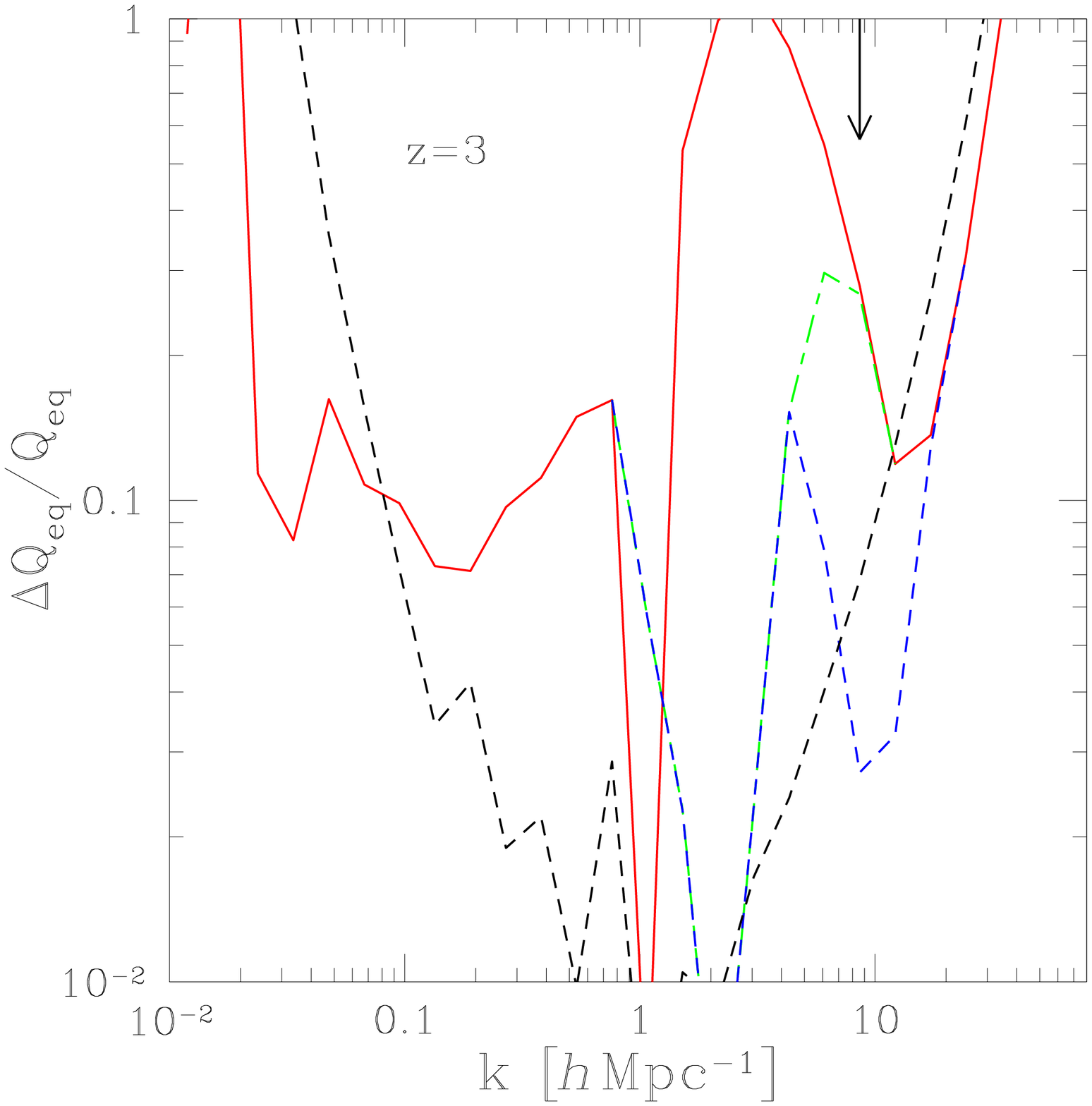}}
\end{center}
\caption{Accuracy of our model and our numerical simulations at redshifts
$z=0.35$, $1$, and $3$ for the reduced bispectrum on equilateral configurations.
We plot our ``direct'' model (red solid line) and its geometrical modifications
``tang.'' (green dot-dashed line) and ``flat'' (blue dashed line), shown in
Fig.~\ref{fig_Qk_eq_mod}. The black dashed line shows the statistical and
shot-noise errors of the simulations.
The vertical arrow in the upper right part
shows the wavenumber beyond which the simulation shot noise is greater than
$10\%$.}
\label{fig_dQk_eq}
\end{figure*}

Next, we show in Fig.~\ref{fig_dBk_eq} the relative difference between our model
and the numerical simulations for the equilateral bispectrum,
\beq
\frac{\Delta B_{\rm eq}}{B_{\rm eq}}(k) = \frac{|B_{\rm model}(k,k,k)
-B_{\rm N-body}(k,k,k)|}{B_{\rm N-body}(k,k,k)} .
\label{Delta-B}
\eeq
We can see that we reach a typical accuracy of $10\%$ on most scales.
At low $k$ we are dominated by the error introduced by the finite size of the
simulation box, as shown by the rise of the statistical error, and the theoretical
predictions are actually more accurate than appears in the figure
(they actually become increasingly good on larger scales where 1-loop perturbation
theory is increasingly accurate).
At high $k$ we are dominated by the error from the finite resolution of
the simulations. As for the power spectrum, studied in \citet{Valageas2010b}
and shown in Fig.~\ref{fig_dP}, the accuracy of the ``direct'' model is
worst on transition scales, especially at $z=3$. 
This agrees with the behavior found in Figs.~\ref{fig_Bk_eq} and
\ref{fig_Qk_eq}. The modified bispectrum $B^{\rm tang.}_{\rm eq}(k)$,
shown in Fig.~\ref{fig_lBk_eq}, provides a modest improvement on the transition
scales, except at $z=0.35$ where in our case it happens to give a slightly larger
error that remains on the order of $10\%$, which is typical in this range, so that
the change caused by the geometrical modification to the bispectrum is not
meaningful here.

Although this $10\%$ accuracy (and better on very large scales) is already
a satisfactory result, it is significantly below what can be achieved for the
power spectrum, where an accuracy on the order of $1\%$ is reached in the
weakly nonlinear regime, as seen in Fig.~\ref{fig_dP}.
From the comparison between different resummation schemes discussed in
Sect.~\ref{Pert-num}, it is not clear whether going to higher orders of perturbation
theory would provide a significant improvement. It may happen that on these
scales, especially in the transition regime, the bridge between the perturbative
3-halo term and the non-perturbative 2-halo and 1-halo terms is the main
source of error because of the intrinsic limitations of a description in terms of
relaxed spherical halos.

On the other hand, the error bars of the simulations are much larger for the
bispectrum than for the power spectrum, and at $k< 0.1 h$Mpc$^{-1}$
the level of $10\%$ seen in Fig.~\ref{fig_dBk_eq} is mostly set by the
simulations. The comparison with Fig.~\ref{fig_dP} suggests that on these
large scales the accuracy of the theoretical model is actually much better,
on the order of $1\%$, because it is determined by the systematic perturbation
theory. 
Thus, our model appears to be competitive with numerical simulations
because its fares as well or better on both large and small scales, but worse
on the intermediate mildly nonlinear scales. Of course, for practical purposes,
the main advantage of the analytical model is that it allows much faster
computations, as well as a greater flexibility.

Finally, we plot in Fig.~\ref{fig_dQk_eq} the relative difference between our model
and the numerical simulations for the equilateral reduced bispectrum, 
$(\Delta Q_{\rm eq}/Q_{\rm eq})(k)$, defined in a fashion similar to
Eqs.(\ref{Delta-P}) and (\ref{Delta-B}).
As in Fig.~\ref{fig_Qk_eq}, the vertical arrow in the
upper right part shows the wavenumber where the shot noise of the simulations
becomes large.

As for the bispectrum, we obtain a typical accuracy on the order of $10\%$
(and better on larger scales), except on the transition scales where the prediction
given by the ``direct'' model can be larger than the numerical results by up to
a factor two. This corresponds to the artificial ``bump'' found in
Fig.~\ref{fig_Qk_eq}. The modified predictions $Q_{\rm eq}^{\rm tang.}$ and
$Q_{\rm eq}^{\rm flat}$ allow us to recover an accuracy on the order of $10\%$
in this transition regime. On smaller scales the error bars of the numerical
simulations are too large to obtain a precise estimate of the accuracy of these
models because the rise seen in Fig.~\ref{fig_dQk_eq} is due to the finite resolution.

\section{Conclusion}
\label{Conclusion}

Extending a previous work dedicated to the matter power
spectrum, we have explained how to combine perturbation theories with
halo models to build unified models that can describe all scales, from
large linear scales to small highly nonlinear scales.
Starting again from a Lagrangian point of view, instead of the usual
Eulerian point of view, we have shown how to recover the decomposition
into 3-halo, 2-halo, and 1-halo contributions, which we relate to perturbative
and non-perturbative terms. This explains how one can build a model that
agrees with perturbation theory at all orders
because the 1-halo and 2-halo terms are non-perturbative corrections that vanish
at all orders over $P_L$. 
Moreover, we explained how new counterterms appear in the
1-halo and 2-halo contributions. This ensures that these contributions vanish
at low $k$, as required by physical constraints on the power generated by
small-scale redistributions of matter. This improves previous models
that displayed an unphysical constant asymptote at low $k$, and allows us
to reach a higher accuracy.
In addition to standard perturbation theory, we described two alternative
perturbative schemes, also complete up to 1-loop order, which can be used for the
perturbative 3-halo contribution. They contain infinite partial resummations
of higher order diagrams. 

Combining the halo model used in \citet{Valageas2010b} for the 
matter power spectrum with the 1-loop standard perturbation theory, we obtain
a good agreement with numerical simulations for the bispectrum without
any new free parameter. We consider the bispectrum as well as the
reduced bispectrum, using for the latter the power spectrum predicted by
the same approach (but with the more accurate steepest-descent resummation
instead of standard perturbation theory). We checked that this reasonably 
good match to simulations holds for equilateral as well as isosceles triangles,
from large to small scales. However, as for the power spectrum, the intermediate
mildly nonlinear scales are not as well reproduced by this direct implementation
of our approach.

The comparison between the three perturbative schemes investigated in this
article shows that the standard 1-loop perturbation theory is actually the
most accurate one. Because it is also simpler and faster to compute, this is also
the most efficient one. This is quite different from the power spectrum, where
the standard 1-loop perturbation theory is not the most accurate on large scales
and behaves badly on small scales (it grows too fast at high $k$), so that
it cannot be used in unified models unless one adds at least an external
high-$k$ cutoff. This problem does not occur for the bispectrum because the
standard 1-loop contribution is now negligible on small scales compared with
the 1-halo contribution. However, if one pushes the perturbative contribution
to higher orders, it may start being too large at high $k$ so that one would need
to resort to alternative, better behaved, perturbative approaches, or to add 
high-$k$ cutoffs.

Next, we have shown how to improve our predictions on the transition scales
for the matter power spectrum and bispectrum, using a simple interpolation
scheme instead of modifications to halo parameters, which would involve
new parameters and do not allow significant improvements.
This method automatically detects this transition regime from the
shape of the reduced equilateral bispectrum, and in this fashion adapts to the
change of initial conditions. Thus, for CDM linear power spectra that are not
pure power laws, the transition interval $[k_-,k_+]$ shifts somewhat 
with redshift (i.e., with the local slope $n$ of the linear power spectrum  
on the transition scale) with respect to the interval that would be defined
by constant thresholds, such as $\Delta^2(k_{\pm})=1$ and $30$.
Moreover, the interpolation through tangent lines adapts to the characteristic
bends of the power spectrum and bispectrum, seen for CDM initial conditions
around $\Delta^2(k) \sim 30$. Since this only modifies our model on the
transition interval $[k_-,k_+]$, large scales are still determined by systematic
perturbation theory and small scales by the halo model.
Then, we obtain an accuracy on the order of $10\%$ for the power spectrum and
the bispectrum on nonlinear scales, and $1\%$ on larger weakly nonlinear scales.
The same levels of improvement and final accuracy are obtained for the
real-space two-point correlation function.

Our model can still be improved in various
manners. First, one may investigate other perturbative approaches because other
resummation schemes may prove more accurate than the standard 1-loop
perturbation theory. However, to be more efficient, they should not be
much more difficult to compute than the standard perturbation theory.
Alternatively, one may go to higher orders. For the power spectrum,
higher orders are indeed relevant because various resummation schemes have
already been shown to be more accurate
\citep{Crocce2008,Taruya2009,Valageas2010b} and non-perturbative 
contributions only dominate after many perturbative orders have become
involved \citep{Valageas2010a}. 
For the bispectrum, the failure of the two resummation schemes investigated
in this paper to improve over standard perturbation theory suggests that it may
be more difficult to reach significant improvements.

Second, one may improve the halo model used in our unified approach.
For instance, one could take into account substructures
\citep{Sheth2003,Giocoli2010},
deviations from spherical profiles \citep{Jing2002,Smith2006}, or the effect of
baryons \citep{Guillet2010}.

Our model could be extended to other initial conditions, especially non-Gaussian
ones for which the bispectrum is a very useful and direct probe
\citep{Sefusatti2010}.
It would also be interesting to use this approach to describe velocity fields,
and to take into account redshift-space distortions \citep{Smith2008}.
However, we leave these tasks to future studies.

\begin{acknowledgements}

We thank S. Colombi for useful discussions about the folding procedure for the bispectrum
measurements. 
T. N. is supported by a Grant-in-Aid for Japan Society for the Promotion of
Science (JSPS) Fellows and by World Premier International Research Center Initiative (WPI
Initiative), MEXT, Japan. Numerical computations for the present work have been carried out in
part on Cray XT4 at Center for Computational Astrophysics, CfCA, of National Astronomical
Observatory of Japan, and in part under the Interdisciplinary Computational Science Program
in Center for Computational Sciences, University of Tsukuba.

\end{acknowledgements}

 \appendix

\section{Behavior of the bispectrum for one low wavenumber
in perturbation theory}
\label{bispectrum-for-one-low-wavenumber}

\begin{figure}[htb]
\begin{center}
\epsfxsize=7 cm \epsfysize=4.5 cm {\epsfbox{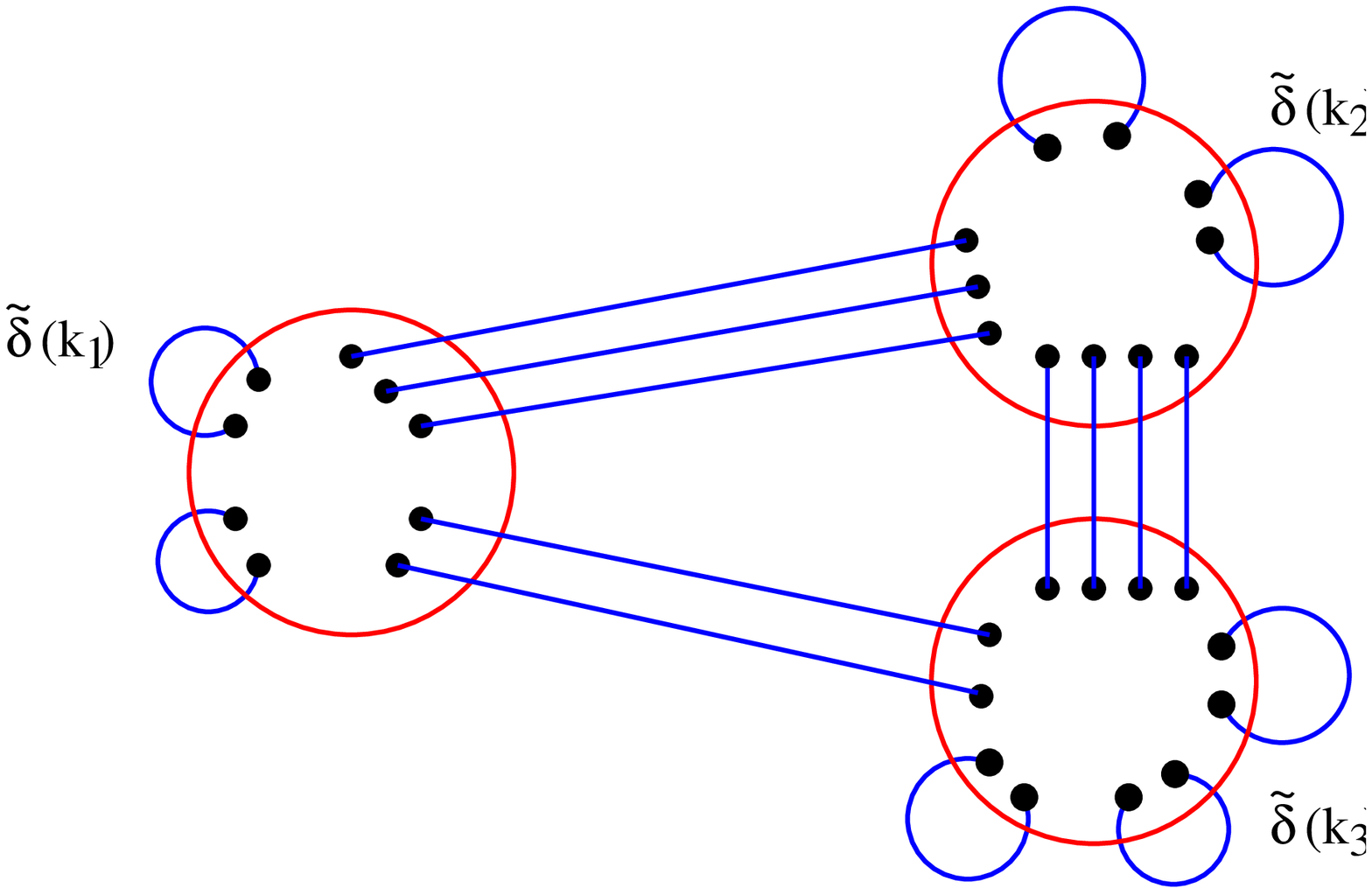}}
\end{center}
\caption{A perturbative contribution to the three-point connected correlation
$\lag\tdelta(\vk_1)\tdelta(\vk_2)\tdelta(\vk_3)\rag_c$, as in Eq.(\ref{B3-pert-F-1}).
The big red circles corresponds to the three nonlinear density contrasts
$\tdelta(\vk_1)$, $\tdelta(\vk_2)$, and $\tdelta(\vk_3)$. The small black dots
represent the linear fields $\tdelta_L(\vk_j')$ and the Gaussian average amounts
to connect them by the linear correlation
$\lag\tdelta_L(\vk_j')\tdelta_L(\vk_j'')\rag= \delta_D(\vk_j'+\vk_j'')P_L(k_j')$,
shown by
the blue solid lines. There are $n_{11}$, $n_{22}$, and $n_{33}$ internal
lines within the circles $\tdelta(\vk_1)$, $\tdelta(\vk_2)$, and $\tdelta(\vk_3)$.
There are $n_{12}$, $n_{13}$, and $n_{23}$ lines connecting the three circles.
Here we have $\{n_{11},n_{22},n_{33}\}=\{2,2,3\}$ and
$\{n_{12},n_{13},n_{23}\}=\{3,2,4\}$. This is part of
$\lag\tdelta^{(9)}(\vk_1)\tdelta^{(11)}(\vk_2)\tdelta^{(12)}(\vk_3)\rag_c$,
which involves the kernels $\tF_9$, $\tF_{11}$, and $\tF_{12}$.}
\label{fig_B_pert}
\end{figure}

We show in this appendix how the large-scale behavior (\ref{B-2H-kj}) also arises
at all orders of perturbation theory.
As is well known, within the standard perturbation theory the density contrast field
$\tdelta(\vk)$ is written as a perturbative expansion over powers of the linear
density contrast $\tdelta_L(\vk)$,
\beq
\tdelta(\vk) = \sum_{n=1}^{\infty} \tdelta^{(n)}(\vk) ,
\label{delta_n}
\eeq
with
\beqa
\tdelta^{(n)}(\vk) & = & \int \dd\vk_1 .. \dd\vk_n \, \delta_D(\vk_1+..+\vk_n-\vk)
\, \tF_n(\vk_1,..,\vk_n) \nonumber \\
&& \times \, \tdelta_L(\vk_1) .. \tdelta_L(\vk_n) .
\label{Fn-def}
\eeqa
The kernels $F_n$ can be obtained from recursion relations, which are derived
from the equations of motion \citep{Goroff1986,Bernardeau2002}, and can be
chosen as symmetric over the wavenumbers $\{\vk_1,..,\vk_n\}$.
Then, the bispectrum can be obtained within perturbation theory from the
three-point correlation
\beq
\lag\tdelta(\vk_1)\tdelta(\vk_2)\tdelta(\vk_3)\rag_c \! = \!\!\!\!\!\!\!\!
\sum_{n_1,n_2,n_3} \!\!\!\!\! \lag \tdelta^{(n_1)}(\vk_1)  \tdelta^{(n_2)}(\vk_2) 
\tdelta^{(n_3)}(\vk_3)  \rag_c ,
\label{B-def-pert}
\eeq
where the subscript $c$ recalls that we only take the connected part of the
Gaussian average. Using Wick's theorem and defining the linear power spectrum
as in Eq.(\ref{Pkdef}) gives
\beqa
\lag\tdelta(\vk_1)\tdelta(\vk_2)\tdelta(\vk_3)\rag_c & = & \sum_{n_*} \int
\!\!\prod_j \! \dd\vk_j^{(n_*)}
\delta_D(\vk_j^{(n_{12})}\!\!+\!\vk_j^{(n_{13})}\!\!-\!\vk_1) \nonumber \\
&& \hspace{-2.5cm} \times \,
\delta_D(-\vk_j^{(n_{12})}\!\!+\!\vk_j^{(n_{23})}\!\!-\!\vk_2)
\delta_D(-\vk_j^{(n_{13})}\!\!-\!\vk_j^{(n_{23})}\!\!-\!\vk_3) \nonumber \\
&& \hspace{-2.5cm} \times \, \tF_{2n_{11}+n_{12}+n_{13}}
(\vk_j^{(n_{11})},-\vk_j^{(n_{11})},\vk_j^{(n_{12})},\vk_j^{(n_{13})}) \nonumber \\
&& \hspace{-2.5cm} \times \, \tF_{2n_{22}+n_{12}+n_{23}}
(\vk_j^{(n_{22})},-\vk_j^{(n_{22})},-\vk_j^{(n_{12})},\vk_j^{(n_{23})}) \nonumber \\
&& \hspace{-2.5cm} \times \, \tF_{2n_{33}+n_{13}+n_{23}}
(\vk_j^{(n_{33})},-\vk_j^{(n_{33})},-\vk_j^{(n_{13})},-\vk_j^{(n_{23})}) \nonumber \\
&& \hspace{-2.5cm} \times \, \prod_j P_L(k_j^{(n_*)}) ,
\label{B3-pert-F-1}
\eeqa
where we used a short-hand notation for the diagram shown in
Fig.~\ref{fig_B_pert},
and the Dirac factors contain sums over the wavenumbers $\vk_j^{(n_*)}$.
Of course, the three Dirac factors can be combined to factorize out a Dirac
prefactor $\delta_D(\vk_1+\vk_2+\vk_3)$, in agreement with Eq.(\ref{Bkdef}).
Taking the connected part means that each of the three circles in
Fig.~\ref{fig_B_pert} is connected to at least one other circle (for instance the case
$n_{12}=n_{13}=n_{22}=0$ is excluded).

We are interested in the limit $k_1\rightarrow 0$, at fixed $k_2$ and $k_3$.
A well-known property of the kernels $\tF_n$, which arises from momentum
conservation \citep{Peebles1974,Goroff1986}, is that
$\tF_n(\vk_1+..+\vk_n)\propto k^2$ as $\vk=\vk_1+..+\vk_n$ goes to zero
while the individual $\vk_j$ do not, for all $n\geq 2$. Then, from the
first Dirac factor in Eq.(\ref{B3-pert-F-1}) we can see that the first kernel
$\tF$ behaves as $k_1^2$ if $2n_{11}+n_{12}+n_{13}\geq 2$.
Therefore, in the limit $k_1\rightarrow 0$ the dominant contributions come
from the diagrams with $2n_{11}+n_{12}+n_{13}=1$, since $F_1\equiv 1$
and for CDM cosmologies $P_L(k_1) \propto k_1^{n_s}$ with $n_s\simeq 1$
at low $k_1$.
This corresponds to either $\{n_{11}=0,n_{12}=1,n_{13}=0\}$ or 
$\{n_{11}=0,n_{12}=0,n_{13}=1\}$ because we only take the connected part of
(\ref{B3-pert-F-1}).
Thus, for $k_1 \rightarrow 0$ the dominant contribution to the perturbative
bispectrum reads as
\beqa
B_{\rm pert}(k_1,k_2,k_3) & \sim & P_L(k_1) \, \lim_{k_1\rightarrow 0}
\biggl \lbrace \sum_{n_*} \int \!\!\prod_j \! \dd\vk_j^{(n_*)}
\nonumber \\
&& \hspace{-2.5cm} \times \, \delta_D(\vk_j^{(n_{23})}-\vk_2) \,
\tF_{2n_{33}+n_{23}}
(\vk_j^{(n_{33})},-\vk_j^{(n_{33})},-\vk_j^{(n_{23})}) \nonumber \\
&& \hspace{-2.5cm} \times \, \tF_{2n_{22}+1+n_{23}}
(\vk_j^{(n_{22})},-\vk_j^{(n_{22})},-\vk_1,\vk_j^{(n_{23})}) \nonumber \\
&& \hspace{-2.5cm} \times \, \prod_j P_L(k_j^{(n_*)}) + 1 \; {\rm perm.}
\biggl \rbrace ,
\label{B3-pert-F-2}
\eeqa
where we note by ``1 perm'' the symmetric contribution with respect to
$\{\vk_2\leftrightarrow\vk_3\}$.
There is a subtlety in Eq.(\ref{B3-pert-F-2}), because the limit at low $k_1$ of
the kernel $\tF_{2n_{22}+1+n_{23}}
(\vk_j^{(n_{22})},-\vk_j^{(n_{22})},-\vk_1,\vk_j^{(n_{23})})$ contains divergent
terms of the form $(\vk_1\cdot\vk_j)/k_1^2$ \citep{Goroff1986,Scoccimarro1996}.
However, taking the limit $k_1\rightarrow 0$ elsewhere, we obtain
$\vk_3=-\vk_2$ so that the symmetric term with respect to
$\{\vk_2\leftrightarrow\vk_3\}$ actually corresponds to a change of sign of all
wavenumbers. This cancels the divergences of the form
$(\vk_1\cdot\vk_j)/k_1^2$ so that the limit $k_1\rightarrow 0$ is finite.
The fact that such infrared divergences cancel out for equal-time statistics
can be traced to the Galilean invariance of the equations of motion
\citep{Jain1996,Valageas2004}.

Terms at all orders of perturbation theory contribute to the
partial large-scale limit (\ref{B3-pert-F-2}) through the nonlinear corrections
to $\tdelta(\vk_2)$ and $\tdelta(\vk_3)$. In contrast, as we have seen above,
only the linear term $\tdelta_L(\vk_1)$ contributes.
The behavior (\ref{B3-pert-F-2}) can also be understood from the physical
arguments discussed in Sect.~\ref{2-halo}, and the explicit result (\ref{B3-pert-F-2})
confirms that this asymptotic behavior should be preserved by both high-order
perturbative terms and non-perturbative corrections.

It is interesting to note that the ``renormalization'' of the prefactor to
the $P(k_1)$ tail by the higher order terms in Eq.(\ref{B3-pert-F-2}) never
occurs for the power spectrum $P(k)$, where we recover $P(k) \rightarrow P_L(k)$
at low $k$. There, as seen from the analysis described in \citet{Valageas2002V},
perturbative terms beyond linear order scale at least as $k^2P_L(k)$ at low $k$.
This can also be understood from the fact that for the power spectrum there
are no partial large-scale limits. On the other hand, if all wavenumbers
go to zero, as $k_1\sim k_2\sim k_3\sim k$ with $k\rightarrow 0$, higher order 
contributions to the bispectrum scale at least as $k^2 P_L(k)^2$ while
the ``tree-order'' result scales as $P_L(k)^2$, so that in the simultaneous
large-scale limit we also recover the lowest order result.

\section{Two-point correlation function}
\label{correlation-function}

\begin{figure}
\begin{center}
\epsfxsize=8.3 cm \epsfysize=6.2 cm {\epsfbox{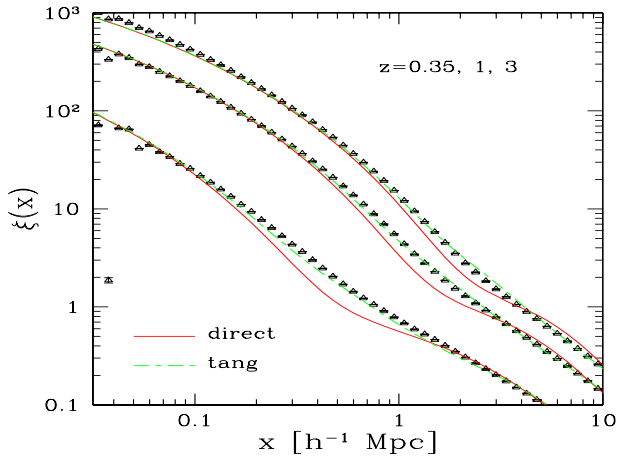}}
\end{center}
\caption{Real-space two-point correlation function $\xi(x)$, at redshifts
$z=0.35$, $1$, and $3$ (from top to bottom). We show the ``direct'' model
(red solid line) and its geometrical modification ``tang'' (green dot-dashed line).
They are the Fourier transforms of the corresponding power spectra shown in
Fig.~\ref{fig_lDk}.}
\label{fig_lxix}
\end{figure}

\begin{figure}
\begin{center}
\epsfxsize=8.3 cm \epsfysize=6.2 cm {\epsfbox{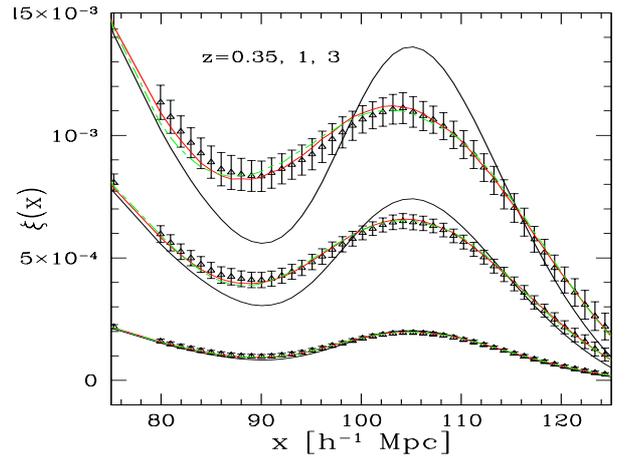}}
\end{center}
\caption{Real-space two-point correlation function $\xi(x)$ at redshifts
$z=0.35$, $1$, and $3$ (from top to bottom) as in Fig.~\ref{fig_lxix} but on
larger scales. In addition to the ``direct'' model (red solid line) and its geometrical
modification ``tang'' (green dot-dashed line), we also show the linear two-point
correlation (black solid line).}
\label{fig_xix}
\end{figure}

\begin{figure*}
\begin{center}
\epsfxsize=6.1 cm \epsfysize=5.4 cm {\epsfbox{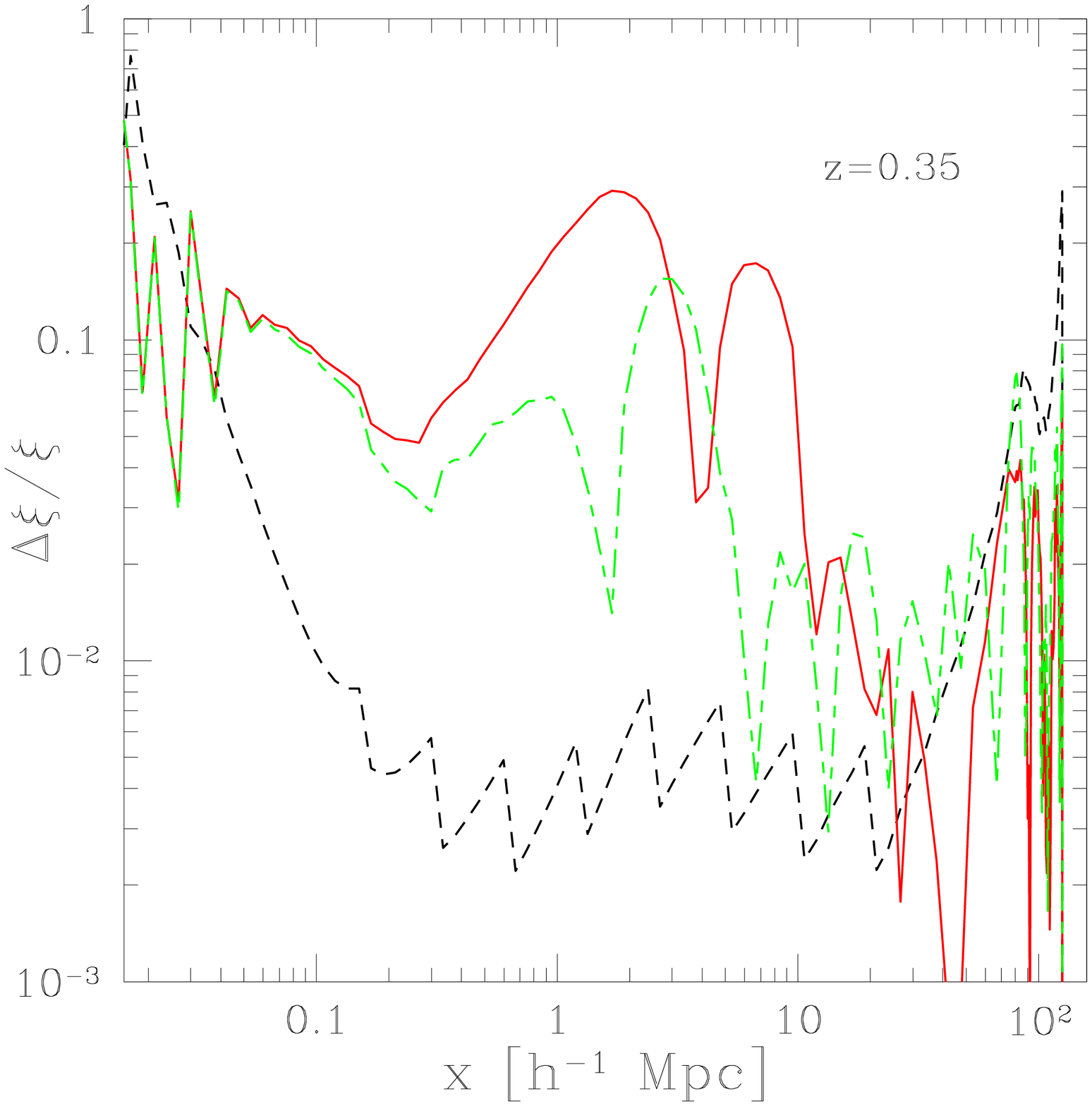}}
\epsfxsize=6.05 cm \epsfysize=5.4 cm {\epsfbox{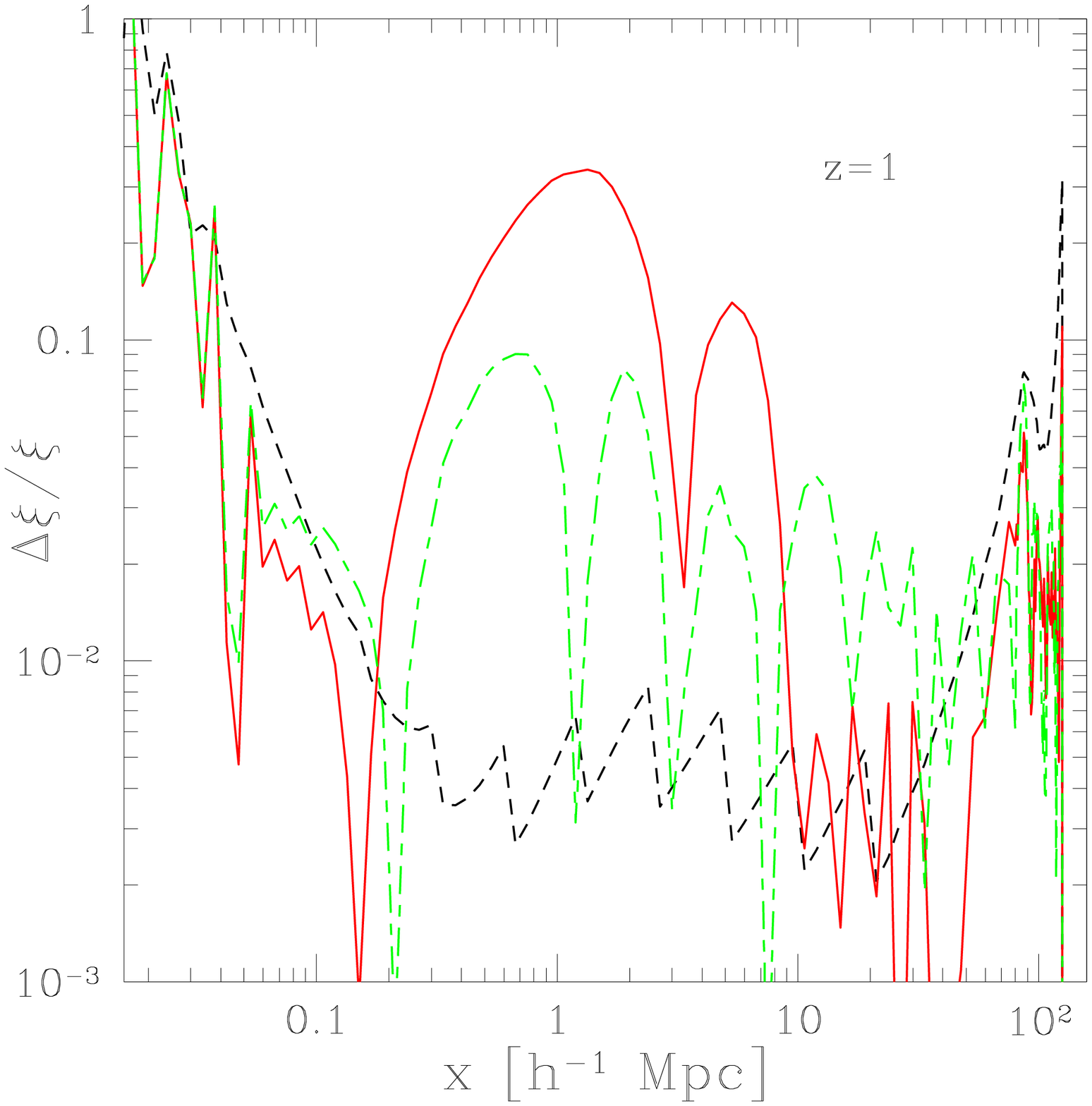}}
\epsfxsize=6.05 cm \epsfysize=5.4 cm {\epsfbox{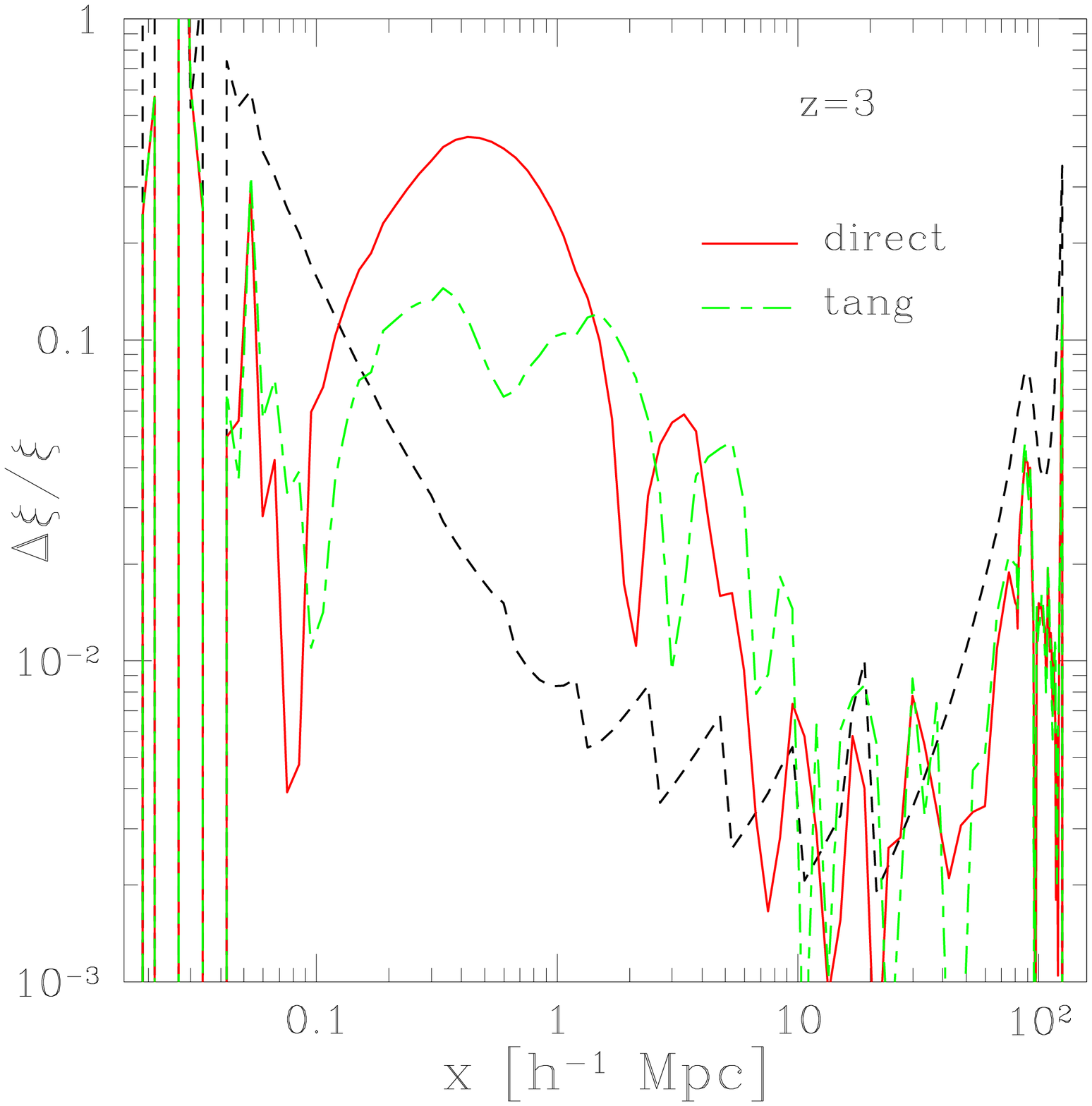}}
\end{center}
\caption{Accuracy of our models and our numerical simulations at redshifts
$z=0.35$, $1$, and $3$ for the real-space two-point correlation function.
As in Figs.~\ref{fig_lxix} and \ref{fig_xix},
the red solid line ``direct'' is the relative difference (\ref{Delta-P}) between the
simulations and our model as described in \citet{Valageas2010b}, while
the green dashed line ``tang.'' corresponds to the geometrical modification 
shown in Fig.~\ref{fig_lDk}.
The black dashed line shows the relative statistical error (which grows at large\
$x$) and shot-noise error (which grows at small $x$) of the simulations.}
\label{fig_dxi}
\end{figure*}

Although the main focus of this paper is the computation of the bispectrum,
we have seen in Sects.~\ref{Improving} and \ref{Typical-accuracy} that the
shape of the reduced bispectrum can be used to improve the model devised
for the power spectrum. This leads in turn to an improved model for the
real-space two-point correlation function, which is given by
\beqa
\xi(x) & = & 4\pi \int_0^{\infty} \dd k \; k^2 \, P(k) \, \frac{\sin(kx)}{kx} \\
& = & \int_0^{\infty} \frac{\dd k}{k} \; \Delta^2(k)  \, \frac{\sin(kx)}{kx} .
\label{xi-def}
\eeqa
Therefore, we compare in this appendix the two-point
correlation functions obtained either with our ``direct'' model, already
studied in \citet{Valageas2010b}, or with the geometrical modification
``tang'' to the power spectrum explained in Fig.~\ref{fig_lDk}.
Thus, we plot in Figs.~\ref{fig_lxix} and \ref{fig_xix} the two-point correlation
functions defined by the two power spectra of Fig.~\ref{fig_lDk} through
the Fourier transform (\ref{xi-def}).
As expected, we can see in Fig.~\ref{fig_lxix} that, as for the power $\Delta^2(k)$
in Fig.~\ref{fig_lDk}, the geometrical modification ``tang'' corrects most of the
artificial ``dip'' that was shown by the ``direct'' model on transition scales
(i.e., the mildly nonlinear regime). The curves even look closer to the
simulation results and more regular because the integral (\ref{xi-def}) has
regularized the geometrical modification. Indeed, while the power
$\Delta^2_{\rm tang}(k)$ shown in Fig.~\ref{fig_lDk} was only continuous,
with a discontinuous derivative at $k_-$, the correlation $\xi_{\rm tang}(x)$
has a well-defined and finite derivative over all $x>0$. 

Contrary to the power spectrum, the modification
``tang'' to the two-point correlation is not restricted to a finite range
$[x_-,x_+]$ because the modification of the power on the wavenumber interval
$[k_-,k_+]$ contributes to all distances $x$ through the integral (\ref{xi-def}).
However, as expected, we can check in Fig.~\ref{fig_lxix} that
on very small and very large scales the modified correlation
$\xi_{\rm tang}(x)$ becomes increasingly close to the original model prediction
$\xi_{\rm direct}(x)$, and as such shows a good agreement with numerical
simulations.

Nevertheless, to check in more detail that the modification ``tang'' does not
spread on large scales, where the initial power is much weaker, we show in
Fig.~\ref{fig_xix} the correlation functions of Fig.~\ref{fig_lxix} on larger
scales, around the baryon acoustic oscillation. We can see that both correlations,
$\xi_{\rm tang}(x)$ and $\xi_{\rm direct}(x)$, are very close and within the
error bars of the numerical simulations. In particular, they capture very well
the departure from the linear correlation, mostly due to the 1-loop
perturbative contribution. Therefore, the modified correlation
$\xi_{\rm tang}(x)$ provides a good model from small to very large scales.

As in Fig.~\ref{fig_dP} for the power spectrum, we show in Fig.~\ref{fig_dxi} the
relative accuracy of our analytical models and numerical simulations.
(The curve obtained for the ``direct'' model was already shown in
\citet{Valageas2010b}.)
In agreement with Figs.~\ref{fig_dP} and \ref{fig_lxix}, we can see that the
modification ``tang'' provides a significant improvement on transition scales.
Thus, it ensures an accuracy of $10\%$ or better on nonlinear scales, while
an accuracy on the order of $1\%$ is again reached on quasilinear scales,
associated for instance with the baryon acoustic oscillation of Fig.~\ref{fig_xix}.

\bibliographystyle{aa} 
\bibliography{16638}

\begin{thebibliography}{78}
\expandafter\ifx\csname natexlab\endcsname\relax\def\natexlab#1{#1}\fi

\bibitem[{{Anselmi} {et~al.}(2010){Anselmi}, {Matarrese}, \&
  {Pietroni}}]{Anselmi2011}
{Anselmi}, S., {Matarrese}, S., \& {Pietroni}, M. 2010, ArXiv e-prints

\bibitem[{Bartolo {et~al.}(2010)Bartolo, Almeida, Matarrese, Pietroni, \&
  Riotto}]{Pietroni2010}
Bartolo, N., Almeida, J. P.~B., Matarrese, S., Pietroni, M., \& Riotto, A.
  2010, JCAP, 3, 11

\bibitem[{{Bartolo} {et~al.}(2010){Bartolo}, {Beltr{\'a}n Almeida},
  {Matarrese}, {Pietroni}, \& {Riotto}}]{Bartolo2010}
{Bartolo}, N., {Beltr{\'a}n Almeida}, J.~P., {Matarrese}, S., {Pietroni}, M.,
  \& {Riotto}, A. 2010, \jcap, 3, 11

\bibitem[{Bernardeau {et~al.}(2002{\natexlab{a}})Bernardeau, Colombi,
  {Gazta\~naga}, \& Scoccimarro}]{Bernardeau2002}
Bernardeau, F., Colombi, S., {Gazta\~naga}, E., \& Scoccimarro, R.
  2002{\natexlab{a}}, Phys. Rep., 367, 1

\bibitem[{Bernardeau {et~al.}(2008)Bernardeau, Crocce, \&
  Scoccimarro}]{Bernardeau2008}
Bernardeau, F., Crocce, M., \& Scoccimarro, R. 2008, Phys. Rev. D, 78, 103521

\bibitem[{{Bernardeau} {et~al.}(2010){Bernardeau}, {Crocce}, \&
  {Sefusatti}}]{Bernardeau2010}
{Bernardeau}, F., {Crocce}, M., \& {Sefusatti}, E. 2010, \prd, 82, 083507

\bibitem[{Bernardeau {et~al.}(2002{\natexlab{b}})Bernardeau, Mellier, \& van
  Waerbeke}]{Bernardeau2002b}
Bernardeau, F., Mellier, Y., \& van Waerbeke, L. 2002{\natexlab{b}}, Astron.
  Astrophys., 389, L28

\bibitem[{Bernardeau \& Valageas(2008)}]{BernardeauVal2008}
Bernardeau, F. \& Valageas, P. 2008, Phys. Rev. D, 78, 083503

\bibitem[{Bernardeau \& Valageas(2010)}]{BernardeauVal2010a}
Bernardeau, F. \& Valageas, P. 2010, Phys. Rev. D, 81, 043516

\bibitem[{Cole {et~al.}(2005)Cole, Percival, \& et~al.}]{Cole2005}
Cole, S., Percival, W.~J., \& et~al., J. A.~P. 2005, Mon. Not. R. Astron. Soc.,
  362, 505

\bibitem[{Colombi {et~al.}(2009)Colombi, Jaffe, Novikov, \&
  Pichon}]{Colombi2009}
Colombi, S., Jaffe, A., Novikov, D., \& Pichon, C. 2009, Mon. Not. R. Astron.
  Soc., 393, 511

\bibitem[{Cooray \& Sheth(2002)}]{Cooray2002}
Cooray, A. \& Sheth, R. 2002, Phys. Rep., 372, 1

\bibitem[{Crocce \& Scoccimarro(2006{\natexlab{a}})}]{Crocce2006b}
Crocce, M. \& Scoccimarro, R. 2006{\natexlab{a}}, Phys. Rev. D, 73, 063520

\bibitem[{Crocce \& Scoccimarro(2006{\natexlab{b}})}]{Crocce2006a}
Crocce, M. \& Scoccimarro, R. 2006{\natexlab{b}}, Phys. Rev. D, 73, 063519

\bibitem[{Crocce \& Scoccimarro(2008)}]{Crocce2008}
Crocce, M. \& Scoccimarro, R. 2008, Phys. Rev. D, 77, 023533

\bibitem[{Dolag {et~al.}(2004)Dolag, Bartelmann, \& et~al.}]{Dolag2004}
Dolag, K., Bartelmann, M., \& et~al., F.~P. 2004, Astron. Astrophys., 416, 853

\bibitem[{Duffy {et~al.}(2008)Duffy, Schaye, Kay, \& Vecchia}]{Duffy2008}
Duffy, A.~R., Schaye, J., Kay, S.~T., \& Vecchia, C.~D. 2008, Mon. Not. R.
  Astron. Soc., 390, L64

\bibitem[{Eisenstein {et~al.}(1998)Eisenstein, Hu, \& Tegmark}]{Eisenstein1998}
Eisenstein, D.~J., Hu, W., \& Tegmark, M. 1998, Astrophys. J. Lett., 504, 57

\bibitem[{Eisenstein {et~al.}(2005)Eisenstein, Zehavi, \&
  et~al.}]{Eisenstein2005}
Eisenstein, D.~J., Zehavi, I., \& et~al., D. W.~H. 2005, Astrophys. J., 633,
  560

\bibitem[{Fosalba {et~al.}(2005)Fosalba, Pan, \& Szapudi}]{Fosalba2005}
Fosalba, P., Pan, J., \& Szapudi, I. 2005, Astrophys. J., 632, 29

\bibitem[{Frieman \& Gaztanaga(1994)}]{Frieman1994}
Frieman, J.~A. \& Gaztanaga, E. 1994, Astrophys. J., 425, 392

\bibitem[{Frieman \& Gaztanaga(1999)}]{Frieman1999}
Frieman, J.~A. \& Gaztanaga, E. 1999, Astrophys. J., 521, L83

\bibitem[{Giocoli {et~al.}(2010)Giocoli, Bartelmann, Sheth, \&
  Cacciato}]{Giocoli2010}
Giocoli, C., Bartelmann, M., Sheth, R.~K., \& Cacciato, M. 2010, Mon. Not. R.
  Astron. Soc.,

\bibitem[{Goroff {et~al.}(1986)Goroff, Grinstein, Rey, \& Wise}]{Goroff1986}
Goroff, M.~H., Grinstein, B., Rey, S.-J., \& Wise, M.~B. 1986, Astrophys. J.,
  311, 6

\bibitem[{Guillet {et~al.}(2010)Guillet, Teyssier, \& Colombi}]{Guillet2010}
Guillet, T., Teyssier, R., \& Colombi, S. 2010, Mon. Not. R. Astron. Soc., 405,
  525

\bibitem[{Guo \& Jing(2009)}]{Guo2009}
Guo, H. \& Jing, Y.~P. 2009, Astrophys. J., 698, 479

\bibitem[{Hamilton {et~al.}(1991)Hamilton, Kumar, Lu, \&
  Matthews}]{Hamilton1991}
Hamilton, A. J.~S., Kumar, P., Lu, E., \& Matthews, A. 1991, Astrophys. J.
  Lett., 374, 1

\bibitem[{Jain \& Bertschinger(1996)}]{Jain1996}
Jain, B. \& Bertschinger, E. 1996, Astrophys. J., 456, 43

\bibitem[{Jing \& Suto(2002)}]{Jing2002}
Jing, Y.~P. \& Suto, Y. 2002, Astrophys. J., 574, 538

\bibitem[{Kayo {et~al.}(2004)Kayo, Suto, Nichol, Pan, Szapudi, \&
  et~al.}]{Kayo2004}
Kayo, I., Suto, Y., Nichol, R.~C., {et~al.} 2004, Publ. Astron. Soc. Japan, 56,
  415

\bibitem[{Komatsu {et~al.}(2009)Komatsu, Dunkley, \& et~al.}]{Komatsu2009}
Komatsu, E., Dunkley, J., \& et~al., M. R.~N. 2009, Astrophys. J. Suppl., 180,
  330

\bibitem[{{Lesgourgues} {et~al.}(2009){Lesgourgues}, {Matarrese}, {Pietroni},
  \& {Riotto}}]{Lesgourgues2009}
{Lesgourgues}, J., {Matarrese}, S., {Pietroni}, M., \& {Riotto}, A. 2009,
  \jcap, 6, 17

\bibitem[{Ma \& Fry(2000b)}]{Ma2000b}
Ma, C.-P. \& Fry, J.~N. 2000b, Astrophys. J., 543, 503

\bibitem[{Massey {et~al.}(2007)Massey, Rhodes, \& et~al.}]{Massey2007}
Massey, R., Rhodes, J., \& et~al., A.~L. 2007, Astrophys. J. Supp., 172, 239

\bibitem[{Matarrese \& Pietroni(2007)}]{Matarrese2007}
Matarrese, S. \& Pietroni, M. 2007, JCAP, 6, 26

\bibitem[{{Matsubara}(2008{\natexlab{a}})}]{Matsubara2008a}
{Matsubara}, T. 2008{\natexlab{a}}, \prd, 78, 083519

\bibitem[{{Matsubara}(2008{\natexlab{b}})}]{Matsubara2008}
{Matsubara}, T. 2008{\natexlab{b}}, \prd, 77, 063530

\bibitem[{McClelland \& Silk(1977)}]{McClelland1977}
McClelland, J. \& Silk, J. 1977, Astrophys. J., 217, 331

\bibitem[{Munshi {et~al.}(2008)Munshi, Valageas, van Waerbeke, \&
  Heavens}]{Munshi2008}
Munshi, D., Valageas, P., van Waerbeke, L., \& Heavens, A. 2008, Phys. Rep.,
  462, 67

\bibitem[{Navarro {et~al.}(1997)Navarro, Frenk, \& White}]{NFW1997}
Navarro, J.~F., Frenk, C.~S., \& White, S. D.~M. 1997, Astrophys. J., 490, 493

\bibitem[{Nishimichi {et~al.}(2007)Nishimichi, Kayo, Hikage, Yahata, Taruya, \&
  et~al.}]{Nishimichi2007}
Nishimichi, T., Kayo, I., Hikage, C., {et~al.} 2007, Publ. Astron. Soc. Japan,
  59, 93

\bibitem[{Nishimichi {et~al.}(2009)Nishimichi, Shirata, Taruya, Yahata, Saito,
  \& et~al.}]{Nishimichi2009}
Nishimichi, T., Shirata, A., Taruya, A., {et~al.} 2009, Publ. Astron. Soc.
  Japan, 61, 321

\bibitem[{Nishimichi {et~al.}(2010)Nishimichi, Taruya, Koyama, \&
  Sabiu}]{Nishimichi2010}
Nishimichi, T., Taruya, A., Koyama, K., \& Sabiu, C. 2010, JCAP, 7, 2

\bibitem[{Pan {et~al.}(2007)Pan, Coles, \& Szapudi}]{Pan2007}
Pan, J., Coles, P., \& Szapudi, I. 2007, Mon. Not. R. Astron. Soc., 382, 1460

\bibitem[{Peebles(1974)}]{Peebles1974}
Peebles, P. J.~E. 1974, Astron. Astrophys., 32, 391

\bibitem[{Peebles(1980)}]{Peebles1980}
Peebles, P. J.~E. 1980, The large scale structure of the universe (Princeton:
  Princeton university press)

\bibitem[{Peebles(1982)}]{Peebles1982}
Peebles, P. J.~E. 1982, Astrophys. J. Lett., 263, 1

\bibitem[{Pietroni(2008)}]{Pietroni2008}
Pietroni, M. 2008, JCAP, 10, 36

\bibitem[{Scherrer \& Bertschinger(1991)}]{Scherrer1991}
Scherrer, R.~J. \& Bertschinger, E. 1991, Astrophys. J., 381, 349

\bibitem[{Schneider \& Bartelmann(1995)}]{Schneider1995}
Schneider, P. \& Bartelmann, M. 1995, Mon. Not. R. Astron. Soc., 273, 475

\bibitem[{Scoccimarro(1997)}]{Scoccimarro1997}
Scoccimarro, R. 1997, Astrophys. J., 487, 1

\bibitem[{Scoccimarro {et~al.}(1998)Scoccimarro, Colombi, Fry, Fireman, Hivon,
  \& Melott}]{Scoccimarro1998b}
Scoccimarro, R., Colombi, S., Fry, J.~N., {et~al.} 1998, Astrophys. J., 496,
  586

\bibitem[{Scoccimarro \& Couchman(2001)}]{Scoccimarro2001a}
Scoccimarro, R. \& Couchman, H. M.~P. 2001, Mon. Not. R. Astron. Soc., 325,
  1312

\bibitem[{Scoccimarro \& Frieman(1996)}]{Scoccimarro1996}
Scoccimarro, R. \& Frieman, J. 1996, Astrophys. J. Supp., 105, 37

\bibitem[{Scoccimarro {et~al.}(2001)Scoccimarro, Sheth, Hui, \&
  Jain}]{Scoccimarro2001}
Scoccimarro, R., Sheth, R.~K., Hui, L., \& Jain, B. 2001, Astrophys. J., 546,
  20

\bibitem[{Sefusatti {et~al.}(2010)Sefusatti, Crocce, \&
  Desjacques}]{Sefusatti2010}
Sefusatti, E., Crocce, M., \& Desjacques, V. 2010, Mon. Not. R. Astron. Soc.,
  406, 1014

\bibitem[{Sefusatti {et~al.}(2007)Sefusatti, Crocce, Pueblas, \&
  Scoccimarro}]{Sefusatti2006}
Sefusatti, E., Crocce, M., Pueblas, S., \& Scoccimarro, R. 2007, Phys. Rev. D,
  74, 023522

\bibitem[{Sefusatti \& Komatsu(2007)}]{Sefusatti2007}
Sefusatti, E. \& Komatsu, E. 2007, Phys. Rev. D, 76, 083004

\bibitem[{Sheth \& Jain(2003)}]{Sheth2003}
Sheth, R.~K. \& Jain, B. 2003, Mon. Not. R. Astron. Soc., 345, 529

\bibitem[{Smith {et~al.}(2003)Smith, Peacock, Jenkins, White, Frenk, Pearce,
  Thomas, Efstathiou, \& Couchman}]{Smith2003}
Smith, R.~E., Peacock, J.~A., Jenkins, A., {et~al.} 2003, Mon. Not. R. Astron.
  Soc., 341, 1311

\bibitem[{Smith {et~al.}(2008)Smith, Sheth, \& Scoccimarro}]{Smith2008}
Smith, R.~E., Sheth, R.~K., \& Scoccimarro, R. 2008, Phys. Rev. D, 78, 023523

\bibitem[{Smith {et~al.}(2006)Smith, Watts, \& Sheth}]{Smith2006}
Smith, R.~E., Watts, P. I.~R., \& Sheth, R.~K. 2006, Mon. Not. R. Astron. Soc.,
  365, 214

\bibitem[{Takada \& Jain(2003)}]{Takada2003}
Takada, M. \& Jain, B. 2003, Mon. Not. R. Astron. Soc., 340, 580

\bibitem[{Taruya \& Hiramatsu(2008)}]{Taruya2008}
Taruya, A. \& Hiramatsu, T. 2008, Astrophys. J., 674, 617

\bibitem[{Taruya {et~al.}(2009)Taruya, Nishimichi, Saito, \&
  Hiramatsu}]{Taruya2009}
Taruya, A., Nishimichi, T., Saito, S., \& Hiramatsu, T. 2009, Phys. Rev. D, 80,
  123503

\bibitem[{Taylor \& Hamilton(1996)}]{Taylor1996}
Taylor, A.~N. \& Hamilton, A. J.~S. 1996, Mon. Not. R. Astron. Soc., 282, 767

\bibitem[{Tegmark {et~al.}(2006)Tegmark, Eisenstein, \& et~al.}]{Tegmark2006}
Tegmark, M., Eisenstein, D., \& et~al., M.~S. 2006, Phys. Rev. D, 74, 123507

\bibitem[{Valageas(1999)}]{Valageas1999}
Valageas, P. 1999, Astron. Astrophys., 347, 757

\bibitem[{Valageas(2002)}]{Valageas2002V}
Valageas, P. 2002, Astron. Astrophys., 382, 477

\bibitem[{Valageas(2004)}]{Valageas2004}
Valageas, P. 2004, Astron. Astrophys., 421, 23

\bibitem[{Valageas(2007{\natexlab{a}})}]{Valageas2007a}
Valageas, P. 2007{\natexlab{a}}, Astron. Astrophys., 465, 725

\bibitem[{Valageas(2007{\natexlab{b}})}]{Valageas2007b}
Valageas, P. 2007{\natexlab{b}}, Astron. Astrophys., 476, 31

\bibitem[{Valageas(2008)}]{Valageas2008}
Valageas, P. 2008, Astron. Astrophys., 484, 79

\bibitem[{Valageas(2009)}]{Valageas2009d}
Valageas, P. 2009, Astron. Astrophys., 508, 93

\bibitem[{Valageas(2011)}]{Valageas2010a}
Valageas, P. 2011, Astron. Astrophys., 526, A67

\bibitem[{{Valageas} \& {Nishimichi}(2011)}]{Valageas2010b}
{Valageas}, P. \& {Nishimichi}, T. 2011, \aap, 527, A87+

\bibitem[{Verde {et~al.}(2002)Verde, Heavens, Percival, Matarrese, Baugh, \&
  et~al.}]{Verde2002}
Verde, L., Heavens, A.~F., Percival, W.~J., {et~al.} 2002, Mon. Not. R. Astron.
  Soc., 335, 432

\bibitem[{Wang {et~al.}(2004)Wang, Yang, Mo, van~den Bosch, \& Chu}]{Wang2004}
Wang, Y., Yang, X., Mo, H.~J., van~den Bosch, F.~C., \& Chu, Y. 2004, Mon. Not.
  R. Astron. Soc., 353, 287

\end{thebibliography}

\end{document}